\documentclass[structabstract]{aa}   
\usepackage{graphicx}
\usepackage{txfonts}
\usepackage[round]{natbib}

\def\Vs{V_{\rm s}}
\def\dr{{\rm d}}
\def\Vi{V_{\rm i}}
\def\Vn{V_{\rm n}}
\def\kms{km\,s${}^{-1}$ }
\def\lesssim{\mathrel{\hbox{\rlap{\hbox{\lower4pt\hbox{$\sim$}}}\hbox{$<$}}}}
\def\gtrsim{\mathrel{\hbox{\rlap{\hbox{\lower4pt\hbox{$\sim$}}}\hbox{$>$}}}}

\begin{document}
   \title{Shocks in dense clouds}

   \subtitle{IV. Effects of grain-grain processing on molecular line emission}

   \author{S. Anderl,
          \inst{1}
          V. Guillet,
           \inst{2}
          G. Pineau des For\^ets
          \inst{2,3}
          \and
          D. R. Flower
          \inst{4}
          }

   \institute{Argelander-Institut f\"ur Astronomie, Universit\"at Bonn, Auf dem H\"ugel 71, 53121 Bonn, Germany\\
              \email{sanderl@astro.uni-bonn.de}
         \and
             Institut d'Astrophysique Spatiale (IAS), UMR 8617, CNRS, B\^atiment 121, Universit\'e Paris Sud 11, 91405 Orsay, France
             \and
             LERMA (UMR 8112 du CNRS), Observatoire de Paris, 61 Avenue de l'Observatoire, 75014 Paris, France
              \and
             Physics Department, The University of Durham, Durham DH1 3LE, UK
             }

   \date{Received 4 March 2013; accepted 19 June 2013}

 
  \abstract
   {Grain-grain processing has been shown to be an indispensable ingredient of shock modelling in high density environments. For densities higher than $\sim$10${}^5$ cm${}^{-3}$, shattering becomes a self-enhanced process that imposes severe chemical and dynamical consequences on the shock characteristics. Shattering is accompanied by the vaporization of grains, which can, in addition to sputtering, directly release SiO to the gas phase. Given that SiO rotational line radiation is used as a major tracer of shocks in dense clouds, it is crucial to understand the influence of vaporization on SiO line emission.}
   {We extend our study of the impact of grain-grain processing on C-type shocks in dense clouds. Various values of the magnetic field are explored. We study the corresponding consequences for molecular line emission and, in particular, investigate the influence of shattering and related vaporization on the rotational line emission of SiO.}
   {We have developed a recipe for implementing the effects of shattering and vaporization into a 2-fluid shock model, resulting in a reduction of computation time by a factor $\sim$100 compared to a multi-fluid modelling approach. This implementation was combined with an LVG-based modelling of molecular line radiation transport. Using this combined model we calculated grids of shock models to explore the consequences of different dust-processing scenarios. }
   {Grain-grain processing is shown to have a strong influence on C-type shocks for a broad range of magnetic fields: the shocks become hotter and thinner. The reduction in column density of shocked gas lowers the intensity of molecular lines, at the same time as higher peak temperatures increase the intensity of highly excited transitions compared to shocks without grain-grain processing. For OH the net effect is an increase in line intensities, while for CO and H${}_2$O it is the contrary. The intensity of H${}_2$ emission is decreased in low transitions and increased for highly excited lines. For all molecules, the highly excited lines become sensitive to the value of the magnetic field. 
   Although vaporization increases the intensity of SiO rotational lines, this effect is weakened by the reduced shock width. The release of SiO early in the hot shock changes the excitation characteristics of SiO radiation, although it does not yield an increase in width for the line profiles. To significantly increase the intensity and width of SiO rotational lines, SiO needs to be present in grain mantles.
   
   }
   {}


   \maketitle
%

\section{Introduction}

Shocks are ubiquitous in the interstellar medium, occurring when matter moves into a more rarefied medium at a velocity that exceeds the local sound speed. Depending on the value of the local magnetosonic speed, different types of shocks can be distinguished. The classical shocks are faster than any signal speed in the shocked medium, so the preshock medium is not able to dynamically respond to the shockwave before it arrives. This type of shock is called "J-type" (see e.g. \citet{Hollenbach:1979p19151,McKee:1980p19192}). A different situation can occur if a low degree of ionization and the presence of a magnetic field allow magnetosonic waves to precede the shock. Ions then decouple from the neutrals and are already accelerated in the preshock gas, so that there is no longer a discontinuity in the flow of the ion fluid. In the preshock gas, the ions heat and accelerate the neutrals and broaden the heating region, so that heating and cooling take place simultaneously. The shock transition can then become continuous in the neutral fluid ("C-type shocks") also. Shocks of this type are thicker and less hot than J-type shocks (see \citealt{Draine:1980p539,Draine:1993p1759}). Observations often reveal shocks as bow-shaped structures, with the ambient material being compressed and pushed aside (e.g. \citealt{Nissen:2007p19095,Davis:2009p19094}).

Shocks play an important role in the energy budget of the interstellar medium by determining the energetic feedback of events such as supernovae, stellar winds, cloud-cloud collisions, or expanding HII regions. On the other hand, shocks have a major influence on the chemistry of the interstellar medium. Among the most characteristic chemical tracers of shock waves are species that are typically heavily depleted on dust grains, such as Fe, Mg, or Si. Dust processing occurring in the violent environment of shocked media is able to release these species into the gas phase (e.g. \citealt{Liffman:1989p17725,ODonnell:1997p19216}). The understanding of dust processing in shocks is therefore intimately linked with the theoretical interpretation of characteristic emission lines in environments where shocks occur. 

The processing of dust in shocks can have two different consequences. It either changes the dust-to-gas mass ratio or alters the dust size distribution. The former occurs in the processes of sputtering and vaporization, while the latter is also found with shattering. Sputtering denotes energetic impacts of gas particles on grains that can release species from the grain surfaces into the gas phase. This can happen either at very high temperatures (thermal sputtering) or at high relative gas--grain velocities (inertial sputtering). Sputtering of dust grains has been the subject of many theoretical studies (\citealt{Barlow:1978p17402,Tielens:1994p2924,Jones:1994p2862,May:2000p14538,VanLoo:2012p18556}) and has become an established ingredient of shock models. Shattering, which is the fragmentation of grains due to grain-grain collisions, has been identified as a crucial process for determining the grain-size spectrum of interstellar grains (\citealt{Biermann:1980p17507,Borkowski:1995p17506}). Together with vaporization, which describes the return of grain material to the gas phase following grain-grain collisions, shattering needs to be included in order to account for UV extinction curves (\citealt{Seab:1983p18312}). Moreover, recent infrared and sub-mm observations hint at an overabundance of small dust grains relative to the expected size-distribution in parts of the interstellar medium and thereby stress the importance of dust processing (\citealt{Andersen:2011p17918,PlanckCollaboration:2011p17886}), which has also been associated with MHD turbulence acceleration of dust grains or charge-fluctuation induced acceleration (e.g. \citealt{Ivlev:2010p17923,Hirashita:2010p17926}).

A rigorous theoretical description of grain-grain collisions requires a detailed description of shock waves in solids, which had not been undertaken before the work of \citet{Tielens:1994p2924}, earlier studies having relied on much simplified models (e.g. \citealt{Seab:1983p18312}). The effect of the microphysics of grain shattering and its effects on the grain size distribution in J-type shocks in the warm intercloud medium was studied by \citet{Jones:1996p2864}. \citet{Slavin:2004p3051} extended this work by explicitly following individual trajectories of the grains, considered as test particles.  

In a series of papers, \citet[hereafter Papers I, II and III]{Guillet:2007p2767,Guillet:2009p3498,Guillet:2011p15512} have shown that a multi-fluid approach to the dust dynamics, together with a detailed calculation of the grain charge distribution, shows shattering and the accompanying vaporization to be indispensable ingredients of shock models. For C-type shocks at preshock densities higher than $\sim$ 10${}^5$ cm${}^{-3}$, shattering becomes dramatically self-enhanced, due to feedback processes: electrons are heavily depleted on to small grain fragments, and the lack of electrons in the gas phase affects the grain dynamics, resulting in even more shattering and production of small grains. Owing to the increase of the total geometrical grain cross-section, these shocks become much hotter and narrower, and vaporization becomes important for the release of depleted species into the gas phase. Both the dynamical and the chemical consequences of shattering affect the predicted observational characteristics, compared with models in which shattering and vaporization are neglected. Given that conditions favoring C-shocks are frequently found in dense clouds and Bok globules, it is necessary to evaluate the observational consequences of shock models including grain-grain processing for a proper understanding of massive star formation and interactions of supernova remnants with molecular cloud cores (\citealt{Cabrit:2012p18561}).

The aforementioned studies leave open a series of issues that we aim to address in this paper. The three main questions are:
\begin{itemize}
\item  How do the effects of shattering and vaporization change if the magnetic field strength is varied?
\item What are the consequences of shattering for molecular line emission?
\item How do shattering and associated vaporization of SiO and C influence the SiO and atomic carbon line emission?
\end{itemize}

In order to answer the last two questions, it is necessary to introduce a detailed treatment of radiative transport. However, due to the numerical complexity of the multifluid treatment of Papers I--III, a self-consistent merging of this multifluid model with an LVG treatment of molecular line emission would be technically difficult. It would also prevent the resulting model from being a practical analysis tool whenever large grids of models are required.
Therefore, we have developed a method for implementing the effects of shattering and vaporization into a 2-fluid shock model that is sufficiently general to be applied to any similar model. We incorporated the main features, neglecting all the minor details of grain-grain processing; the resulting saving in computation time amounts to a factor of  $\sim$100. With this model, the consequences of shattering and vaporization on the molecular line emission of C-type shocks can be studied in detail, using a self-consistent treatment of the line transfer (\citealt{Flower:2010p18961}). In the context of the emission of SiO from C-type shocks (\citealt{Schilke:1997p2037,Gusdorf:2008p974,Gusdorf:2008p1617}), the inclusion of grain shattering and vaporization, along with the line transfer, should improve significantly our ability to interpret the observations correctly.

Our paper is structured as follows. In Sect. \ref{Sect2}, we introduce our model, which extends the work of \citet{Flower:2010p18961} by including the shattering and vaporization of grains and molecular line transfer. We use this model to demonstrate the effect of shattering on the shock structure for different values of the magnetic field in the preshock gas (Sect. \ref{Sect3}). In Sect. \ref{Sect4}, we consider observational diagnostics, with the main emphasis being on the rotational line emission of SiO. In addition,  we investigate the influence of vaporization on the [C~I] emission lines. The results are discussed and summarized in Sect. \ref{Sect5}.


\section{Our model}
\label{Sect2}

In order to study the effect of shattering and vaporization on molecular line emission, we have built on the findings of Papers I - III, where shattering and vaporization are described within a multi-fluid formalism for the dust grains. These results needed to be transferred to a 2-fluid formulation, as used in the model of \citealt{Flower:2010p18961} (hereafter FPdF10), whose model includes a detailed treatment of molecular line radiative transfer, in the presence of the cosmic microwave background radiation. In the present Section, we first describe the treatment of dust in the model of FPdF10 and then summarize the multi-fluid treatment of dust and its implications, finally outlining how we introduced the effects of shattering and vaporization into the model of FPdF10.

\subsection{Two-fluid treatment of dust}
\label{subsec2.1}

The way in which dust is treated in the LVG-model of FPdF10 derives from the work of \citealt{Flower:2003p1558} (hereafter FPdF03). This one-dimensional, steady-state, 2-fluid model\footnote{Although electrons and ions are treated as one dynamical fluid, their temperatures are calculated separately.} of plane--parallel C-type shocks solves the magneto-hydrodynamical equations in parallel with a large chemical network, comprising more than 100 species and approximately 1000 reactions, including ion-neutral and neutral-neutral reactions, charge transfer, radiative and dissociative recombination. Furthermore the populations of the H${}_2$ ro-vibrational levels are computed at each integration step, along with the populations of the rotational levels of other important coolants, such as CO, OH and H${}_2$O. In these cases, the molecular line transfer is treated by means of the large velocity gradient approximation (see also Sect. \ref{Sect4}). `Dust' is included in the forms of polycyclic aromatic hydrocarbons (PAHs), represented by C${}_{54}$H${}_{18}$, and large grains from bulk carbonaceous material and silicates (specifically olivine, MgFeSiO${}_4$). The large grains are assumed to have a power law size distribution, $\dr n_{\rm g}(a)/\dr a \propto a^{-3.5}$ (Mathis et al. 1977), and radii in the range 100 to 3000 \r{A} (0.3~${\rm \mu}$m). In the preshock medium, the grain cores are covered by ice mantles, consisting of the chemical species listed in Table 2 of FPdF03 and including H${}_2$O, CO and CO${}_2$. Both PAHs and large grains exist as neutral, singly positively and singly negatively charged species. The physical treatment of dust comprises:
\begin{itemize}
\item determination of the grain charge distribution, limited to $Z~=-1, 0, 1$;
\item sputtering of grain cores and mantles due to grain-gas collisions;
\item removal of mantles by cosmic ray desorption;
\item build-up of mantles by adsorption of gas-phase species in the postshock gas.
\end{itemize}

\subsection{Multi-fluid treatment of dust}
\label{subsec2.2}

To properly account for the effects of grain-grain collisions in shocks, it is necessary to introduce a multi-fluid description of the dust dynamics, in which dust grains are treated as test particles. In Papers I-III the dust size distribution was modeled by the use of discrete bins, with grain sizes ranging from 5 to 3000 \r{A}. PAHs were not included as separate species, distinct from the dust grains (as in the two-fluid model), but were incorporated into the dust size distribution. The equations describing the 2-D grain dynamics and complete charge distribution were integrated for each individual bin size, independently for silicate and carbon grains. 

While the shock structure is not much affected by this more accurate treatment of the grain dynamics and charging (Paper I, Appendix D\footnote{Although changes to the re-accretion of mantle species in the multi-fluid treatment give rise to a slightly different temperature profile in the postshock gas.}), the introduction of dust shattering in grain-grain collisions and the corresponding changes to the grain size distribution have major effects on shocks in dense clouds with a low degree of ionization (Paper III). 
When grains collide with a velocity greater than 1.2~\kms  (carbon grains) or 2.7~\kms  (silicate grains), a fraction of their mass, which increases with velocity, is shattered into smaller fragments. At higher velocities, another fraction is vaporized and released into the gas phase (the numerical treatment follows the models of \citet{Tielens:1994p2924} and \citet{Jones:1996p2864}, where the vaporization threshold is $\sim$~19~km\,s${}^{-1}$). 

   \begin{figure}
   \centering
   \includegraphics[width=9cm]{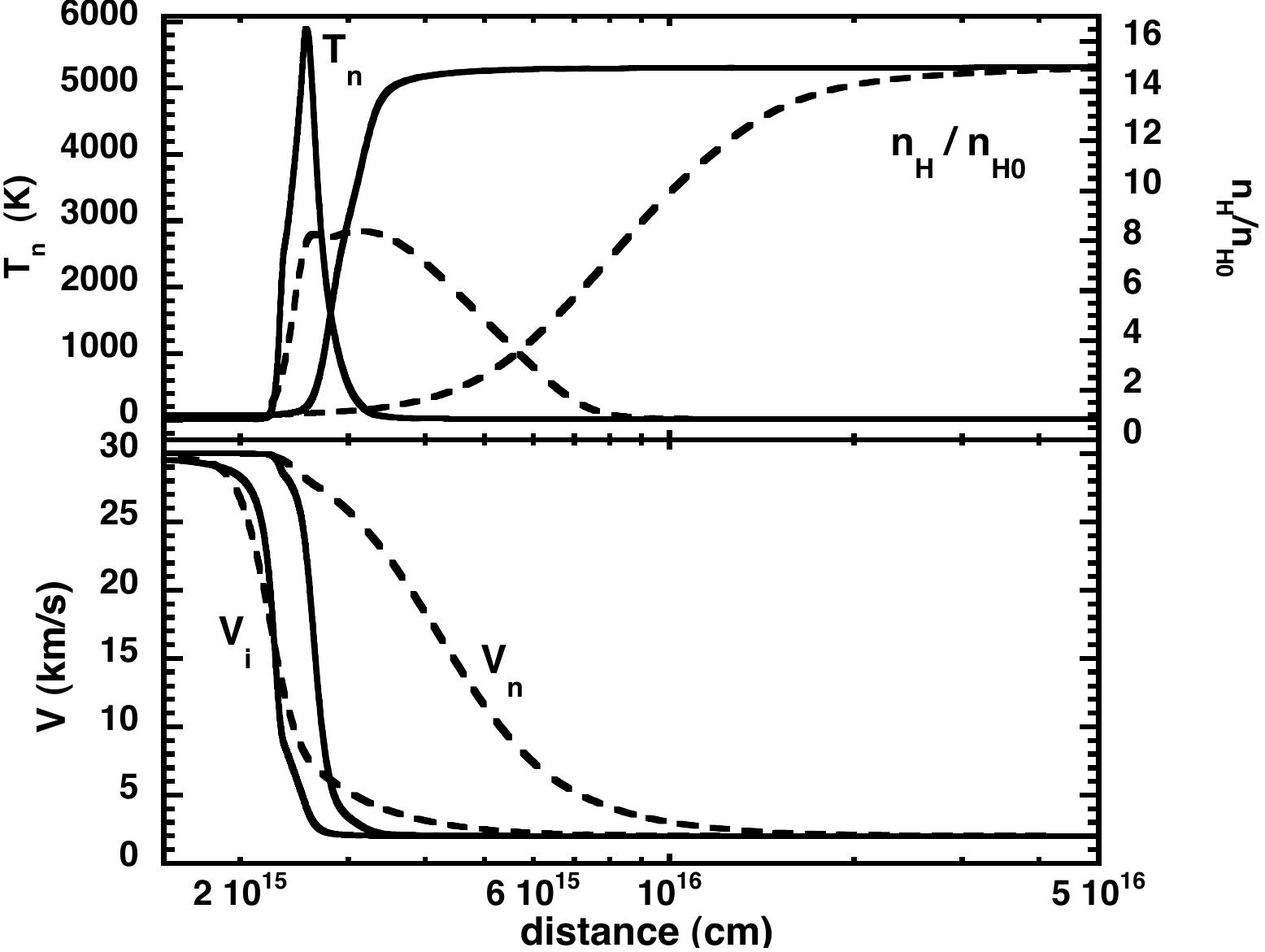} 
      \caption{Temperature and density (upper panel) and velocity profiles in the shock frame (lower panel) for a 30 km s$^{-1}$ C-type shock with a magnetic field parameter $b=1.5$ and a preshock density of $n{}_{\rm H}$ = 10${}^5$ cm${}^{-3}$. The full curves show the results obtained when grain-grain processing is treated as in Paper III; the broken curves are obtained by following the approach of FPdF03. 
              }
         \label{Fig1_dist}
   \end{figure}

Paper III demonstrated that there is a marked shift of the dust size distribution towards smaller grains when the grain dynamics, charging and evolution are coupled self-consistently with the shock dynamics and chemistry. The most dramatic consequence is an increase of the total geometrical grain cross-section -- which governs the coupling between the neutral and charged fluids -- by a factor\footnote{This factor refers to the total cross-section of grain cores. The re-accretion of grain mantles in the postshock gas leads to a much larger increase in the total cross-section, by a factor of $\sim$100 (see. Fig.~\ref{FigA1_shattering}). However, this large increase is irrelevant to the shock dynamics, as it happens where the momentum transfer between the ionized and the neutral fluids is almost complete ($T\sim 100\,$K).} of $\sim$10. The shock becomes narrower by a factor of $\sim$4, and the peak temperature increases by a factor of 1.5--2 (see Fig.~\ref{Fig1_dist}). The small grain fragments deplete ions and electrons from the gas until grains become the dominant charge carriers, a situation known as a {\it dusty plasma} (\citealt{Fortov:2005}). The strong effect of shattering, for preshock densities of 10${}^5$ cm${}^{-3}$ and higher, can be understood in terms of the feedback processes described in Paper III. At these high preshock densities, it turns out that vaporization related to shattering becomes important, if not dominant, compared to sputtering.

   \begin{figure}
   \centering
   \includegraphics[width=9cm]{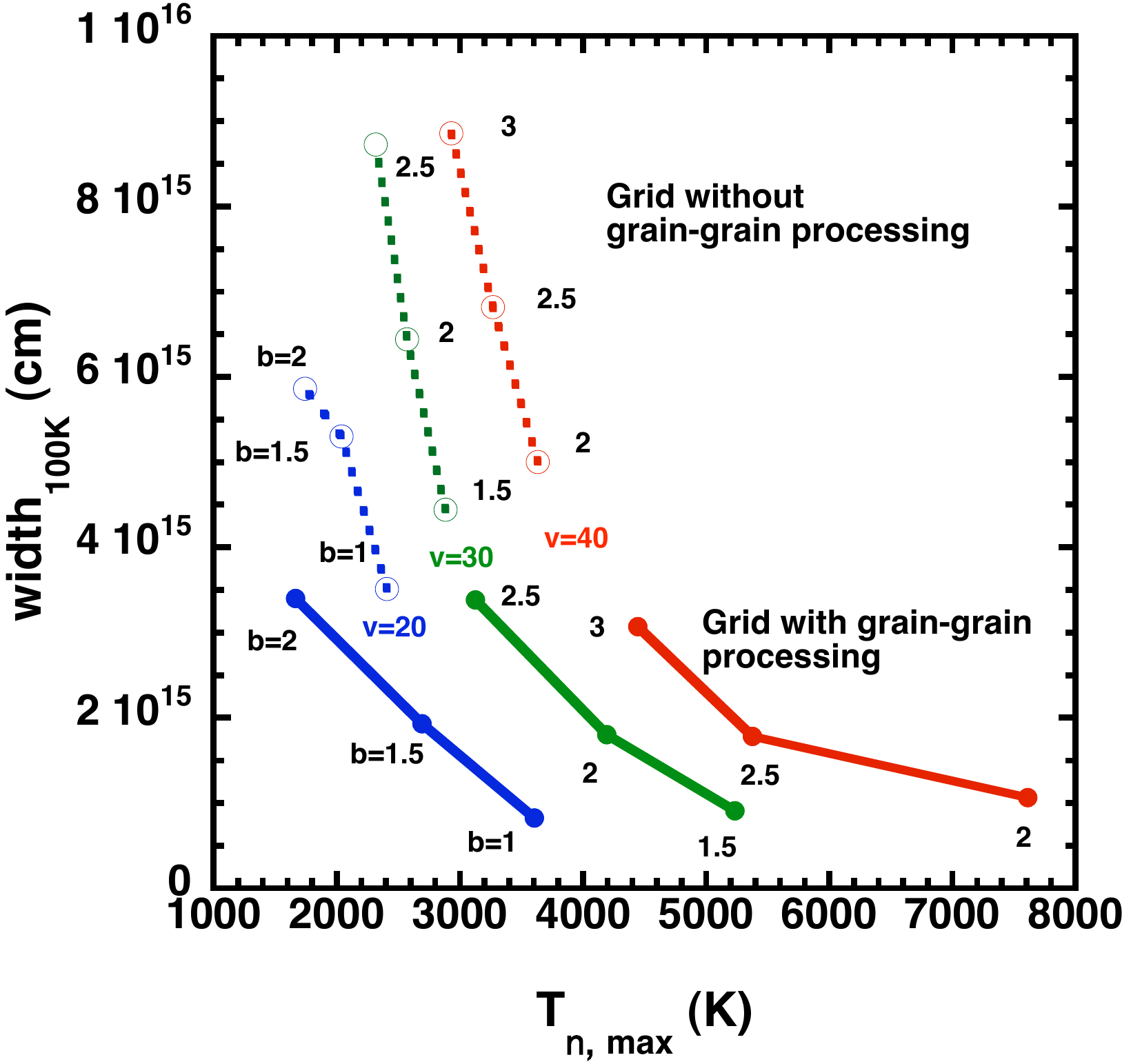} 
      \caption{Peak temperature of the neutral fluid and the shock width (in cm), to $T = 100$~K in the cooling flow, for the grid of models without (broken lines) and with (full lines) grain-grain processing. The preshock density is 10${}^5$ cm${}^{-3}$, and the shock velocities are 20 \kms (blue), 30 \kms (green) and 40 \kms (red) for various values of the magnetic field parameter, $b$, in $B(\mu {\rm G})=b\sqrt{n_{\rm H}({\rm cm}^{-3})}$. 
              }
         \label{Fig2_Twidth}
   \end{figure}

 \subsection{Implementation of shattering}
 \label{subsec2.3}
 
 As described in Sect. \ref{subsec2.2}, there are two ways in which shattering affects the shock wave: first it increases the collisional dust cross-section, and second it lowers the degree of ionization of the gas. In the model of FPdF10, dust as a dynamical species and dust as a chemically reacting species were treated separately. Our implementation and validation of shattering in the FPdF10 model are described in Appendix A.1. Here we only summarize our basic approach.
 
\begin{itemize}
\item The dynamical effects of shattering can be simulated by incorporating the change in the total grain cross section $\langle n\sigma \rangle$, computed with the model of Paper III; this is done by multiplying $\langle n\sigma \rangle$ by an additional factor that varies with the spatial coordinate, $z$, through the shock wave. This factor is modelled as an implicit function of the compression of the ionized fluid; its value is 1 in the preshock gas, and its maximum value -- derived from the multi-fluid model -- is attained when shattering is completed. It is possible to use linear fits of the parameters that are required.
\item The chemical effects of shattering, i.e. the consequences for grains as charged species, as determined by grain-charging reactions, need to be considered separately. In order to model the influence of shattering on the abundances of charged grains, a shattering source-term is introduced, whose value is consistent with $\langle n\sigma \rangle$. 
\end{itemize}

\subsection{Implementation of vaporization}
\label{subsec2.4}

Grain-grain collisions lead not only to shattering of grains, and consequent changes in the shock structure and degree of ionization, but also to their vaporization, when the impact velocity is higher than the vaporization threshold of $\sim$ 19 km\,s${}^{-1}$. We assume that the silicon released through vaporization is in the form of SiO (\citealt{1996GeCoA..60.1445N,Wang:1999}).
We have implemented the effect of SiO vaporization in a similar way as for shattering\footnote{We emphasize the effects of SiO vaporization because the fine-structure lines of atomic carbon are optically thin and do not require an LVG treatment.}; the details are given in Appendix A.2. Our procedure involves introducing an additional term in the rate of creation of gas-phase SiO from Si and O in the grain cores. Thus, vaporization is incorporated as a new type of pseudo-chemical reaction. The rate of creation of SiO through vaporization is determined by parameterizing the results of the multi-fluid models. 

Another modification of the FPdF10 model concerns the mantle thickness, which is now calculated as described in Paper I, Appendix B. Our simplified, but self-consistent, treatment of shattering and vaporization reduces the computation time by a factor of more than 100.


\section{The influence of grain-grain processing on the shock structure}
\label{Sect3}

\subsection{The grid of models}
\label{subsec3.1}

Having implemented shattering and vaporization in the model of FPdF10, we first study the dependence of grain-grain processing and feedback on the strength of the transverse magnetic field. The analysis of Paper III was restricted to shocks propagating at the critical velocity, which defines the fastest possible C-type shock for a given magnetic field (or, conversely, the lowest value of the magnetic field possible for a given shock velocity). This critical velocity is determined by the condition that the shock velocity should be only slightly smaller than the velocity of magnetosonic waves,
\begin{displaymath}
V_{\rm ms}=\frac{B}{\sqrt{4\pi \rho_{\rm c}}}
\end{displaymath}
where $\rho_{\rm c}$ is the mass density of matter that is strongly coupled to the magnetic field. We assumed that the magnetic field strength, $B$, in dense clouds scales with the total proton density, $n{}_{\rm H}$, as $B(\mu {\rm G})=b\sqrt{n_{\rm H}({\rm cm}^{-3})}$ (\citealt{Crutcher:1999p19138}), which implies that the magnetic energy density is proportional to $n{}_{\rm H}$. The assumed power-law exponent of 0.5 is somewhat lower than the more recent value of 0.65$\pm$0.05, given by \citet{Crutcher:2012p19030}, but is probably consistent with the uncertainties associated with its deduction from measurements of Zeeman splitting. The restriction to critical shocks was justified in Paper III by the fact that the observations will be dominated by the fastest shocks, if present. However, to assess the importance of shattering in the dense interstellar medium, it is necessary to establish whether shattering is still a relevant factor when the magnetic field is higher and the (charged) grains are better protected by their coupling to the magnetic field.
Shattering is sensitive to the preshock density because the drag force becomes more important, relative to the Lorentz force, as the density increases. At high densities, large grains decouple from the magnetic field and exhibit a large velocity dispersion in the initial part of the shock wave, where they undergo destruction. Furthermore, the degree of ionization is lower in higher density gas, so the feedback due to the depletion of electrons is enhanced. 

We have studied shocks with a preshock density of 10$^5$ cm${}^{-3}$. We consider only this value because, on the one hand, it was shown that the change in the grain size distribution due to shattering is negligible for a preshock density of 10$^4$ cm${}^{-3}$, and, on the other hand, the strength of the feedback from shattering at higher densities prevents the multi-fluid model of Paper III from converging at a preshock density of 10$^6$ cm${}^{-3}$, and hence there are no results with which to compare at this density. We have restricted our calculations to shocks that do not fully dissociate H${}_2$ (shock velocity $\Vs \le$ 50km\,s${}^{-1}$), in practice to 20, 30 and 40 km\,s${}^{-1}$. 

The values of the magnetic field that we considered were determined by two considerations. First, in order to study critical shocks, we adopted the corresponding (minimum) values of $b$ (see Paper III); and, second, we varied $b$ for a given shock velocity in order to study the influence of variations in the magnetic field strength. Thus, we consider three values of the $b$ parameter for each velocity, in steps of 0.5, where the lowest value is close to that for a critical shock; the lowest values are $b=1.0$ for 20 km\,s${}^{-1}$, $b=1.5$ for 30 km\,s${}^{-1}$, and $b=2.0$ for 40 km\,s${}^{-1}$. The corresponding grid of nine models is summarized in Table~\ref{t1}. We note that a comparison of models with different shock velocities and the same magnetic field is possible for the case $b=2.0$.

\begin{table} 
 \caption[]{Parameters defining our grid of models.
 }

\begin{center}
\begin{tabular}{llll}
\hline
            \noalign{\smallskip}

$\Vs$ [km\,s${}^{-1}$] &  $b$ & $B$ [$\mu $G] & $n{}_{\rm H}$ [cm$^{-3}$]\\ 
            \noalign{\smallskip}
\hline
            \noalign{\smallskip}
20 & 1.0 & 316 & 10${}^{5}$ \\
20 & 1.5 & 474 & 10${}^{5}$ \\
20 & 2.0 & 632 & 10${}^{5}$\\
30 & 1.5 &474 & 10${}^{5}$\\
30 & 2.0 & 632 & 10${}^{5}$\\
30 & 2.5 &791 & 10${}^{5}$\\
40 & 2.0 & 632 & 10${}^{5}$\\
40 & 2.5 & 791 & 10${}^{5}$\\
40 & 3.0 & 949 & 10${}^{5}$\\
\hline
\end{tabular}

\end{center}
\label{t1}
\end{table}

\subsection{Hotter and thinner}
\label{subsec3.2}

The most dramatic effect of shattering is the change in the overall shock structure, owing to the increase in the collision cross-section of the dust and the corresponding increase in the coupling between the neutral and charged fluids. Figure 2 shows the peak temperature of the neutral fluid and the shock width (up to the point at which the temperature has fallen to 100~K) for our grid of nine models, as compared with the results obtained without grain-grain processing.

The peak temperature of the models that include shattering increases by a factor of 1.5--2 for the shocks closest to the critical velocity, as in Paper III. Our models show that this increase is smaller for slower shocks, and for shocks with higher magnetic field strengths, because the charged grains are more strongly bound to and protected by the magnetic field. Thus, shocks with $\Vs$~=~20 \kms and $b$=2 have a peak temperature comparable to that found without grain-grain processing. However, the shock width, as determined at the point in the cooling flow at which the gas temperature has fallen to 100~K, is always much smaller for models in which grain-grain processing is included. For the shocks closest to the critical velocity, the width is reduced by a factor of 4--5. The trend is the same as for the peak temperature: the shock widths, as computed with and without grain-grain processing, are most similar for lower velocities and higher magnetic fields, differing by a factor of only 1.7 for $\Vs$~=~20~\kms and $b$=2. 

It is interesting to see the dependence of the shock structure on the transverse magnetic field strength. In models that neglect grain-grain processing, a change in the magnetic field modifies the shock width at almost constant peak temperature, whereas, when grain-grain processing is included, a variation in the magnetic field has a strong affect on the peak temperature. We return to this point in Sect. 4, where we consider the differences between the spectra predicted by these two categories of model; but it is already clear that shattering is significant, even for non-critical shocks, and should be included in steady-state C-type shock models with high preshock densities ($n{}_H\ge$ 10$^5$ cm${}^{-3}$).


 \section{Observational  consequences}
 \label{Sect4}
 
In our model, the molecular line transfer is treated using the large velocity gradient (LVG) approximation (e.g. \citealt{1977A&A....60..303S}), allowing for self-absorption via the escape probability formalism. The LVG method is well adapted to the conditions of shocks, where flow velocities change rapidly. The computation of the molecular energy level populations is performed in parallel with the integration of the chemical and dynamical rate equations, as introduced for CO by \citet{Flower:2009p18313}. The molecular line transfer of H${}_2$O, CH${}_3$OH, NH${}_3$, OH and SiO has since been added (\citealt{Flower:2010p19006,Flower:2010p18961,Flower:2012p18432}. 

The modelling of interstellar shock waves requires many molecular energy levels to be considered. As described in FPdF10, we include levels of H${}_2$O up to an energy of approximately 2000~K above ground. Although this is less than the maximum temperatures that are attained, the high values of the radiative (electric dipole) transition probabilities ensure that the populations of higher, neglected levels remain small, and hence the associated errors in the computed line intensities are modest. Above the maximum temperatures for which the rotational de-excitation coefficients have been calculated, their values are assumed to remain constant (cf. \citealt{Flower:2012p18432}, Appendix A). Rate coefficients for excitation are obtained from the detailed balance relation.

The consequences of grain-grain processes for the molecular line radiation of H${}_2$, H${}_2$O and OH for representative shocks of  $\Vs$~=~30 \kms are briefly discussed in Sect. \ref{subsec4.1} and in more detail in Appendix B. In Sect. \ref{subsec4.2}, we consider the rotational line emission of SiO and, in Sect. \ref{subsec4.3}, the vaporization of carbon.

\subsection{Molecular line emission} 
\label{subsec4.1}

We have seen that including grain-grain processing leads to an increase in the peak shock temperature; this affects the chemistry and gives rise to higher fractional abundances of molecules in excited states. The intensities of lines emitted by highly excited states are thereby enhanced. At the same time, the column density of shocked gas decreases, which tends to reduce the intensity of molecular line emission. The net effect depends on the chemical and spectroscopic properties of the individual molecules.

In Appendix B, we show that the intensities of all transitions of OH increase in shocks faster than 30 \kms that incorporate grain-grain processing; this is due to the temperature sensitivity of OH formation. On the other hand, the intensities of the emission lines of H${}_2$, CO, and H${}_2$O decrease because of the reduction in the column density of shocked material. Furthermore, the inclusion of grain-grain processing introduces a dependence of the intensities of lines from highly excited states on the magnetic field, owing to the variation of the peak shock temperature with the field strength (see Sect. \ref{subsec3.2} and Fig.~\ref{Fig2_Twidth}).

\subsection{The effect of vaporization on SiO emission}
\label{subsec4.2}

       \begin{figure}
   \centering
   \includegraphics[width=9cm]{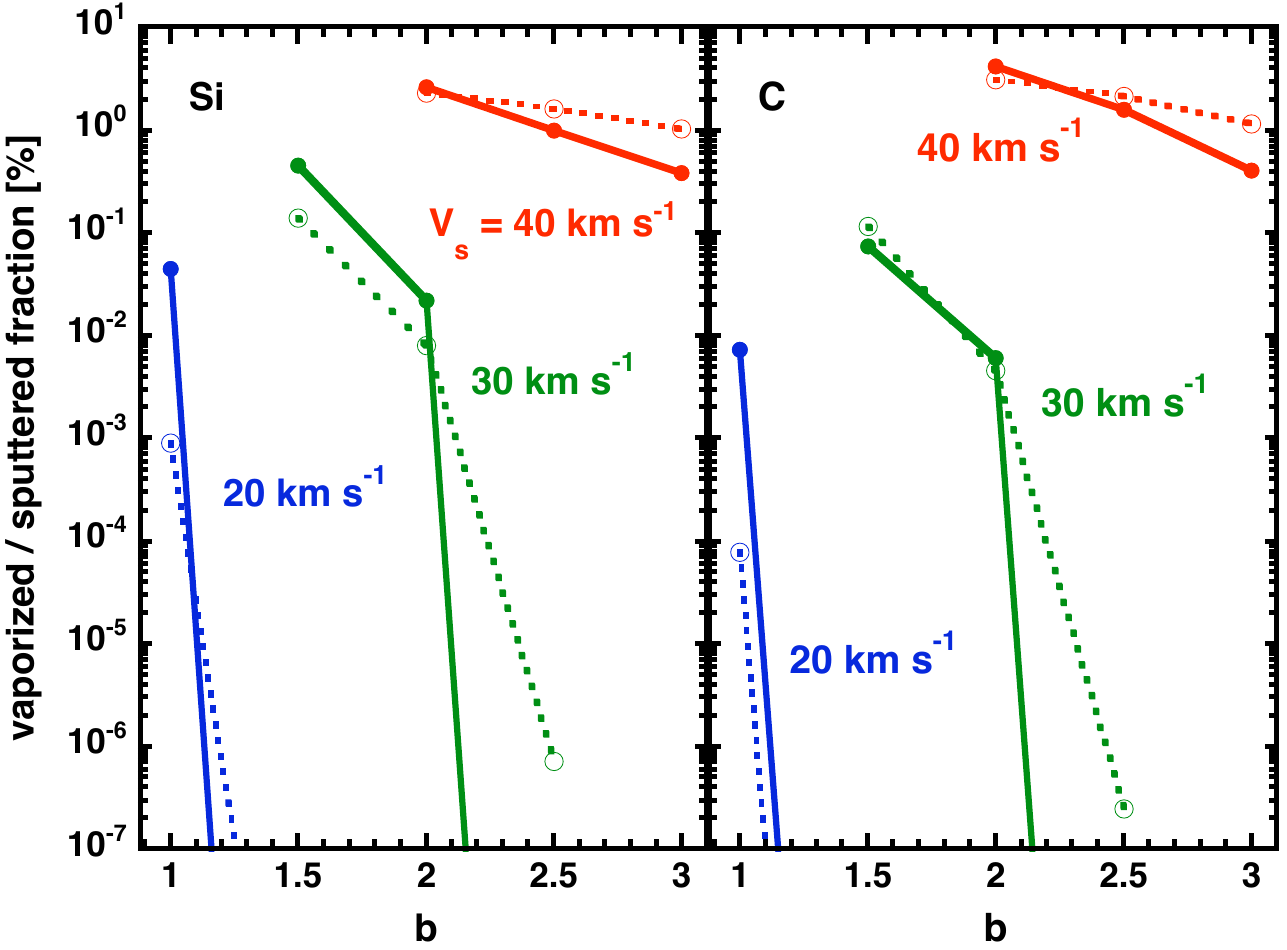}
      \caption{
    The fractions of Si (left panel)  and C (right panel) released from grain cores in model M1 of Table~\ref{t2} by vaporization (full lines) and sputtering (broken lines),  for $\Vs$~=~20~\kms (blue),  $\Vs$~=~30~\kms (green), and $\Vs$~=~40~\kms (red), as functions of the magnetic field parameter, $b$, in $B(\mu {\rm G})=b\sqrt{n_{\rm H}({\rm cm}^{-3})}$.
            }
         \label{Fig3_CSifrac}
   \end{figure}

 SiO is a prominent indicator of shock processing in dense clouds associated with jets and molecular outflows (e.g. \citealt{Bachiller:1991p19103,MartinPintado:1992p19152,1998A&A...333..287G,Nisini:2007p18347,Cabrit:2007p19113}). The first chemically adequate theoretical study of SiO production by sputtering in C-type shocks was conducted by  \citet{Schilke:1997p2037}; this work was pursued subsequently by \citet{Gusdorf:2008p974,Gusdorf:2008p1617}. These studies considered the release, by sputtering, of Si from grain cores and SiO from grain mantles, but not the process of vaporization, which could modify the predicted SiO rotational line profiles. We note that \citet{Cabrit:2012p18561} found that, in the protostellar jet HH212, at least 10\% of elemental Si could be present as gas-phase SiO, if the wind is dusty. To explain such a high value, they hint at the possible importance of grain-grain processing and the release of SiO by vaporization. The sputtering of Si from the grain cores seems unable to account for the observations in this case, as the dynamical timescale of 25~yr is too short for the chemical conversion, in the gas phase, of the sputtered Si into SiO.
 
 \begin{table} 
 \caption[]{Summary of the three dust processing scenarios investigated in order to study the release of SiO into the gas phase. The percentage refers to elemental silicon in the form of SiO in the mantles.}

\begin{center}
\begin{tabular}{llll}
\hline
            \noalign{\smallskip}

Scenario & Si in cores & SiO in mantles & Grain-grain processing\\ 
            \noalign{\smallskip}
\hline
            \noalign{\smallskip}
M1  & yes & no & yes \\
M2  & yes & no & no \\
M3  & yes & 10\% & no \\
\hline
\end{tabular}

\end{center}
\label{t2}
\end{table}

The influence of grain-grain processing on the SiO line emission is twofold. On the one hand, vaporization strongly increases the amount of SiO released from grain cores, at sufficiently high shock speeds. On the other hand, as the process of vaporization is necessarily related to shattering of dust grains, shocks with grain-grain processing have a different structure (cf. Sect. \ref{subsec3.2}). The consequences of the latter effect for the emission by molecules whose abundances are not directly influenced by vaporization are summarized in Sect. \ref{subsec4.1}. 

In order to address the question of how the related effects of shattering and vaporization affect the emission of SiO in C-type shocks, we compare results of shock models, obtained including and excluding grain-grain processing, for three different scenarios, summarized in Table~\ref{t2}. Models corresponding to scenario 1 (M1) were calculated using the implementation of shattering and vaporization described in Sect. 2.3 and 2.4 with the parameters of Table~\ref{ta1}. Models corresponding to scenario 2 (M2) do not include grain-grain processing, but Si is still released by the sputtering of grain cores. Scenario 3 (M3) differs from M2 in that 10\% of the elemental Si is assumed to be in the form of SiO in the grain mantles. In total, there are $9\times3=27$ individual models to be computed.

\subsubsection{The release of SiO through dust processing}
\label{subsubsec4.2.1}

In order to compare the relative importance of vaporization and sputtering, we investigate the release of the Si in grain cores into the gas phase by each of these processes during the passage of a shock wave. The left-hand panel of Figure~\ref{Fig3_CSifrac} shows the fraction of Si eroded from grain cores through sputtering and vaporization in model M1. There is negligible sputtering for $\Vs \le20$ km\,s${}^{-1}$, because the adopted sputtering threshold for refractory grain material is $\sim$ 25 \kms (\citealt{May:2000p14538}). Both the sputtering of Si and the vaporization of SiO are inhibited by the magnetic field. In the case of sputtering, the maximum ion-neutral drift velocity decreases with increasing field strength, whereas, for vaporization, the charged grains are more strongly coupled to the field and hence to the charged fluid (we assume that the magnetic field is `frozen' in the charged fluid). For models M3 (SiO in mantles) with velocities sufficient for mantle sputtering, all the SiO initially in the mantles is released into the gas phase. The rotational line emission of SiO depends not only on the amount of silicon released from the grains but also on the location of its release and the physical conditions prevailing where SiO is present within the shock wave.

    \begin{figure}
   \centering
   \includegraphics[width=9cm]{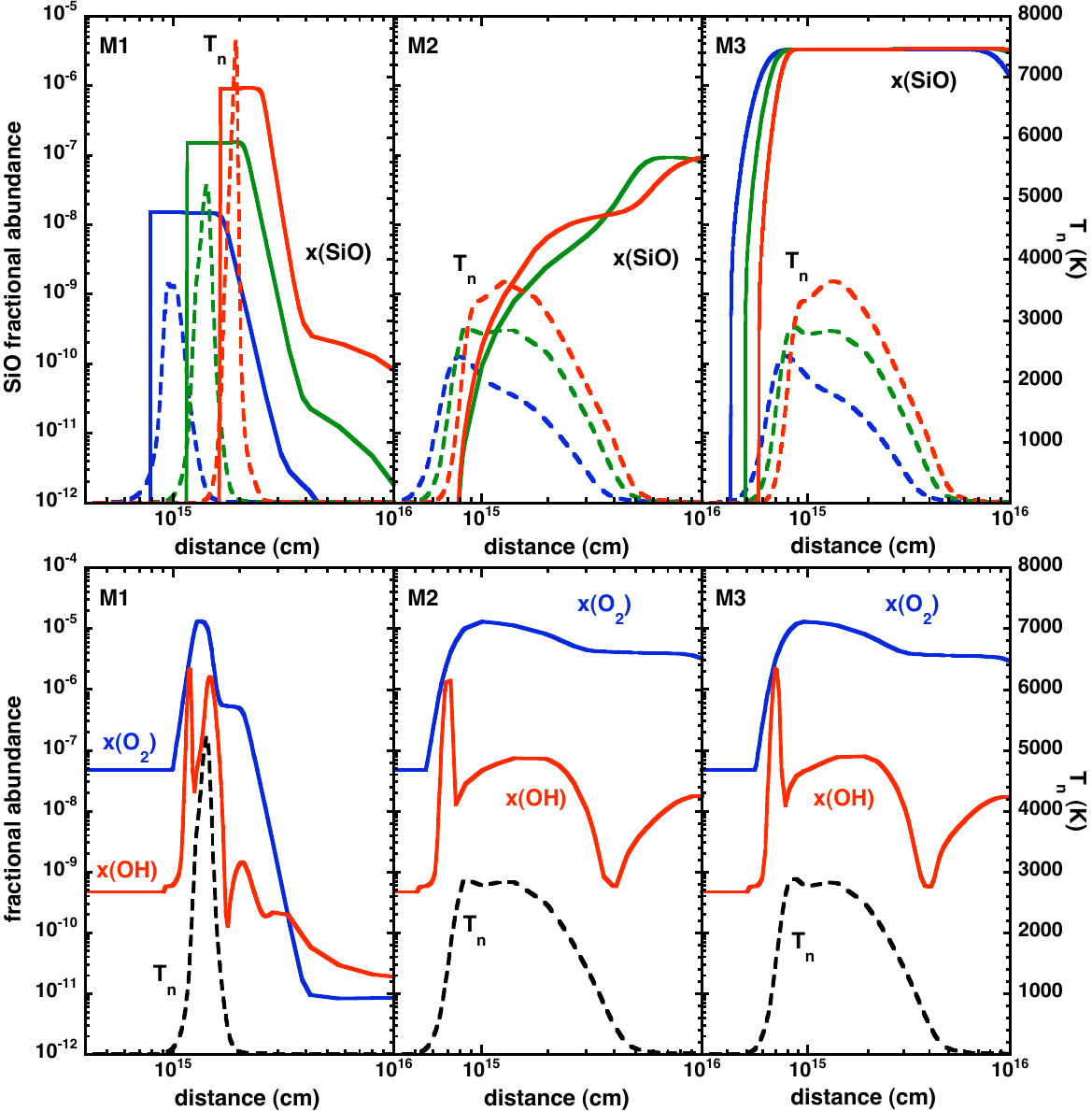}
      \caption{
   Upper panel: Evolution of the fractional abundance of SiO in the gas phase (left ordinate, full curves) and temperature of the neutral fluid (right ordinate, broken curves). From left to right: dust modelling scenarios M1, M2, and M3. The models shown in all three panels are for $\Vs$~=~20~\kms with $b$ = 1.0 (blue), $\Vs$~=~30~\kms with $b$ = 1.5 (green), and $\Vs$~=~40~\kms with $b$ = 2.0 (red). The fractional abundance of SiO in the gas phase is negligible for $\Vs$~=~20~\kms  with $b$ = 1.0 in scenario M2. Lower panel: variation through the shock wave of the gas-phase fractional abundances of O${}_2$ (left ordinate, full blue curves) and OH (left ordinate, full red curves), together with the temperature of the neutral fluid (right ordinate, black broken curves), for $\Vs$~=~30~\kms  and $b$ = 1.5 and each of the dust models M1, M2 and M3 (from left to right).
            }
         \label{Fig4_xSiO}
   \end{figure}

The upper panel of Figure~\ref{Fig4_xSiO} shows the fractional abundance of SiO in the gas phase, for  dust models M1--3 of Table~\ref{t2} and three different values of the shock speed. In model M1, vaporization releases SiO directly into the gas phase in the region where the temperature of the neutral fluid is rising steeply. The higher the shock speed, the more SiO is produced. The large increase in the total grain cross section in the postshock gas enhances the rate of formation of mantles and removes SiO from the gas phase. 

The variation of the fractional abundance of SiO is very different if only the sputtering of grain cores is considered (M2), because the Si that is released by sputtering has to be transformed into SiO by gas-phase chemical reactions, predominantly oxidation by O${}_2$ and OH. The corresponding reactions are

\begin{equation}
  {\rm Si +  O}_2 \longrightarrow {\rm SiO + O}
\end{equation}

\begin{equation}
 {\rm Si +  OH} \longrightarrow {\rm SiO + H} .
\end{equation}
For reaction (1) the rate coefficient (cm${}^3$ s${}^{-1}$)
\begin{displaymath}
k = 1.72 \times 10^{-10} (T/300)^{-0.53}\exp(-17/T)
\end{displaymath}
was measured by \citet{LePicard:2001p19024}. The rate coefficient for reaction (2), which is not measured, was adopted to be the same.

Thus, the fractional abundance of SiO in the gas phase depends not only on the amount of Si sputtered from the grain cores but also on the abundances of O${}_2$ and OH, displayed in the lower panel of Fig.~\ref{Fig4_xSiO}. The chemical delay in SiO production is apparent in the upper panel of Fig.~\ref{Fig4_xSiO}; the abundance of SiO peaks in the cool and dense postshock region. If SiO is initially in the grain mantles (M3), its release is rapid and complete even before the temperature of the neutral fluid rises significantly.\footnote {The threshold velocity for sputtering of the mantles is $\sim 10 \mu ^{-\frac {1}{2}}$\,km\,s${}^{-1}$, where $\mu$ is the reduced mass of the colliding species, relative to atomic hydrogen (\citealt{Barlow:1978p17402}).} For the shock models considered, all the SiO in the mantles is released into the gas phase. We have assumed that 10\% of elemental silicon is initially in the form SiO in the mantles; this is the largest of the values considered by \citet{Gusdorf:2008p974}. If the same fraction of silicon is initially in the form of SiH${}_4$ in the mantles, there occurs the same chemical delay in the production of SiO in the gas phase as in models M2.

\subsubsection{Excitation of the SiO rotational lines}
\label{subsubsec4.2.2}

\begin{figure}
   \centering
   \includegraphics[width=9cm]{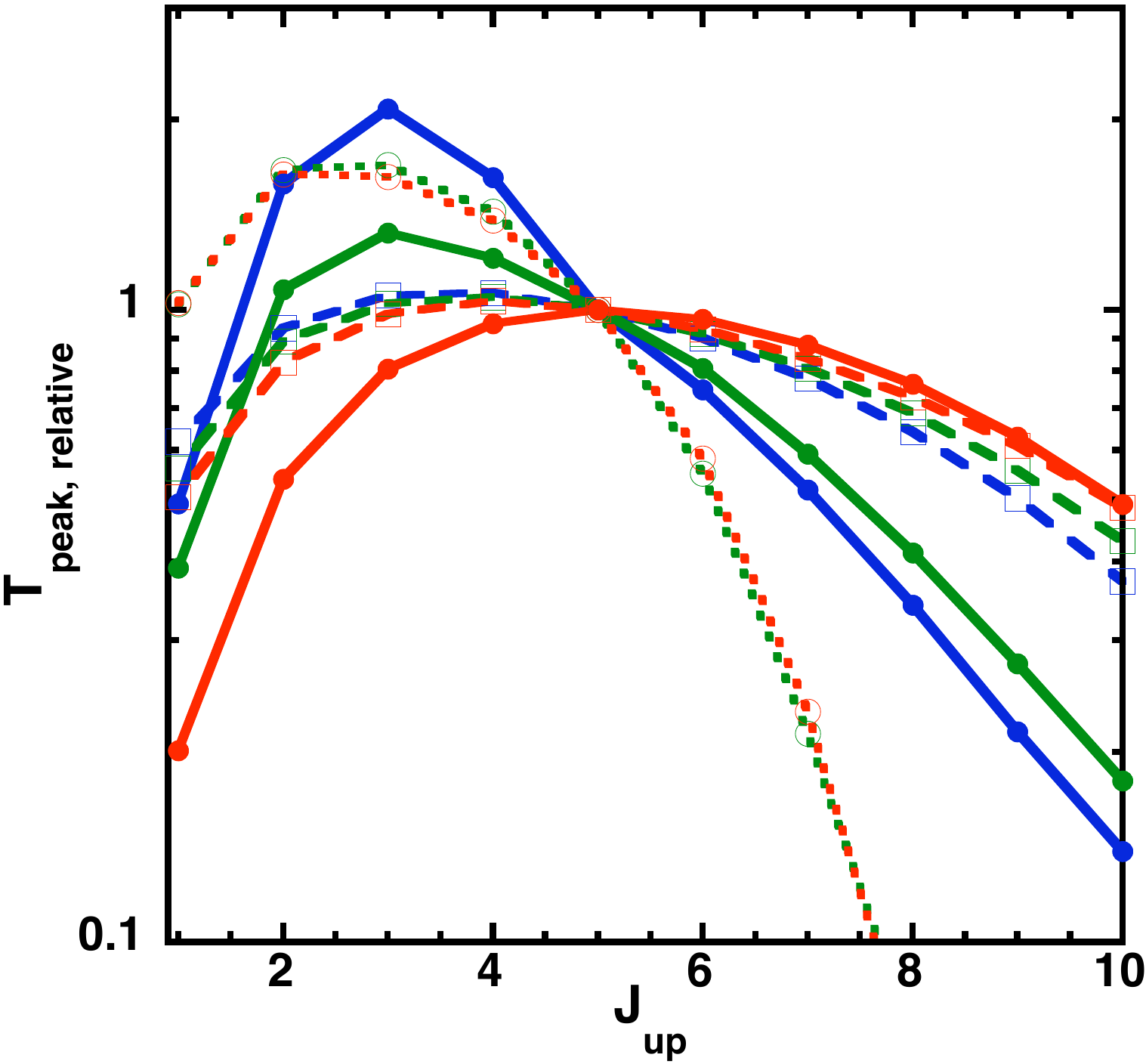}
      \caption{
    The peak temperatures (K) of the rotational emission lines of SiO, relative to the $J = 5$--$4$ transition, as functions of the rotational quantum number of the upper level of the transition $J_{\rm up}$. Displayed are the shock models in our grid that are most strongly influenced by vaporization, i.e. $\Vs$~=~20~\kms with $b$ = 1 (blue), $\Vs$~=~30~\kms with $b$ = 1.5 (green), and $\Vs$~=~40~\kms with $b$ = 2 (red). Full lines: M1; dotted lines: M2; broken lines: M3. 
           }
         \label{Fig5_Tmaxrel}
   \end{figure}

The differences in the SiO abundance profiles in the three cases M1--3 -- specifically, whether SiO is already present in the hot gas, early in the shock wave, or only in the cold postshock gas -- have consequences for the relative intensities of the SiO rotational lines. To illustrate this point, we compare, in Fig.~\ref{Fig5_Tmaxrel}, the peak temperatures of the rotational lines of SiO, relative to the $J = 5$--$4$ transition, calculated for each of these three cases. The results in Fig.~\ref{Fig5_Tmaxrel} may be better understood by referring to the line profiles, shown for the $\Vs$~=~30~\kms shock in Fig.~\ref{Fig7_SiOlinesorigin}.

The first thing to note is the similarity of the relative peak line temperatures for the $\Vs$~=~30~\kms and $\Vs$~=~40~\kms  models with only core sputtering (M2); this can be ascribed to the chemical delay in SiO formation. The rotational lines up to the 8--9 transition in these models mostly stem from the cold postshock gas, where almost maximum compression is reached. Therefore, the lines have large optical depths and near-LTE excitation conditions, which makes the relative peak temperatures of the transitions displayed independent of the shock velocity. 
The models with vaporization (M1), on the other hand, show a clear variation of excitation conditions with shock speed. The lines are formed mainly in the cooling flow, where the temperature of the neutral fluid is a few hundred K (the case for $\Vs$~=~30~km\,s${}^{-1}$, $J_{\rm up}$ $>$ 4 and for $\Vs$~=~40~km\,s${}^{-1}$, $J_{\rm up} > 2$), and are sensitive to the velocity dependence of both the temperature in the cooling flow and the amount of SiO that is produced by vaporization. 

For the models with SiO in the mantles (M3), the differences in the relative peak line temperatures are much less between shocks of different velocities than is the case for models with vaporization. The same amount of SiO is released for both velocities and the column density of the cooling gas is greater than in models that include vaporization. Most of the radiation arises in the cooling flow, where the temperature profiles are very similar and the optical depths are large: for model M2, with $\Vs$~=~30~\kms and $b$ = 1.5, the transitions up to $J = 7$--$6$ become optically thick, with optical depths at the line centres reaching $1 \lesssim \tau \lesssim 6$, depending on the transition. We conclude that clear variations of the relative line intensities with the shock velocity are characteristic of those  models in which grain vaporization occurs.

       \begin{figure*}
   \centering
   \includegraphics[width=\textwidth]{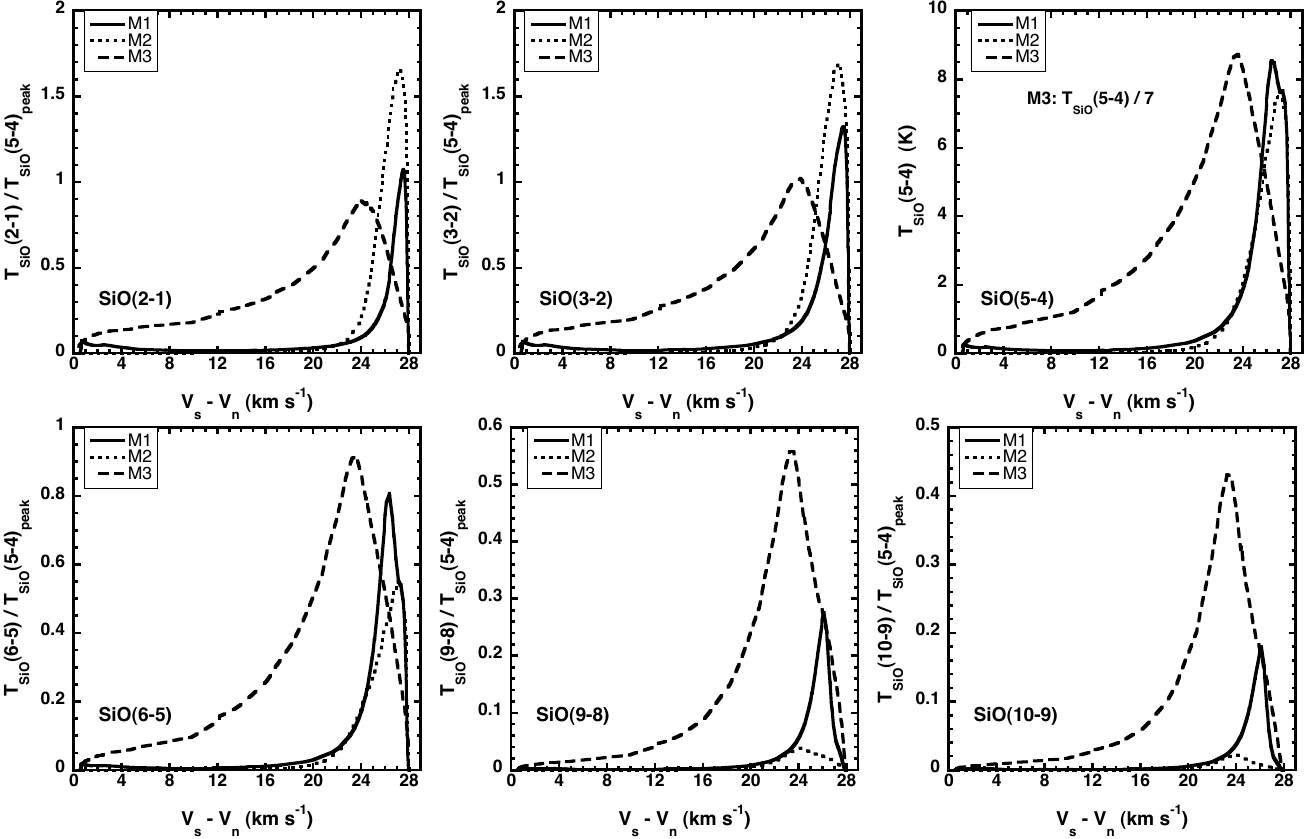}
      \caption{
      Profiles of the SiO rotational transitions (2--1), (3--2), (5--4), (6--5), (9--8), and (10--9), for $\Vs$~=~30~\kms and $b$ = 1.5. The line temperatures have been divided by the peak temperature of the (5--4) transition, except for the (5--4) transition itself, for which the absolute line profiles are given (for models M1 and M2; for model M3, the line temperature has been divided by a factor of 7 for the purposes of presentation). Full lines: M1; dotted lines: M2; broken lines: M3.
              }
         \label{Fig6_SiOlines}
  \end{figure*}

      \begin{figure}
   \centering
   \includegraphics[width=9cm]{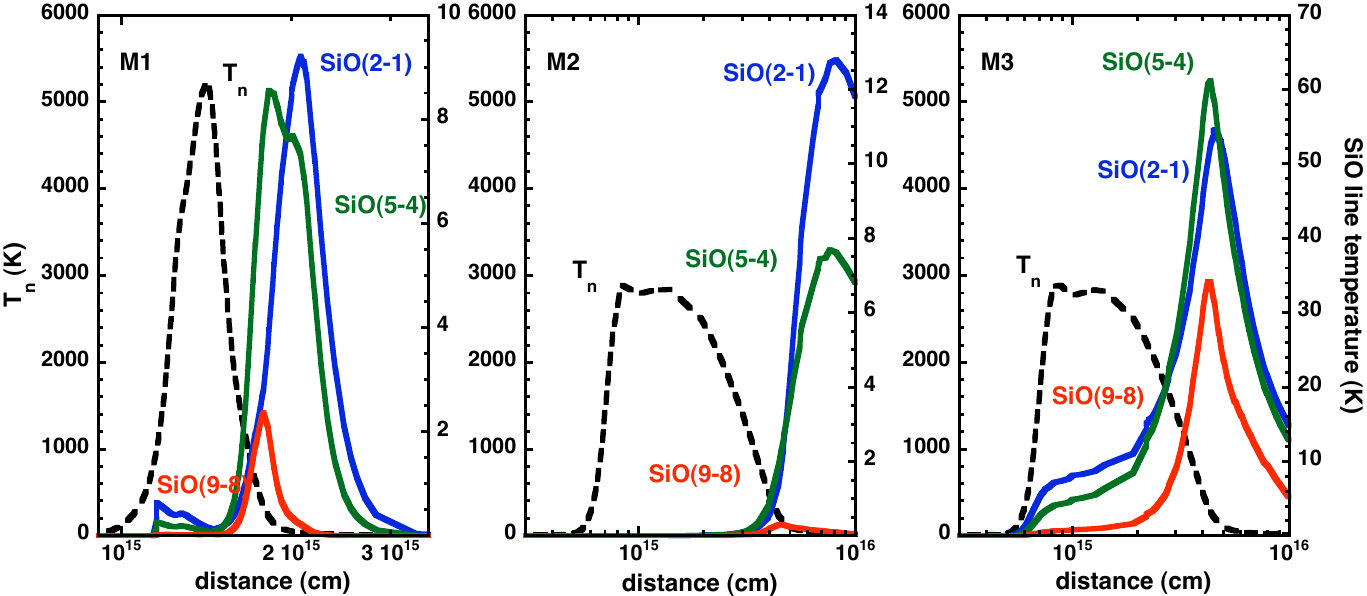}
      \caption{
       Temperature profiles of the neutral fluid (left ordinate, broken curves) and SiO line temperatures (right ordinate, full curves) of the rotational transitions $J = 2$--$1$ (blue), 5--4 (green), and 9--8 (red) for $\Vs$~=~30~\kms, $b$ = 1.5; models M1 (left panel), M2 (centre panel), and M3 (right panel). The abscissa is the distance through the shock wave.               }
         \label{Fig7_SiOlinesorigin}
  \end{figure}

 \subsubsection{SiO rotational line profiles}
 \label{subsubsec4.2.3}

Given that the excitation conditions of the SiO lines differ in the three scenarios M1--3, we expect the SiO spectral line profiles differ also, with respect to line width, location of the peak, and integrated flux. Line profiles for $\Vs$~=~30~km\,s${}^{-1}$, divided by the peak temperature of the $J = 5$--$4$ transition (except for the 5--4 transition itself), are shown in Fig.~\ref{Fig6_SiOlines}.  Model M1 predicts narrower lines for low-$J$ transitions, but similar widths for high-$J$ transitions, as model M2, and much narrower lines than model M3. There is a weak variation with $J_{\rm up}$ of the location of the line peak in models M1 and M2. For M1, the peak in the profile of the lowest transition occurs in the cold posthock gas, where the neutral fluid is moving at a velocity $\Vn$~$\sim$~3~km\,s${}^{-1}$, whereas, for higher transitions, the peak moves to hotter gas, at $\Vn$~$\sim$~4~km\,s${}^{-1}$. A similar trend is seen for model M2, but, in this case, it is only the highest (and very weak) transitions that peak earlier in the shock wave. In contrast, the line peaks for model M3 are always located at the same velocity of $\Vn$~$\sim$~24~km\,s${}^{-1}$.
 Thus, the differences in shock structure and spatial distribution of SiO between models M1 and M2 do not have a strong effect on the SiO rotational line profiles. \emph{In particular, the release of SiO through vaporization early in the shock wave does not lead to significant broadening of the lines}. However, the lines are strongly broadened if SiO is present in the grain mantles (scenario M3), and this seems to be the only way of accounting for observed line widths of several tens of \kms  if they are to be explained by one single shock. Alternatively, broad SiO lines can be explained by the existence of several shocks inside the telescope beam, such that the individual narrow line profiles appear spread out in radial velocity due to different velocities and inclination angles, in the observer's frame. Similarly, the lines would be broadened by the velocity profile associated with a bow shock (e.g. \citealt{1989MNRAS.237.1009B}).

     \begin{figure}
   \centering
   \includegraphics[width=9cm]{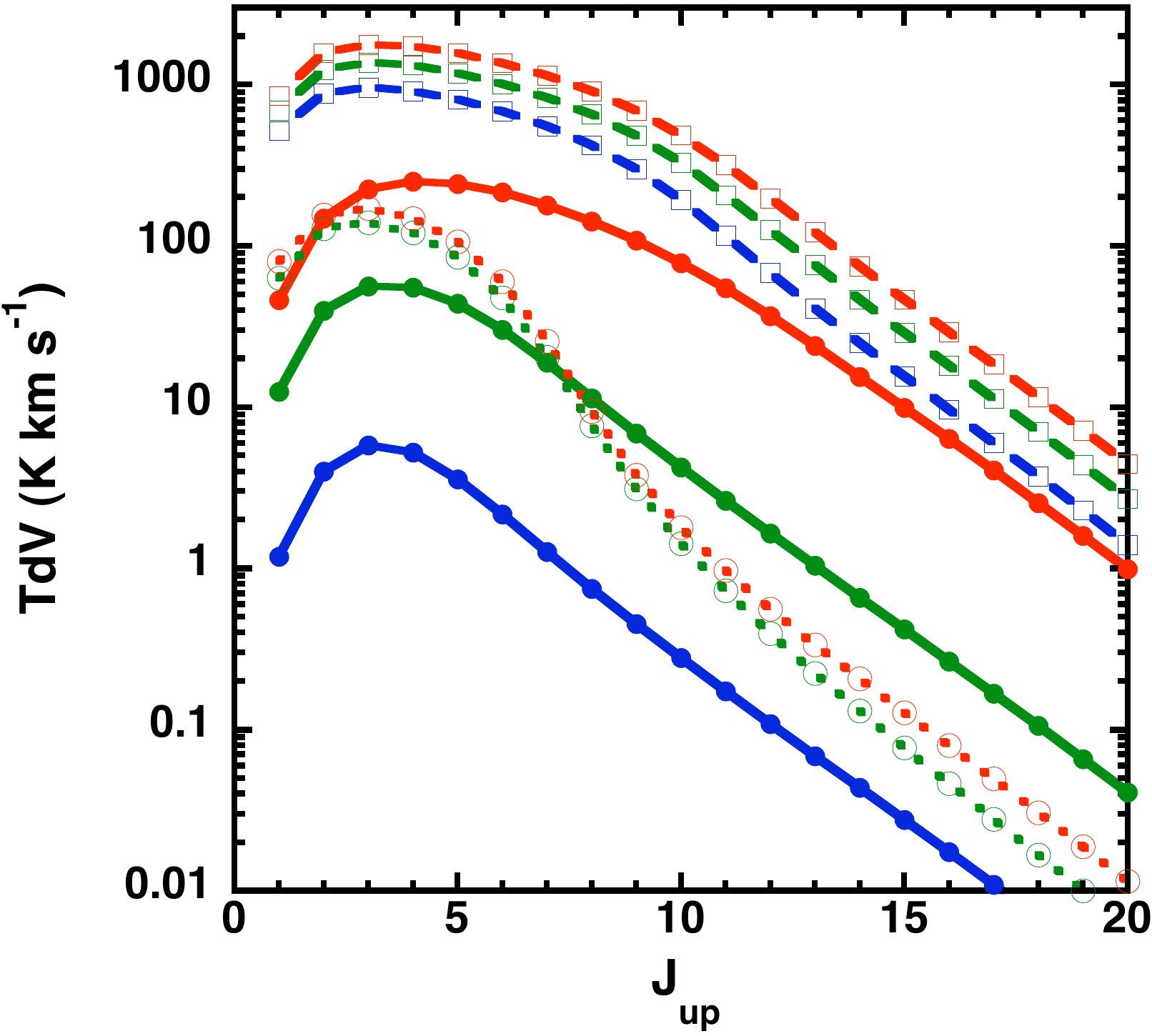}
      \caption{Integrated intensities of the rotational transitions $J_{\rm up} \rightarrow J_{\rm up}-1$ of SiO for shocks with $\Vs$~=~20~km\,s${}^{-1}$, $b$ = 1 (blue), $\Vs$~=~30~km\,s${}^{-1}$, $b$ = 1.5 (green), and $\Vs$~=~40~km\,s${}^{-1}$, $b$ = 2 (red). Full lines: model M1; dotted lines: model M2; broken lines: model M3. 
              }
         \label{Fig8_SiOTdV}
   \end{figure}

       \begin{figure}
   \centering
   \includegraphics[width=9cm]{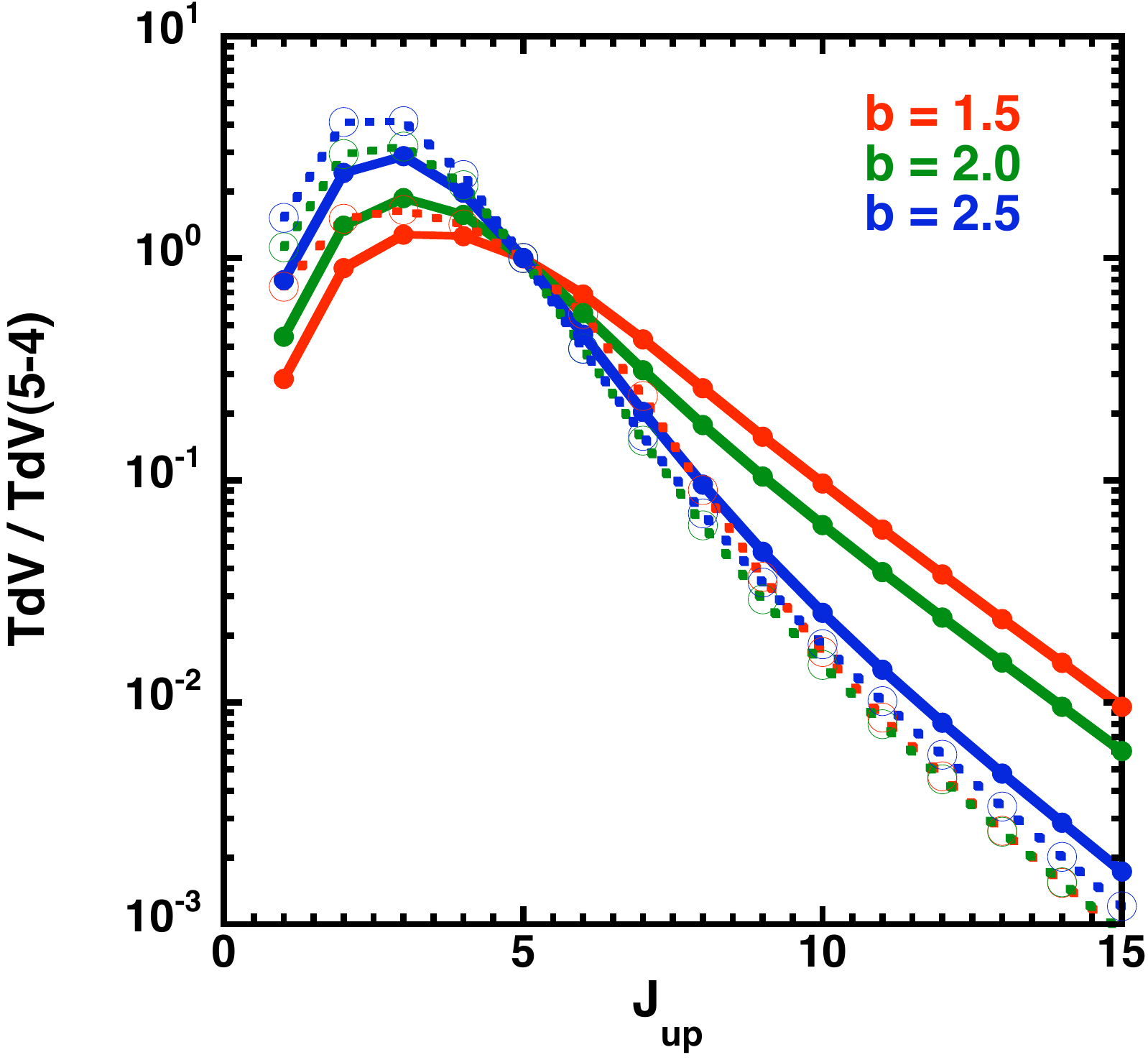}
      \caption{Integrated intensities of the rotational transitions $J_{\rm up} \rightarrow J_{\rm up}-1$ of SiO, relative to the intensity of the transition with $ J_{up}$ = 5. The shock velocity is $\Vs$~=~30~km\,s${}^{-1}$, with values of the magnetic field parameter $b$ = 1.5 (red), $b$ = 2.0 (green) and $b$ = 2.5 (blue). Full curves denote model M1, dotted curves model M2. 
              }
         \label{Fig9_SiOTdVrel}
   \end{figure}

 \subsubsection{Integrated SiO rotational line intensities}
 \label{subsubsec4.2.4}
 
 An interesting question is whether the release of SiO due to vaporization significantly increases the integrated (along the $z$-direction) SiO rotational line intensities. Fig.~\ref{Fig8_SiOTdV} shows, that indeed there is an increase in the intensities of the highly excited transitions. Furthermore, the slope of the integrated intensity curves differs between models M1 and M2 for transitions $3 \le J_{\rm up} \le 10$. 
 
 The integrated intensities of the lowest transitions -- up to $J_{\rm up} \sim 7$ -- are dependent on the timescale for accretion of gas-phase species on to grains in the cooling flow; this timescale depends on the (uncertain) accretion rates. In addition, 1-dimensional models tend to overestimate the compression of the postshock gas. Previous studies (\citealt{Gusdorf:2008p1617}) have found the accretion timescale to be unimportant for the molecular line emission. However, owing to the large increase in the total grain cross-section, induced by vaporization, we see, for the first time, an observational consequence that may be related to this timescale.
 
Independent of the accretion timescale, the integrated intensities of, in particular, the lowest transitions increase significantly for model M3; but we recall that the amount of SiO in the mantles is treated as a free parameter, and the absolute intensities can be varied accordingly. The intensity curves corresponding to model M3 have similar slopes to the curves of model M1, for transitions from $J_{\rm up} \gtrsim 10$; they peak at $J_{\rm up} = 3$. On the other hand, the peak of the curve corresponding to model M1 at $\Vs$~=~40~\kms is displaced to $J_{\rm up} = 4$, owing to the higher maximum temperature. As may be seen from Fig.~\ref{Fig8_SiOTdV}, the integrated line intensities decrease by a factor $\sim$100 between $5 \le J_{\rm up} \le 10$ for model M2, which sets it apart from models M1 and M3.
 
 Fig.~\ref{Fig9_SiOTdVrel} shows the effect of variations in the magnetic field on the integrated SiO rotational line intensities. In this Figure, the integrated intensities are expressed relative to the 5--4 transition, in order to focus attention on the excitation conditions, rather than the release of SiO into the gas phase. For model M1, there is a clear variation with magnetic field strength, which is not present for model M2. This difference can be understood from Fig.~\ref{Fig2_Twidth}, which shows that the maximum temperature of the neutral fluid hardly varies with the magnetic field strength for model M2, whereas the maximum value changes for model M1. Thus, Fig.~\ref{Fig9_SiOTdVrel} demonstrates that the excitation conditions vary with the strength of the magnetic field when the effects of grain-grain collisions are incorporated.

\subsection{The effect of vaporization on [C~I] emission}
\label{subsec4.3}

Our model, and the models of Papers I--III, includes two different populations of dust grains: silicate and carbon grains. While emission of SiO could be used as a tracer of vaporization in shocks because of the strong depletion of Si and SiO in quiescent gas, the effects of vaporization on [C~I] emission will not be seen as clearly. Nonetheless, the vaporization of graphite grains modifies the emission of atomic carbon. In order to predict the magnitude of this effect, we have used the multi-fluid model of Paper III. 

Fig.~\ref{Fig3_CSifrac} (right-hand panel) shows the fraction of carbon released from grain cores by sputtering (M2) and vaporization (M1). At low velocities, the destruction of carbon grains is less than for silicates because the gyration of large carbon grains is damped more rapidly, owing to their lower specific density. At higher velocities, for which small grains also contribute significantly to vaporization, the fraction of carbon released from graphite grains is higher than the fraction of silicon from silicates; graphite grains are more strongly affected by shattering. Table B.1 shows the intensities of two [C~I] forbidden lines, at 609.8 $\mu$m and 370.4 $\mu$m. The values in these Tables confirm that the effect of vaporization on the [C~I] lines is less than on the SiO rotational transitions.
   

 \section{ Concluding remarks}
 \label{Sect5}
  
In this series of papers, the consequences of grain-grain collisions in shock waves have been investigated; such processes had been ignored in previous studies of C-type shocks. Shattering is a key factor in the production of the large populations of very small grains (mostly carbonaceous: see, for example, \citealt{1984ApJ...285...89D}) in the turbulent ISM (\citealt{Hirashita:2010p17926}). It has been shown, in earlier papers in the series, that C-type shock waves provide similar dynamical conditions to turbulence, and so shattering needs to be included in shock models. 

Our study relies on the shattering model that was developed in Paper III, and our results inherit the dependence of this model on parameters such as the size of the smallest fragments (5\AA~ in Paper III), the slope of the size distribution of the fragments, their charge distribution, and the composition of the dust grains. Whilst shattering almost certainly occurs in the ISM, as a direct consequence of the weak coupling of large grains to the magnetic field in high density clouds, the threshold density at which shattering starts to have a significant effect on the shock dynamics is more uncertain.

In our model, the processes of shattering and vaporization are linked, so that the threshold density for shattering applies also to vaporization. The parameters of the shattering model in Paper III were chosen to minimize the amount of shattering and thereby yield a conservative estimate of the threshold density. It was shown that strong feedback on the shock dynamics was expected only for densities higher than $\sim10^5$ cm${}^{-3}$. The results of the present paper include some observational predictions that should help to constrain the grain-shattering model.
The main conclusions of our study are as follows:

\begin{enumerate}

\item The influence of grain-grain processing on the overall shock structure was found to be significant for the full range of magnetic field strengths that we studied. The maximum temperature increases by a factor of 1.5--2, and the shock width is reduced by a factor 4--5. 
The inclusion of grain-grain processing changes the dependence of the shock structure on the magnetic field strength. While for shocks without grain-grain processing there is only a weak dependence of the peak temperature on the magnetic field, the peak temperature becomes strongly dependent on the magnetic field when grain-shattering and vaporization are incorporated. Consequently, the intensities of, in particular, highly excited molecular transitions become dependent on the strength of the magnetic field. 
While shattering is shown to be important for all models of our grid, the vaporization of SiO from silicate grain cores is significant only for fast shocks and low magnetic fields.
\item There are two consequences of grain-grain processing for the molecular line emission: the reduced shock width results in a lower column density of shock-heated gas, whereas the higher peak temperature can modify the chemistry and enhance the fractional abundances of molecules in highly excited states. Which of these tendencies prevails is decided by the chemical and physical properties of the individual molecules and their transitions. The intensities of all lines of OH increase in shocks with velocities greater than 30 km\,s${}^{-1}$, when grain-grain processing is included, owing to the temperature sensitivity of OH formation. On the other hand, the intensities of the emission of CO and H${}_2$O decrease because of the reduced column density of shocked material. In the case of H${}_2$, the intensities of highly-excited transitions increase, whilst the intensities of lines of lower excitation decrease.
\item The release of SiO through collisional vaporization of silicate grain cores enhances the integrated intensities of SiO rotational lines, mainly from highly-excited levels. However, this effect is counteracted by the reduction in shock width. To obtain significantly higher line intensities, it is necessary to introduce SiO into the grain mantles. The situation is similar with respect to the widths of the SiO rotational lines. Although vaporizations releases SiO early, in the hot part of the shock wave, the reduction in the shock width prevents the lines from becoming significantly broader than in models that neglect grain-grain processing. Therefore, vaporization alone cannot account for broad lines if only one single shock is considered. To obtain broad profiles, it seems necessary that SiO should be present in the grain mantles, such that mantle sputtering releases SiO, also in the early part of the shock wave. It is essential that SiO is released directly, thereby eliminating the chemical delay that would be associated with its production, in gas-phase reactions, from Si or SiH${}_4$.

Due to the fact, that most of the SiO emission stems from the early part of the shock, where the temperature profile depends on the shock velocity, the SiO rotational line ratios vary with the shock velocity; this variation is not present if vaporization is ignored. The effect of vaporization on [C~I] emission lines was found to be less than for the SiO rotational lines.
 
\end{enumerate}

\begin{acknowledgements}
      We are grateful to an anonymous referee for useful comments that helped to strengthen the paper. S.~Anderl acknowledges support by the DFG SFB 956, the International Max Planck Research School (IMPRS) for Astronomy and Astrophysics, and the Bonn-Cologne Graduate School of Physics and Astronomy.
      \end{acknowledgements}
      
\bibliographystyle{aa}
\bibliography{GG_biblio}

\begin{appendix} 

\section{Shattering and vaporization: their implementation and validation}

\subsection{Shattering}

 \subsubsection{Dynamical effects of shattering}

The dynamical effects of shattering on the shock structure can be simulated by ensuring that the total grain cross section, $\langle n\sigma \rangle$, changes in accord with the results of Paper III. We found that it is possible to model the increase of $\langle n\sigma \rangle$ due to shattering by multiplying the total grain cross section, in the absence of grain-grain processing, by an additional factor, which varies with the spatial coordinate, $z$, through the shock wave. This factor is modelled as an intrinsic function of the compression of the ion fluid, normalized to the theoretical postshock compression at infinity, as predicted by the Rankine-Hugoniot relations. This normalized ion compression parameter constitutes a function, $\eta(z)$, varying  from $0$ in the preshock to $1$ in the postshock medium.

The preshock (medium 1) and postshock (medium 2) kinetic temperatures are approximately equal in the C-type shock models of our grid. Thus, the Rankine-Hugoniot continuity relations may be applied across the shock wave, replacing the relation of conservation of energy flux by the isothermal condition,  $T_1 = T_2$. Then, setting the ratio of specific heats $\gamma = 1$, the expression for the compression ratio, $\rho _2/\rho _1$, across the shock wave may be derived (cf. \citealt{Draine:1993p1759}). Under the conditions of our models, which are such that M$_{\rm s} = V_{\rm s}/c_1 \gg~$M$_{\rm A} = V_{\rm s}/V_{\rm A} \gg 1$, where M$_{\rm s}$ is the sonic Mach number of the flow, evaluated in the preshock gas, $c_1$ is the sound speed, and M$_{\rm A}$ is the Alfv\'{e}nic Mach number, also in the preshock gas, the compression ratio reduces to $\rho _2/\rho _1 \approx \sqrt {2}$M$_{\rm A}$, whence

  \begin{equation}
  V_{\rm postshock} = \frac{V_{\rm A}}{\sqrt{2}} = \frac{B}{\sqrt{2\times 4\pi ~1.4~ m_{\rm H} n_{\rm H}}} ~ {\rm  ,}
\end{equation}
where $V_{\rm postshock}$ is the flow speed in the postshock gas, $m_{\rm H}$ the proton mass and $n_{\rm H}~=~n({\rm H})+2~n({\rm H}_2)$ the proton density in the preshock gas. Using
 \begin{equation}
\eta(z)=\frac{\Vs/\Vi(z)-1}{\Vs/V_{\rm postshock}-1}
\end{equation}
(where $\Vs$ is the shock velocity and $\Vi$ the velocity of the ion fluid in the shock frame, respectively), the factor by which the total grain cross section is enhanced is given by
\begin{equation}
\Sigma(z)=\eta(z) \cdot(\Sigma_{\rm max}-1)+1 ~{\rm .}
\end{equation}

The extent of the increase of $\langle n\sigma \rangle$ due to shattering is given by the final value of the shattering-factor, $\Sigma{}_{\rm max}$. This value depends on the shock velocity and the magnetic field and is an external parameter, which needs to be extracted from the multi-fluid models. However, this simple functional form did not reproduce satisfactorily the onset of shattering in the shock. We therefore propose the following refined expression

\begin{equation}
\Sigma=\Big( \eta^{\beta}-\frac{\sin{(2\pi~\eta^{\beta})}}{2\pi}\Big)\cdot \Big(\Sigma_{max}-1\Big)+1
\end{equation}
that introduces another parameter, $\beta$, which needs to be extracted from the corresponding multi-fluid model. $\beta$ describes the delay in the shattering feedback, relative to the ion compression, and is only weakly dependent on the shock velocity.
The parameters, $\beta$ and $\Sigma_{\rm max}$, which correspond to the grid of models introduced in Section \ref{subsec3.1}, are listed in Table~\ref{ta1}. Linear fits in the shock velocity $\Vs$ and  the magnetic field parameter $b$ are
 \begin{equation}
  \Sigma_{max} = 9.5 + 0.4 \cdot \Vs - 4.6\cdot b
  \end{equation}
  and
  \begin{equation}
  \beta = 0.95 - 0.025\cdot \Vs + 0.4\cdot b \textrm{  .}
  \end{equation}
The increase of the total grain collisional cross-section needs to be consistent with the mean square radius of the grains and with their total number density, following the compression of the ions. Analyzing the multi-fluid computations corresponding to Paper III, we find a reasonable approximation to the behaviour of the total grain number density, $n_{\rm G}$, and the mean square radius, $\langle \sigma \rangle_{\rm G}$: the former increases as $\Sigma{}^4$, the latter  decreases as $\Sigma{}^{-3}$, relative to the corresponding values without shattering. These changes affect the rates of grain-catalyzed reactions, adsorption to the grain mantles, excitation of H$_2$ in collisions with grains, and transfer of momentum and thermal and kinetic energy between the neutral and the charged fluids. 

\begin{figure}
  \centering
  \includegraphics[width=9cm]{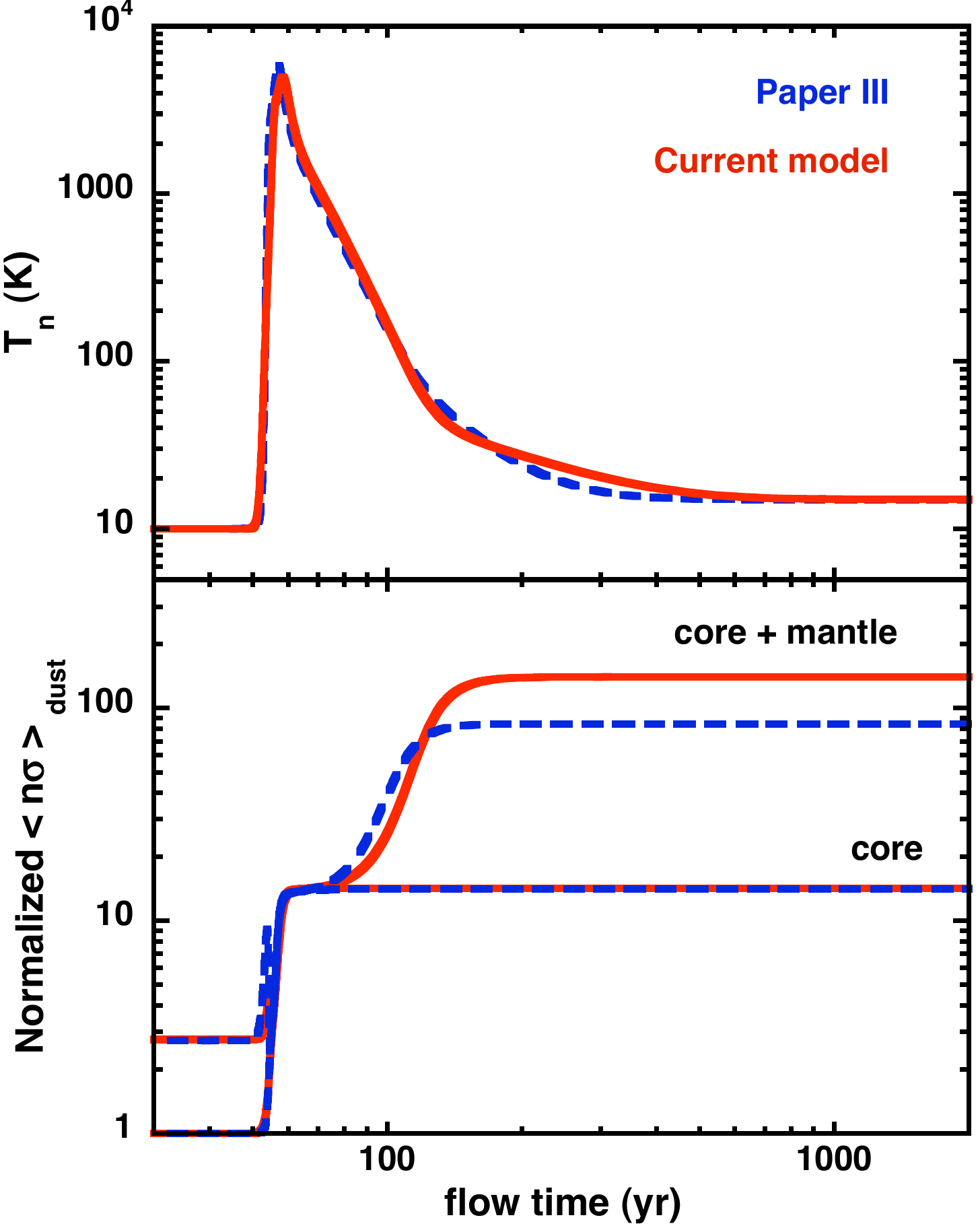}
     \caption{Upper panel: temperature profiles of the neutral fluid for a shock with $n_{\rm H}~=~10^5$~cm$^{-3}$, $\Vs$~=~30 \kms and $b$ = 1.5 corresponding to Paper III (in blue) and our current model (in red). Lower panel: evolution of the total grain cross section, $\langle n\sigma \rangle$, for the same shock models, with and without taking into account the grain mantles and normalized to the preshock values of $\langle n\sigma \rangle$.               }
        \label{FigA1_shattering}
  \end{figure}

Figure~\ref{FigA1_shattering} shows a comparison between the multi-fluid model of Paper III and our current model\footnote{In order to make a direct comparison with the results of Paper III, it is necessary to disable the LVG treatment of molecular line transfer in our current model.} for a representative shock for which $n_{\rm H} = 10^5$~cm$^{-3}$, $\Vs$~=~30~\kms and $b$~=~1.5. The temperature profiles of the neutral fluid agree well. Furthermore, the variation of the total grain-core cross-section is reproduced by our simulation, as may be seen from the lower panel of Figure~\ref{FigA1_shattering}. The total grain cross-section (including mantles) increases somewhat later in our current model, due to small differences -- independent of grain shattering --with the model of Paper III. The discrepancy in the value of the grain cross-section in the postshock medium arises because our simplified treatment of shattering tends to underestimate the final number density of small grains. However, this simplification introduces only small deviations in the hydrodynamic parameters, such that the values of both the peak temperature of the neutral fluid and the shock width (up to 100~K in the cooling flow) agree to within $\pm \sim$15 \% for the entire grid of models incorporating shattering.

\begin{table}
\caption[]{Parameters relating to the modification by shattering of the total grain cross-section, $\langle n\sigma \rangle$ and to SiO vaporization, for the grid of multi-fluid models introduced in Section \ref{subsec3.1}.}

\begin{center}
\begin{tabular}{llllll}
\hline
           \noalign{\smallskip}

$\Vs$ [km\,s${}^{-1}$] & $b$ & $B$ [$\mu {\rm G}$] & $\Sigma{}_{\rm max}$ & $\beta$ & $x$(SiO${}_{\rm peak}$)\\
           \noalign{\smallskip}
\hline
           \noalign{\smallskip}
20 & 1.0 & 316 & 11.7 & 0.8 & 1.53 $\cdot$ 10${}^{-8}$ \\
20 & 1.5 &474 & 10.5 & 1.1 & 1.99 $\cdot$ 10${}^{-17}$ \\
20 & 2.0 & 632 & 8.5 & 2.0 & 2.16 $\cdot$ 10${}^{-19}$\\
30 & 1.5 & 474 & 14.2 & 0.8& 1.55 $\cdot$ 10${}^{-7}$\\
30 & 2.0 & 632 & 11.7 & 0.9& 7.92 $\cdot$ 10${}^{-9}$\\
30 & 2.5 & 791 & 10.5 & 1.2 & 5.63 $\cdot$ 10${}^{-13}$\\
40 & 2.0 & 632 & 18.4 & 0.8& 9.13 $\cdot$ 10${}^{-7}$\\
40 & 2.5 & 791 & 13.7  & 0.9 & 3.43 $\cdot$ 10${}^{-7}$\\
40 & 3.0 & 949 & 11.4 & 1.1 & 1.34 $\cdot$ 10${}^{-7}$\\
\hline
\end{tabular}

\end{center}
\label{ta1}
\end{table}

\subsubsection{Chemical effect of shattering}

The change in the total grain cross-section, $\langle n\sigma \rangle$, owing to shattering, has consequences for the rates of grain-charging reactions. In order to model the effect of shattering on the abundances of charged grains, the corresponding chemical source term needs to be introduced.

The charge distribution of the fragments is essentially unknown and cannot be numerically integrated separately from that of grains already present in the medium. In the present paper\footnote{The situation was more complicated in Paper III, which dealt with a full grain size distribution. Fragments were allocated to the size bins corresponding to their individual mass, while the corresponding mass was removed from the projectile and target size bins. This procedure does not allow for charge conservation, because small grains carry much more (negative) charge per unit mass than large grains. To compensate for this charge excess, the charge distributions of all grain sizes was shifted infinitesimally at each shattering event to ensure charge conservation.}, the charge distribution of fragments is designed to ensure charge conservation. 
We fitted the fragment charge distribution by aligning the shock widths (see Fig.~\ref{FigA1_shattering}, upper panel). This procedure yielded a charge distribution in which 1/2 of the fragments were neutral, and 1/4 were positively and 1/4 negatively charged, following shattering. 
Subsequently, the grain charge distribution evolved on a timescale that can be long -- of the order of the flow time through the shock wave when the mean grain size becomes very small.

The source term for the creation of grains through shattering can be derived from the equation of conservation of the total flux of grains, in the absence of shattering,

\begin{equation} \label{eqnA7}
\sum_{j ~\in \{\rm G0, G+, G-\}} n_{ j} \cdot V_j \cdot \Sigma^{-4}(z) = {\rm constant}
\end{equation}
where $n_{\rm G0}$ is the number density of neutral grains, $n_{\rm G+,G-}$ are the number densities of positively and negatively charged grains, respectively, $V_{\rm G0} = V_{\rm n}$ is the velocity of the neutral grains, in the shock frame, and $V_{\rm G+, G-} = V_{\rm i}$ is the velocity of the charged grains, in the shock frame.
Differentiation of (\ref{eqnA7}), subject to the charge distribution of the fragments, then yields
\begin{equation}
\frac{\dr n_{\rm G0}}{\dr z}\bigg|_{\rm shat}=\frac{1}{2}\cdot4\cdot n_{\rm G,preshock}\frac{\Vs}{\Vn}\cdot \Sigma^3\cdot \frac{\dr \Sigma}{\dr z}
\end{equation}
for the neutral grains, and
\begin{equation}
\frac{\dr n_{\rm G+,G-}}{\dr z}\bigg|_{\rm shat}=\frac{1}{4}\cdot4\cdot n_{\rm G,preshock}\frac{\Vs}{\Vi}\cdot \Sigma^3\cdot \frac{\dr \Sigma}{\dr z}
\end{equation}
for the charged grains, where $n_{\rm G,preshock}$ is the total (charged and neutral) number density of grains in the preshock gas.
The derivative of $\Sigma$ is given by
\begin{equation}
\frac{\dr \Sigma}{\dr z} = \Big( \frac{\dr}{\dr z} \eta^\beta - \cos\big(2\pi \eta^\beta\big)\frac{\dr}{\dr z} \eta^\beta \Big) \Big(\Sigma_{\rm max} - 1 \Big)
\end{equation}
where
\begin{equation}
\frac{\rm d}{\dr z}  \eta^\beta= \beta~ \eta^{(\beta -1)} \frac{\dr \eta}{\dr z}
\end{equation}
and
 \begin{equation}
 \frac{\dr \eta}{\dr z} = -\frac{1}{(\Vs/V_{\rm postshock}-1)}\frac{\Vs}{\Vi^2}\frac{\dr \Vi}{\dr z}.
\end{equation}
With these parameters, our model is able to reproduce the main effects of the increase in the grain cross section, reported in Paper III: the effective rate of recombination of electrons and ions is enhanced; the fractional abundance of free electrons falls by three orders of magnitude; and dust grains become the dominant charge carriers, with equal numbers of positively and negatively charged grains being produced.

\subsection{Vaporization}

The effect of vaporization is modelled as an additional term in the creation rate of gas-phase SiO, from Si and O in grain cores (denoted by **), corresponding to a new type of pseudo-chemical reaction:
\begin{verbatim}
Si** + O** = SiO + GRAIN
\end{verbatim}
Because vaporization sets in suddenly, when the vaporization threshold is reached, the function
\begin{equation}
\Omega(z) = \frac{1}{1+\exp \bigl(-10^3\cdot \bigl(\Vs/\Vi(z)-\Vs/6 V_{\rm postshock}\bigr)\bigr)}
\end{equation}
can be used to approximate the rate of creation of SiO. The function $\Omega(z)$ is centred at the point where the compression of the charged fluid reaches $1/6$ of its final value, as determined by our fitting procedure.
Similarly, the factor $10^3$ in the exponent, which determines the steepness of the function, derives from a fit to the numerical results of Paper III.
Using this function, the creation rate (${\rm cm}^{-3}~{\rm s}^{-1}$) can be expressed as
\begin{eqnarray}
\frac{\dr n({\rm SiO})}{\dr t}\bigg|_{\rm vapo}&=&  \frac{\dr \Omega(z)}{\dr t} \cdot n_{\rm H}\cdot x({\rm SiO})_{\rm peak} \\
&=&  \frac{\dr \Omega(z)}{\dr z} \cdot V_{\rm n}  \cdot n_{\rm H}\cdot x({\rm SiO})_{\rm peak} \\
&=& \frac{\dr \Omega(z)}{\dr z} \cdot V_{\rm s}  \cdot n_{\rm H,preshock}\cdot x({\rm SiO})_{\rm peak} ~{\rm ,}
\end{eqnarray}
where use is made of the conservation of the flux of $n_{\rm H}$, and where
\begin{equation}
\frac{\dr}{\dr z}\Omega(z) = -\Omega(z) \cdot \big( 1 - \Omega(z) \big)\cdot 10^3   \frac{\Vs}{\Vi^2}\cdot \frac{\dr \Vi}{\dr z}~ {\rm .}
\end{equation}
The spatial change in number density of SiO, owing to vaporization, is then given by
\begin{equation}
\frac{\dr n({\rm SiO})}{\dr z}\bigg|_{\rm vapo}= \bigg( \frac{\dr n({\rm SiO})}{\dr t}\bigg|_{\rm vapo} - n({\rm SiO})\frac{\dr \Vn}{\dr z}\bigg)\frac{1}{\Vn}
\end{equation}

The peak fractional abundance $x$(SiO)${}_{\rm peak}$ needs to be computed with the multi-fluid model and constitutes a free parameter, given in Table \ref{ta1} for our grid of models. As can be seen in Table \ref{ta1}, vaporization of SiO is relevant only for the shocks with high velocity and low magnetic field; it can be neglected for the shocks with $\Vs$~=~20~\kms and $b$ = 1.5, $b$= 2, as well as for the model with $\Vs$~=~30~\kms and $b$ = 2.5 .

  \begin{figure}
  \centering
  \includegraphics[width=9cm]{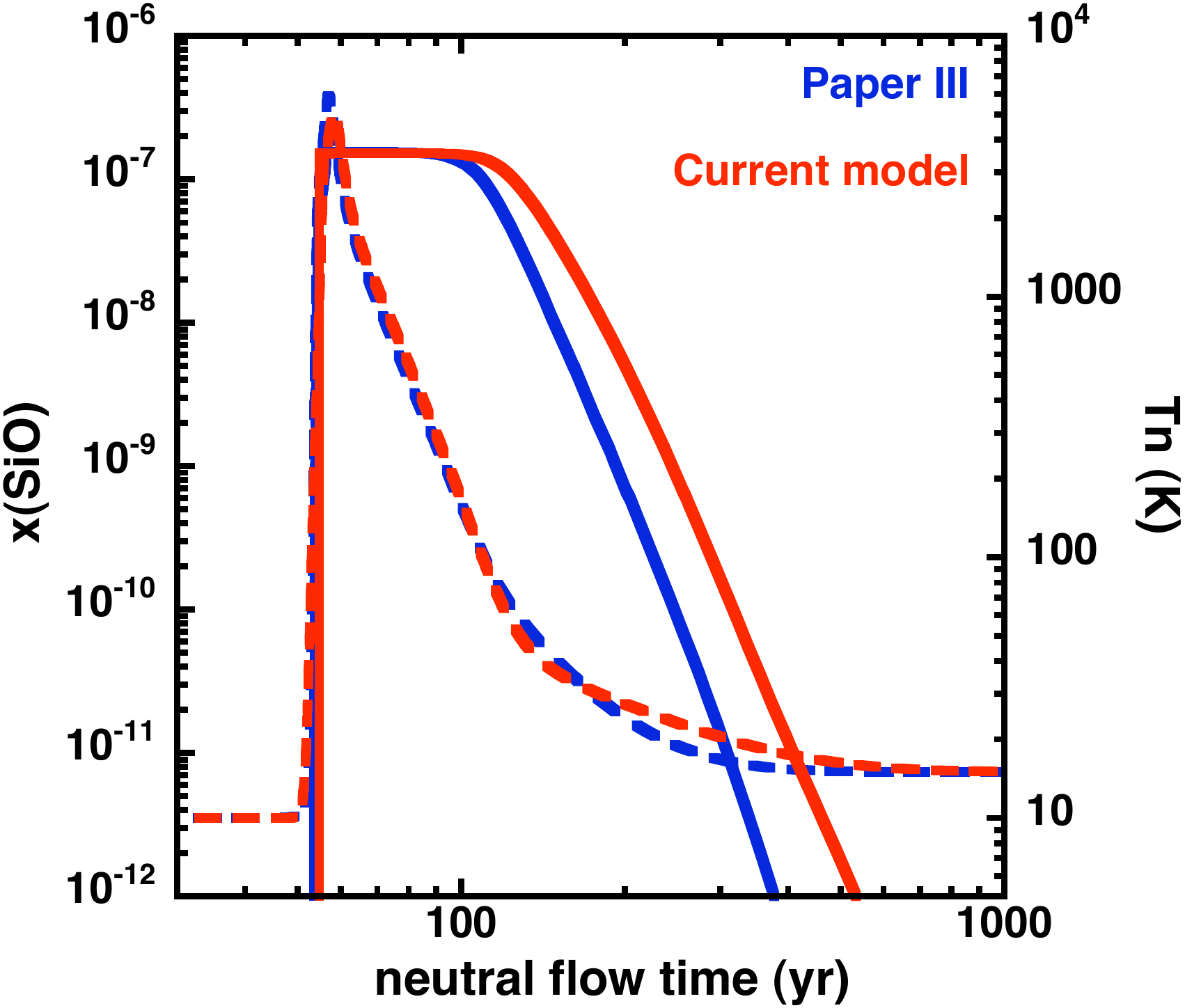}
     \caption{The fractional abundance of SiO (full curves), as determined when including grain-grain processing, using the present model (in red) and the multi-fluid model of Paper III (in blue); the shock parameters are $n_{\rm H}~=~10^5$~cm$^{-3}$, $\Vs$~=~30 \kms and $b$ = 1.5. The temperature of the neutral fluid is shown also (broken curves; right-hand ordinate).}
        \label{FigA2_vaporization}
  \end{figure}

Figure~\ref{FigA2_vaporization} shows the result of our implementation of vaporization. We assume that, initially (in the preshock medium), there are no Si-bearing species, either in the gas phase or in the grain mantles; all the Si is contained in olivine (MgFeSiO${}_4$) grain cores. This Figure compares the fractional abundance of SiO, as computed with our current model (incorporating grain-grain processing) and as predicted by the multi-fluid model of Paper III.
By construction, the peaks of the fractional abundance of SiO agree, whereas the timescale for its accretion on to grains differs between the two models, as is visible in the plot; this discrepancy relates to the imperfect agreement of the total cross-section, $\langle n \sigma \rangle$, in the postshock medium (see Fig.~\ref{FigA1_shattering}). However, we have verified that the timescale for SiO accretion is not critical to our analysis: the complete neglect of accretion on to grains leads to increases in the integrated intensities of the lowest rotational transitions of SiO and CO, by factors of $\sim$2 and $\sim$3, respectively. We note that the intensities of these transitions are, in any case, affected by the foreground emission of ambient, non-shocked gas.

\end{appendix}

\begin{appendix} 

\section{Molecular line emission}

 \begin{figure}
  \centering
  \includegraphics[width=8.5cm]{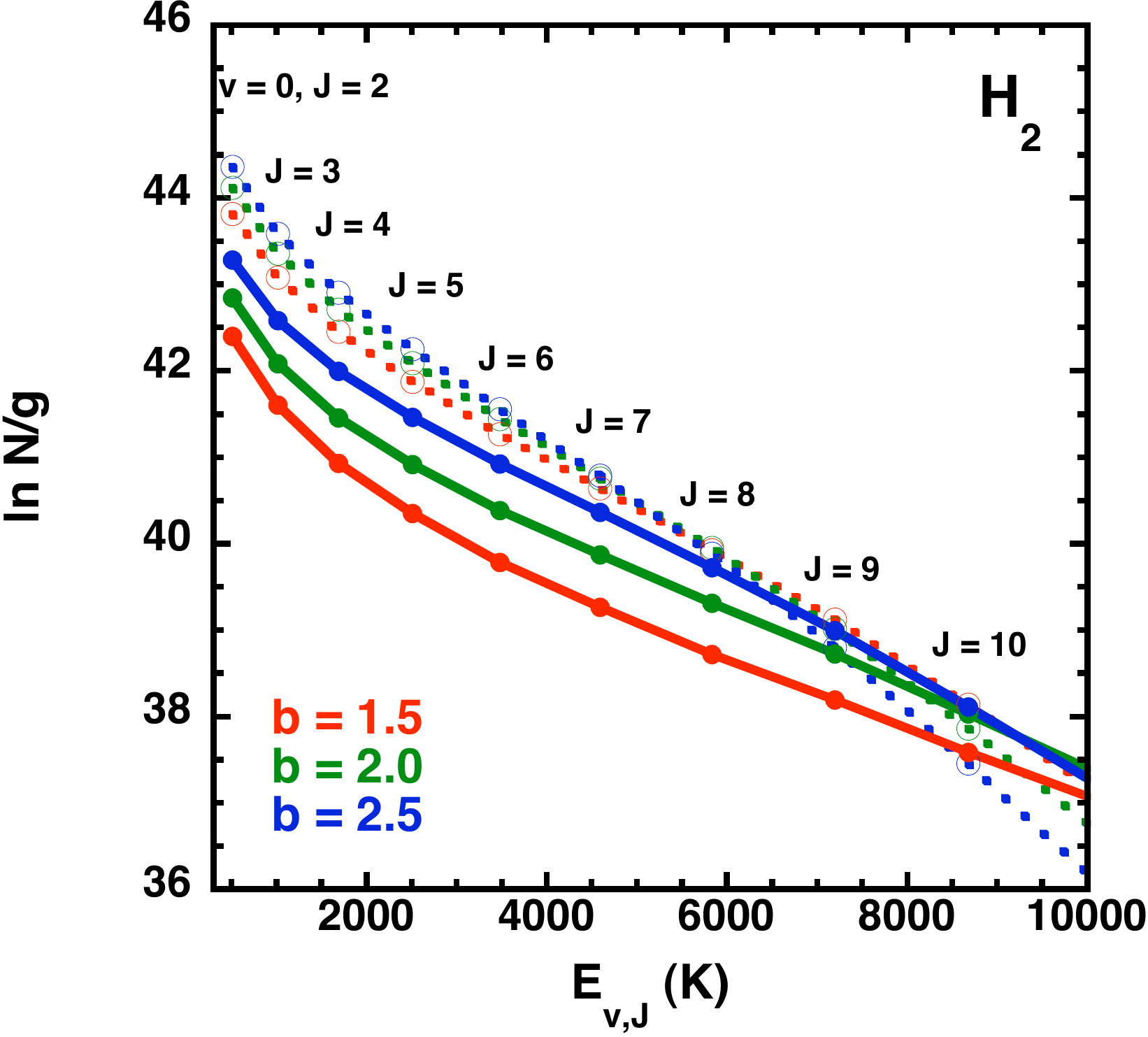}
     \caption{Computed H${}_2$ excitation diagrams for rovibrational levels with energies $E_{{\rm v},J} \le 10~000$~K and for shocks with $n_{\rm H}~=~10^5$~cm$^{-3}$, $\Vs$~=~30~\kms and $b$ = 1.5 (red), $b$ = 2.0 (green) and $b$ = 2.5 (blue). Full lines: model M1; dotted lines: model M2.
             }
        \label{FigB1_H2}
  \end{figure}

  \begin{figure}
  \centering
  \includegraphics[width=8.5cm]{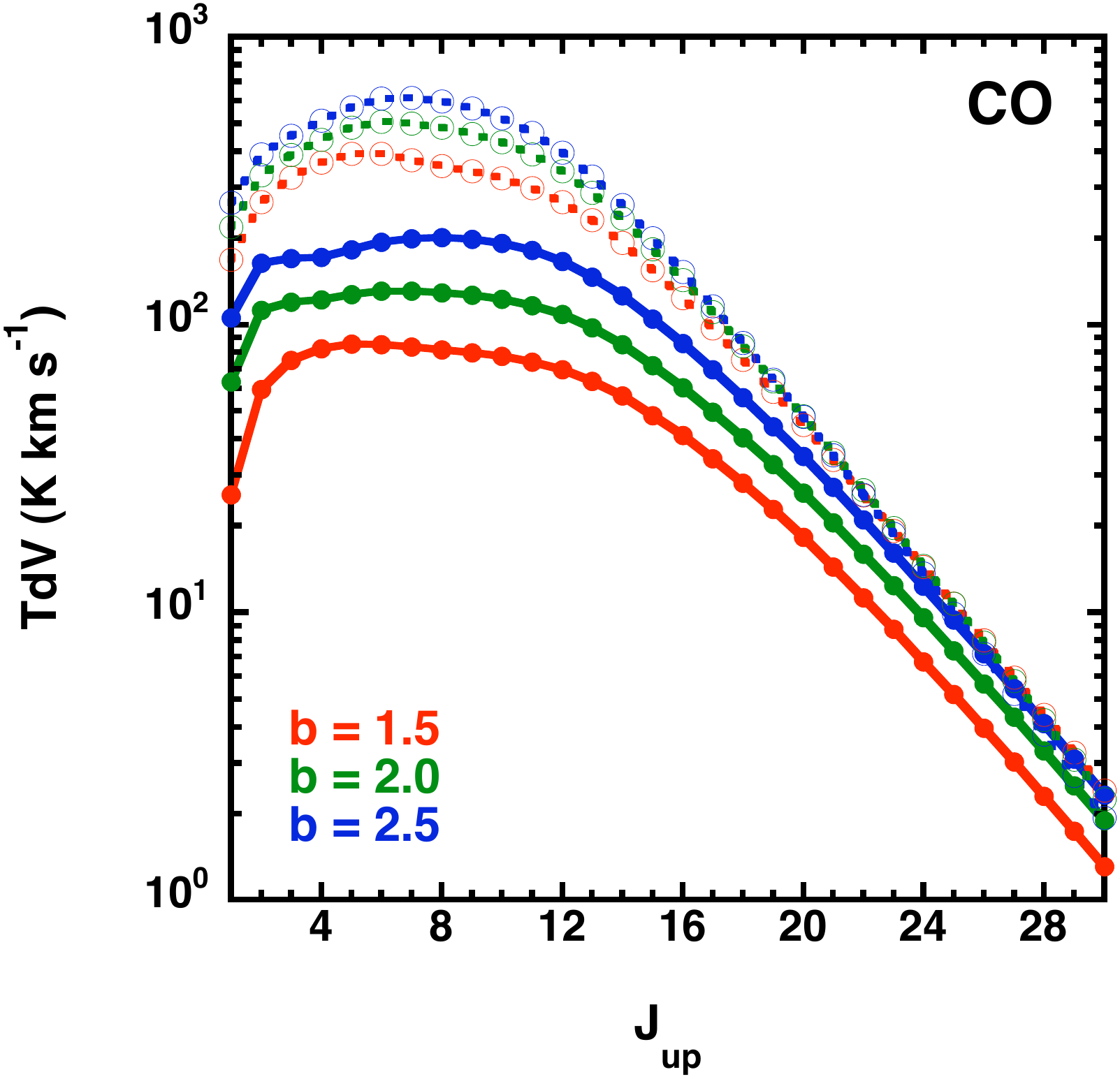}
    \caption{Integrated intensities of the rotational transitions $J_{\rm up} \rightarrow J_{\rm up}-1$ of CO for shocks with $n_{\rm H}~=~10^5$~cm$^{-3}$, $\Vs$~=~30~\kms and $b$~=~1.5 (red), $b$~=~2.0 (green) and $b$~=~2.5 (blue). Full lines: model M1; dotted lines: model M2.
               }
        \label{FigB2_CO}
  \end{figure}

  \begin{figure}
  \centering
  \includegraphics[width=8.5cm]{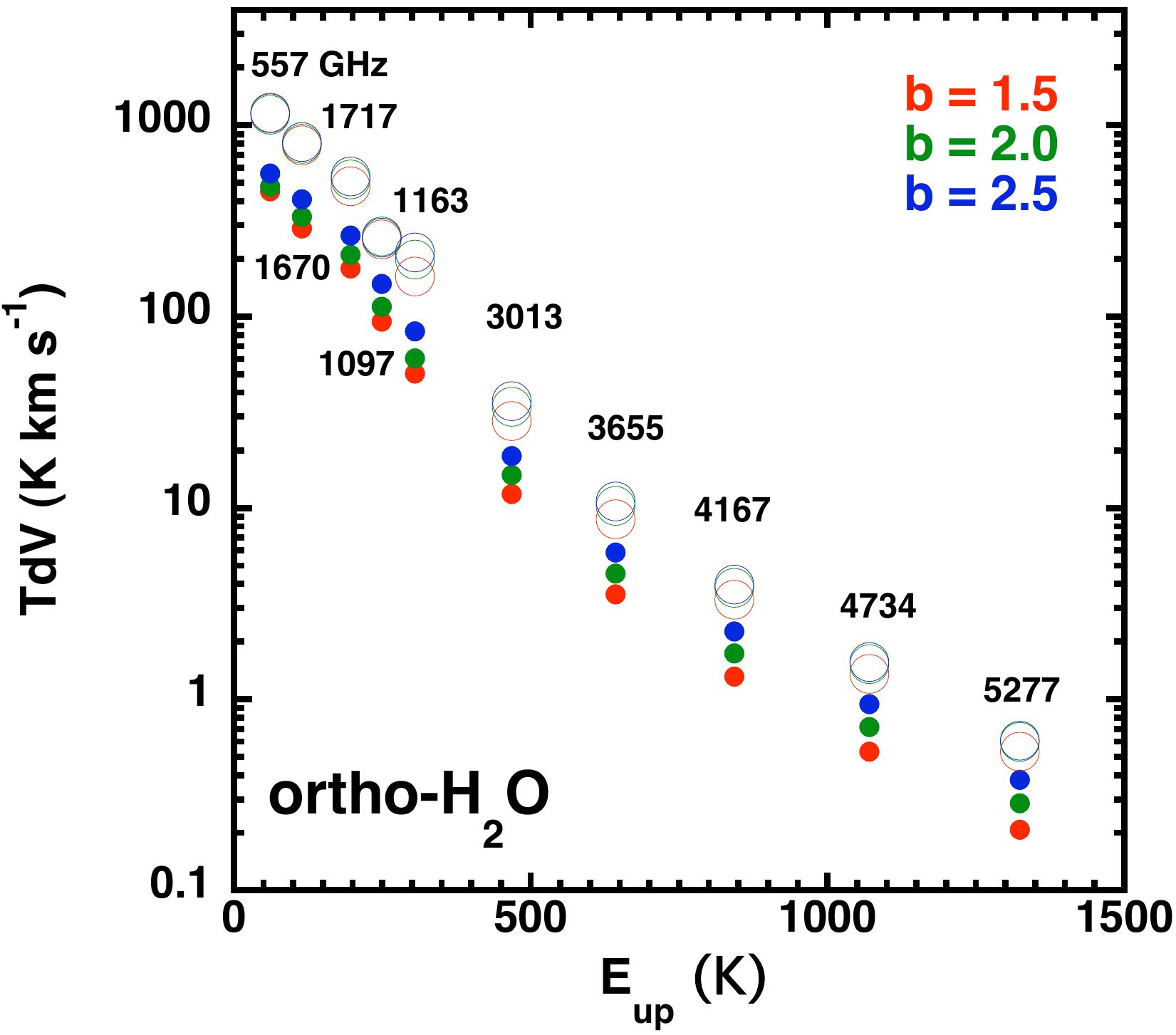}
     \caption{Integrated intensities of selected rotational transitions of ortho-H${}_2$O plotted against the excitation energy of the emitting level, expressed relative to the energy of the $J = 0 = K$ ground state of para-H${}_2$O. Results are shown for shocks with $n_{\rm H}~=~10^5$~cm$^{-3}$, $\Vs$~=~30 \kms and $b$~=~1.5 (red), $b$~=~2.0 (green) and $b$~=~2.5 (blue). Full circles: model M1; open circles: model M2.
                   }
        \label{FigB3_oH2O}
  \end{figure}

  \begin{figure}
  \centering
  \includegraphics[width=8.5cm]{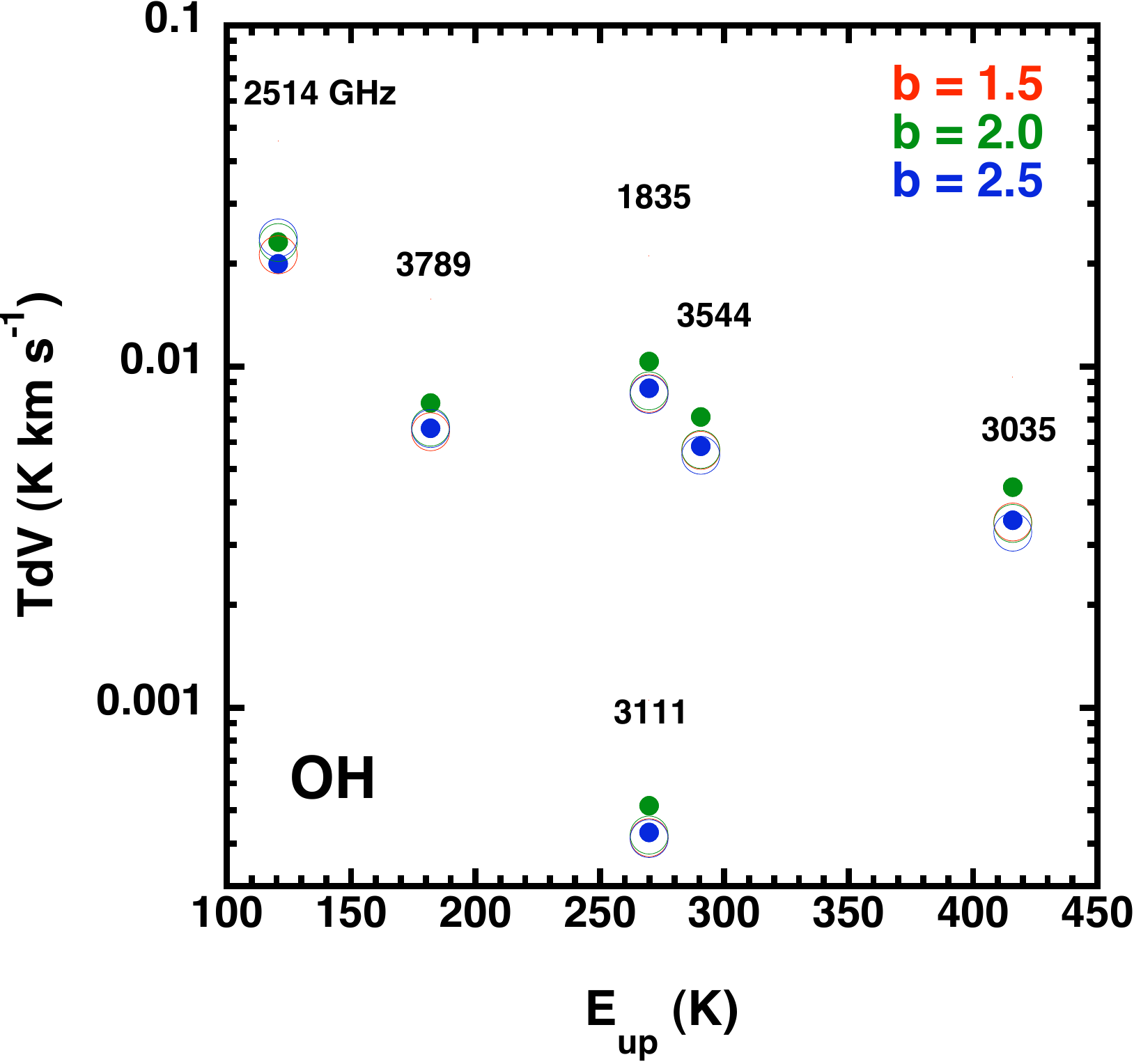}
     \caption{Integrated intensities of the rotational transitions of OH for emitting levels of negative parity, plotted against the excitation energy of the upper level. Results are shown for shocks with $\Vs$~=~30~\kms and $b$~=~1.5 (red), $b$~=~2.0 (green) and $b$~=~2.5 (blue).  Full circles: model M1; open circles: model M2.
             }
        \label{FigB4_OH}
  \end{figure}

The introduction of shattering leads to a reduction of the shock width, and hence to lower column densities of shocked material, and to higher peak temperatures, which affect the chemistry and enhance the fractional abundances of molecules in excited states. Which of these effects prevails is determined by the chemical and spectroscopic properties of the individual molecular species.

\subsection{H${}_2$}

The intensities of pure rotational and ro-vibrational lines of H${}_2$ contain key information on the structure of C-type shock waves, as was demonstrated, for example, by \citet{Wilgenbus:2000p1267}. These lines are optically thin, and their intensities are integrated in parallel with the shock structure, neglecting radiative transfer.

As may be seen from Fig.~\ref{FigB1_H2}, the introduction of shattering leads to a reduction in the computed column densities of the lowest rotational levels of H${}_2$, by approximately an order of magnitude. With increasing energy of the emitting level, the effect of the decrease in the shock width is compensated by the higher peak temperature, and the column densities predicted by the models that include shattering eventually exceed those calculated neglecting shattering. The change-over occurs at energies of the emitting level of $\gtrsim$ 6000 K for $\Vs$~=~20~km\,s${}^{-1}$, $\gtrsim$ 7000 K for $\Vs$~=~30~km\,s${}^{-1}$, and $\gtrsim$ 10,000 K for $\Vs$~=~40~km\,s${}^{-1}$, with the exact values depending on the magnetic field strength. The computed intensities of selected lines of H${}_2$ are given in Table~\ref{tb1}, together with the intensities of forbidden lines of atomic oxygen  (63~$\mu$m and 147~$\mu$m), of atomic sulfur (25~$\mu$m) and of  [C~I] (610~$\mu$m and 370~$\mu$m).

\subsection{CO, H${}_2$O and OH}

Figure~\ref{FigB2_CO} shows the integrated line intensities, $TdV$, of the rotational transitions of CO, plotted against the rotational quantum number, $J$, of the emitting level for shocks with $n_{\rm H}~=~10^5$~cm$^{-3}$ and $\Vs$~=~30~km s${}^{-1}$. The line intensities, which are listed in Tables~\ref{tbCO20}--\ref{tbCO40}, are lower when shattering is included, owing to the reduction in the shock width; this effect is most pronounced at low magnetic field strengths. While the intensities computed with models M2 peak at around $J_{\rm up}$ = 7, those of models M1 peak at higher values of $J_{\rm up}$ and exhibit a plateau extending to $J_{\rm up} \approx 12$. These differences reflect the corresponding excitation conditions. As in the case of SiO (see Sect. 4.2), the peak temperature shows a stronger dependence on the strength of the magnetic field in models that include grain-grain processing. Accordingly, the integrated intensities of highly excited transitions vary with $b$ for models M1.

Similarly to CO, the intensities of lines of H$_2$O also become weaker when shattering is included, owing to the reduced shock width. Figure~\ref{FigB3_oH2O} shows the computed intensities of the lines of ortho-H${}_2$O as a function of the excitation energy of the emitting level. The intensities of all the lines of ortho- and para-H${}_2$O that fall in the Herschel/PACS/HIFI bands are listed in Tables~\ref{tb3}--\ref{tb8}.

      \begin{figure}
   \centering
   \includegraphics[width=9cm]{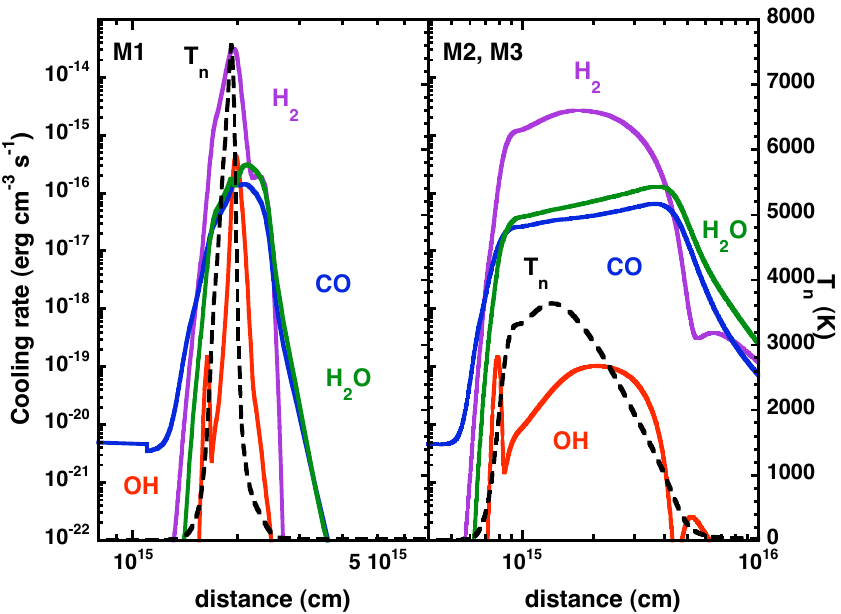}
      \caption{
    The rate of cooling by the principal molecular coolants, H${}_2$ (mauve), H${}_2$O (green), CO (blue) and OH (red) for $\Vs$~=~40~\kms and $b$~=~2 for model M1 (left panel) and M2/M3 (right panel), which are  shown together because the presence of SiO in grain mantles does not affect the cooling of the shock wave. Note the different distance intervals on the $x$-axes.}
                
         \label{Fig_Cooling}
   \end{figure}

Contrary to the behaviour of CO and H${}_2$O,  the line intensities of OH become stronger for the 30 \kms and 40 \kms shocks when shattering is included, as may be seen for $\Vs$~=~30~\kms in Figure~\ref{FigB4_OH}. The lines displayed fall within the Herschel/PACS band; their emitting levels have negative parity. The intensities of all transitions of OH observable with Herschel are listed in Table~\ref{tb2}. The increase in the integrated intensities in models M1 is due to the higher peak temperatures, which favour the conversion of the gas-phase oxygen that is not bound in CO into OH. Again, we see a variation of the integrated line intensities with the magnetic field strength in models M1, associated with the temperature-dependent rate of OH formation.

On the basis of these findings, it is interesting to ask how the allowance for grain-grain processing affects the radiative cooling of the medium. Although the narrower shock width, and hence larger velocity gradients, in scenario M1 might be expected to modify the optical thickness of the lines, and thereby their rate of cooling, we do not detect such an effect in our models, as is demonstrated by Fig.~\ref{Fig_Cooling}. Instead, we see an increase in the contribution of OH to the rate of cooling, owing to the enhanced abundance of gas-phase OH in the hot shocked medium (see the lower panel of Fig.~\ref{Fig4_xSiO}).


\begin{table*}
\caption[]{Selected H${}_2$, [OI], [CI], and [SI] line intensities (erg\,cm${}^{-2}$\,s${}^{-1}$\,sr${}^{-1}$), for shocks with velocities $\Vs$~=~20~\kms (top), $\Vs$~=~30~\kms (middle), and $\Vs$~=~40~\kms (bottom) and the magnetic field strengths listed in Table 1. Results are given for models M1, which include grain-grain processing, and M2 (in parentheses), which neglect grain-grain processing. The preshock density is $n{}_{\rm H}$ = 10${}^5$ cm${}^{-3}$.}

\begin{center}
$$
\begin{tabular}{llllllll}
\hline
           \noalign{\smallskip}

Transition & $\lambda$ ($\mu$m) & v20b1, M1 & (v20b1, M2) & v20b1.5, M1 & (v20b1.5, M2) & v20b2, M1 & (v20b2, M2)\\
           \noalign{\smallskip}
\hline
           \noalign{\smallskip}
H${}_2$ 0-0 S(0) & 28.22 & 1.87e-06 & (6.80e-06) & 3.33e-06 & (9.44e-06) & 5.98e-06 & (1.01e-05) \\
H${}_2$ 0-0 S(1) & 17.04 & 9.59e-05 & (3.48e-04) & 1.74e-04 & (4.50e-04) & 3.31e-04 & (4.84e-04) \\
H${}_2$ 0-0 S(2) & 12.28 & 1.69e-04 & (5.85e-04) & 3.06e-04 & (6.96e-04) & 5.88e-04 & (7.45e-04) \\
H${}_2$ 0-0 S(3) & 9.66 & 1.55e-03 & (4.79e-03) & 2.73e-03 & (5.17e-03) & 4.92e-03 & (5.46e-03) \\
H${}_2$ 0-0 S(4) & 8.02 & 1.11e-03 & (2.89e-03) & 1.85e-03 & (2.79e-03) & 2.93e-03 & (2.85e-03) \\
H${}_2$ 0-0 S(5) & 6.91 & 5.64e-03 & (1.14e-02) & 8.48e-03 & (9.63e-03) & 1.07e-02 & (9.23e-03) \\
H${}_2$ 0-0 S(6) & 6.11 & 2.57e-03 & (3.65e-03) & 3.33e-03 & (2.66e-03) & 2.97e-03 & (2.26e-03) \\
H${}_2$ 0-0 S(7) & 5.51 & 8.75e-03 & (7.71e-03) & 9.09e-03 & (4.78e-03) & 4.90e-03 & (3.35e-03) \\
H${}_2$ 1-0 S(1) & 2.12 & 1.02e-03 & (2.17e-04) & 4.12e-04 & (1.00e-04) & 8.82e-05 & (5.43e-05) \\
          \noalign{\smallskip}
          \hline
          \noalign{\smallskip}

OI 2p${}^4$ ${}^3$P${}_1$ $\rightarrow$ 2p${}^4$ ${}^3$P${}_2$ & 63.1 & 3.72e-06 & (3.93e-06) & 1.03e-05 & (7.57e-06) & 2.02e-05 & (1.27e-05) \\
OI 2p${}^4$ ${}^3$P${}_0$ $\rightarrow$ 2p${}^4$ ${}^3$P${}_1$ & 145.5 & 2.66e-07 & (2.74e-07) & 7.44e-07 & (5.32e-07) & 1.45e-06 & (9.06e-07) \\
          \noalign{\smallskip}
CI ${}^3$P${}_1$ $\rightarrow$ ${}^3$P${}_0$ & 609.75 & 4.84e-11 & (1.11e-10) & 2.21e-10 & (3.36e-10) & 4.68e-10 & (6.45e-10) \\
CI ${}^3$P${}_2$ $\rightarrow$ ${}^3$P${}_1$ & 370.37 & 4.05e-11 & (3.40e-10) & 1.23e-10 & (7.04e-10) & 3.12e-10 & (1.24e-09) \\
           \noalign{\smallskip}
SI 3P J=1 $\rightarrow$ 3P J=2 & 25.25 & 2.81e-05& (7.03e-05) & 3.71e-05 & (9.61e-05) & 8.44e-05 & (1.21e-04)\\

           \noalign{\smallskip}
\hline
	\noalign{\smallskip}
\end{tabular}
$$
\end{center}
\label{tb1}
\end{table*}	


 \begin{table*}

\begin{center}
$$
\begin{tabular}{llllllll}
\hline
           \noalign{\smallskip}

Transition & $\lambda$ ($\mu$m) & v30b1.5, M1& (v30b1.5, M2) & v30b2, M1& (v30b2, M2) & v30b2.5, M1 & (v30b2.5, M2)\\
           \noalign{\smallskip}
\hline
           \noalign{\smallskip}
H${}_2$ 0-0 S(0) & 28.22 & 2.13e-06 & (8.83e-06) & 3.34e-06 & (1.19e-05) & 5.19e-06 & (1.52e-05) \\
H${}_2$ 0-0 S(1) & 17.04 & 1.08e-04 & (4.77e-04) & 1.76e-04 & (6.31e-04) & 2.90e-04 & (7.91e-04) \\
H${}_2$ 0-0 S(2) & 12.28 & 1.90e-04 & (8.79e-04) & 3.24e-04 & (1.13e-03) & 5.54e-04 & (1.38e-03) \\
H${}_2$ 0-0 S(3) & 9.66 & 1.77e-03 & (8.14e-03) & 3.13e-03 & (1.01e-02) & 5.43e-03 & (1.19e-02) \\
H${}_2$ 0-0 S(4) & 8.02 & 1.29e-03& (5.68e-03) & 2.34e-03 & (6.78e-03) & 4.03e-03 & (7.57e-03) \\
H${}_2$ 0-0 S(5) & 6.91 & 6.87e-03 & (2.71e-02) & 1.26e-02 & (3.04e-02) & 2.06e-02 & (3.15e-02) \\
H${}_2$ 0-0 S(6) & 6.11 & 3.31e-03 & (1.10e-02) & 5.96e-03 & (1.13e-02) & 8.99e-03 & (1.06e-02) \\
H${}_2$ 0-0 S(7) & 5.51 & 1.27e-02 & (3.20e-02) & 2.15e-02 & (2.88e-02) & 2.83e-02 & (2.32e-02) \\
H${}_2$ 1-0 S(1) & 2.12 & 6.92e-03& (3.42e-03) & 4.82e-03 & (2.32e-03) & 2.93e-03 & (1.40e-03) \\
          \noalign{\smallskip}
                    \hline
          \noalign{\smallskip}

OI 2p${}^4$ ${}^3$P${}_1$ $\rightarrow$ 2p${}^4$ ${}^3$P${}_2$ & 63.1 & 4.00e-06 & (4.40e-06) & 8.05e-06 & (7.18e-06) & 1.40e-05 & (1.13e-05) \\
OI 2p${}^4$ ${}^3$P${}_0$ $\rightarrow$ 2p${}^4$ ${}^3$P${}_1$ & 145.5 & 2.87e-07 & (2.79e-07) & 5.76e-07 & (4.67e-07) & 1.01e-06 & (7.37e-07) \\
          \noalign{\smallskip}
CI ${}^3$P${}_1$ $\rightarrow$ ${}^3$P${}_0$ & 609.75 & 1.88e-10 & (1.24e-10) & 1.85e-10 & (3.04e-10) & 3.66e-10 & (5.24e-10) \\
CI ${}^3$P${}_2$ $\rightarrow$ ${}^3$P${}_1$ & 370.37 & 1.18e-09 & (3.98e-10) & 1.09e-10 & (6.58e-10) & 2.01e-10 & (9.67e-10) \\
           \noalign{\smallskip}
SI 3P J=1 $\rightarrow$ 3P J=2 & 25.25 & 3.93e-04& (6.10e-04) & 2.56e-04 & (7.34e-04) & 1.56e-04 & (8.24e-04)\\

           \noalign{\smallskip}
\hline
	\noalign{\smallskip}
\end{tabular}
$$
\end{center}
\end{table*}	


 \begin{table*}

\begin{center}
$$
\begin{tabular}{llllllll}
\hline
           \noalign{\smallskip}

Transition & $\lambda$ ($\mu$m) & v40b2, M1 & (v40b2, M2) & v40b2.5, M1 & (v40b2.5, M2) & v40b3, M1 & (v40b3, M2)\\
           \noalign{\smallskip}
\hline
           \noalign{\smallskip}
H${}_2$ 0-0 S(0) & 28.22 & 3.54e-06 & (1.08e-05) & 3.60e-06 & (1.35e-05) & 5.02e-06 & (1.65e-05) \\
H${}_2$ 0-0 S(1) & 17.04 & 1.46e-04 & (5.81e-04) & 1.81e-04 & (7.26e-04) & 2.73e-04 & (8.79e-04) \\
H${}_2$ 0-0 S(2) & 12.28 & 2.03e-04 & (1.07e-03) & 3.21e-04 & (1.34e-03) & 5.21e-04 & (1.62e-03) \\
H${}_2$ 0-0 S(3) & 9.66 & 1.59e-03 & (1.02e-02) & 3.04e-03 & (1.27e-02) & 5.21e-04 & (1.51e-02) \\
H${}_2$ 0-0 S(4) & 8.02 & 1.09e-03& (7.27e-03) & 2.25e-03 & (9.10e-03) & 4.00e-03 & (1.07e-02) \\
H${}_2$ 0-0 S(5) & 6.91 & 5.78e-03 & (3.65e-02) & 1.23e-02 & (4.54e-02) & 2.20e-02 & (5.19e-02) \\
H${}_2$ 0-0 S(6) & 6.11 &2.97e-03 & (1.59e-02) &6.03e-03 & (1.95e-02) & 1.06e-02 & (2.14e-02) \\
H${}_2$ 0-0 S(7) & 5.51 & 1.24e-02 & (5.21e-02) & 2.36e-02 & (6.15e-02) & 3.92e-02 & (6.35e-02) \\
H${}_2$ 1-0 S(1) & 2.12 & 1.34e-02& (1.58e-02) & 1.59e-02 & (1.15e-02) & 1.31e-02 & (8.75e-03) \\
          \noalign{\smallskip}
                    \hline
          \noalign{\smallskip}

OI 2p${}^4$ ${}^3$P${}_1$ $\rightarrow$ 2p${}^4$ ${}^3$P${}_2$ & 63.1 & 3.72e-05 & (3.48e-06) & 7.41e-06 & (5.98e-06) & 1.17e-05 & (9.20e-06) \\
OI 2p${}^4$ ${}^3$P${}_0$ $\rightarrow$ 2p${}^4$ ${}^3$P${}_1$ & 145.5 & 1.36e-06 & (2.44e-07) & 5.34e-07 & (3.96e-07) & 8.42e-07& (5.97e-07) \\
          \noalign{\smallskip}
CI ${}^3$P${}_1$ $\rightarrow$ ${}^3$P${}_0$ & 609.75 & 5.50e-08 & (6.66e-10) & 1.04e-08 & (4.27e-10) & 5.34e-10 & (5.95e-10) \\
CI ${}^3$P${}_2$ $\rightarrow$ ${}^3$P${}_1$ & 370.37 & 4.73e-07 & (2.26e-09) & 8.16e-08 & (1.04e-09) & 1.83e-09 & (1.27e-09) \\
           \noalign{\smallskip}
SI 3P J=1 $\rightarrow$ 3P J=2 & 25.25 & 1.35e-03& (1.43e-03) & 7.07e-04 & (1.37e-03) & 5.97e-04 & (1.40e-03)\\

           \noalign{\smallskip}
\hline
	\noalign{\smallskip}
  \end{tabular}
$$
\end{center}
\end{table*}


 \begin{table*}
\caption[]{ Intensities of CO lines (erg\,cm${}^{-2}$\,s${}^{-1}$\,sr${}^{-1}$) for shocks with velocity $\Vs$~=~20~\kms and the magnetic field strengths listed in Table~1. Results are given for models M1, which include grain-grain processing, and M2 (in parentheses), which neglect grain-grain processing. The preshock density is $n{}_{\rm H}$ = 10${}^5$ cm${}^{-3}$.}

\begin{center}
$$
\begin{tabular}{lllllllll}
\hline
           \noalign{\smallskip}

Transition & Freq. (GHz) & $\lambda$ ($\mu$m) & v20b1, M1 & (v20b1, M2) & v20b1.5, M1 & (v20b1.5, M2) & v20b2, M1 &(v20b2, M2)\\          
 \noalign{\smallskip}
\hline
           \noalign{\smallskip}
CO (1--0) & 115.27 & 2600.7 & 3.47e-08 & (1.85e-07) & 9.49e-08 & (2.63e-07) & 1.47e-07 & (2.69e-07) \\ 
CO (2--1) & 230.54 & 1300.4 & 6.39e-07 & (2.41e-06) & 1.28e-06 & (3.23e-06) & 1.76e-06 & (3.14e-06) \\ 
CO (3--2) & 345.80 & 866.96 & 2.72e-06 & (1.01e-05) & 4.69e-06 & (1.30e-05) & 6.33e-06 & (1.22e-05) \\ 
CO (4--3) & 461.04 & 650.25 & 7.01e-06 & (2.73e-05) & 1.15e-05 & (3.50e-05) & 1.57e-05 & (3.24e-05) \\ 
CO (5--4) & 576.27 & 520.23 & 1.41e-05 & (5.78e-05) & 2.36e-05 & (7.55e-05) & 3.40e-05 & (7.07e-05) \\ 
CO (6--5) & 691.47 & 433.55 & 2.42e-05 & (1.00e-04) & 4.22e-05 & (1.36e-04) & 6.46e-05 & (1.30e-04) \\ 
CO (7--6) & 806.65 & 371.65 & 3.76e-05 & (1.51e-04) & 6.71e-05 & (2.09e-04) & 1.09e-04 & (2.06e-04) \\ 
CO (8--7) & 921.80 & 325.22 & 5.51e-05 & (2.13e-04) & 9.84e-05 & (2.91e-04) & 1.66e-04 & (2.94e-04) \\ 
CO (9--8) & 1036.9 & 289.12 & 7.68e-05 & (2.83e-04) & 1.36e-04 & (3.76e-04) & 2.32e-04 & (3.86e-04) \\ 
CO (10--9) & 1152.0 & 260.24 & 1.02e-04 & (3.54e-04) & 1.76e-04 & (4.52e-04) & 3.01e-04 & (4.70e-04) \\ 
CO (11--10) & 1267.0 & 236.61 & 1.29e-04 & (4.16e-04) & 2.17e-04 & (5.09e-04) & 3.63e-04 & (5.31e-04) \\ 
CO (12--11) & 1382.0 & 216.93 & 1.56e-04 & (4.57e-04) & 2.52e-04 & (5.38e-04) & 4.09e-04 & (5.62e-04) \\ 
CO (13--12) & 1496.9 & 200.27 & 1.78e-04 & (4.72e-04) & 2.75e-04 & (5.33e-04) & 4.32e-04 & (5.58e-04) \\ 
CO (14--13) & 1611.8 & 186.00 & 1.94e-04 & (4.64e-04) & 2.87e-04 & (5.03e-04) & 4.34e-04 & (5.25e-04) \\ 
CO (15--14) & 1726.6 & 173.63 & 2.00e-04 & (4.29e-04) & 2.83e-04 & (4.47e-04) & 4.11e-04 & (4.65e-04) \\ 
CO (16--15) & 1841.3 & 162.81 & 2.02e-04 & (3.94e-04) & 2.75e-04 & (3.95e-04) & 3.84e-04 & (4.10e-04) \\ 
CO (17--16) & 1956.0 & 153.27 & 1.98e-04 & (3.52e-04) & 2.59e-04 & (3.42e-04) & 3.49e-04 & (3.52e-04) \\ 
CO (18--17) & 2070.6 & 144.78 & 1.89e-04 & (3.09e-04) & 2.39e-04 & (2.91e-04) & 3.09e-04 & (2.97e-04) \\ 
CO (19--18) & 2185.1 & 137.20 & 1.78e-04 & (2.67e-04) & 2.17e-04 & (2.44e-04) & 2.70e-04 & (2.48e-04) \\ 
CO (20--19) & 2299.6 & 130.37 & 1.64e-04 & (2.28e-04) & 1.94e-04 & (2.03e-04) & 2.32e-04 & (2.04e-04) \\ 
CO (21--20) & 2413.9 & 124.19 & 1.47e-04 & (1.91e-04) & 1.70e-04 & (1.66e-04) & 1.96e-04 & (1.65e-04) \\ 
CO (22--21) & 2528.2 & 118.58 & 1.31e-04 & (1.59e-04) & 1.47e-04 & (1.34e-04) & 1.63e-04 & (1.33e-04) \\ 
CO (23--22) & 2642.3 & 113.46 & 1.15e-04 & (1.31e-04) & 1.26e-04 & (1.08e-04) & 1.34e-04 & (1.06e-04) \\ 
CO (24--23) & 2756.4 & 108.76 & 9.98e-05 & (1.07e-04) & 1.07e-04 & (8.63e-05) & 1.09e-04 & (8.34e-05) \\ 
CO (25--24) & 2870.3 & 104.44 & 8.58e-05 & (8.66e-05) & 8.98e-05 & (6.85e-05) & 8.83e-05 & (6.54e-05) \\ 
CO (26--25) & 2984.2 & 100.46 & 7.32e-05 & (6.98e-05) & 7.50e-05 & (5.41e-05) & 7.08e-05 & (5.10e-05) \\ 
CO (27--26) & 3097.9 & 96.772 & 6.20e-05 & (5.60e-05) & 6.23e-05 & (4.25e-05) & 5.63e-05 & (3.95e-05) \\ 
CO (28--27) & 3211.5 & 93.348 & 5.21e-05 & (4.46e-05) & 5.13e-05 & (3.32e-05) & 4.44e-05 & (3.04e-05) \\ 
CO (29--28) & 3325.0 & 90.162 & 4.35e-05 & (3.53e-05) & 4.20e-05 & (2.58e-05) & 3.48e-05 & (2.33e-05) \\ 
CO (30--29) & 3438.4 & 87.190 & 3.59e-05 & (2.78e-05) & 3.40e-05 & (1.99e-05) & 2.70e-05 & (1.77e-05) \\ 
CO (31--30) & 3551.6 & 84.410 & 2.95e-05 & (2.17e-05) & 2.74e-05 & (1.52e-05) & 2.08e-05 & (1.33e-05) \\ 
CO (32--31) & 3664.7 & 81.805 & 2.40e-05 & (1.68e-05) & 2.19e-05 & (1.16e-05) & 1.58e-05 & (9.96e-06) \\ 
CO (33--32) & 3777.6 & 79.359 & 1.92e-05 & (1.29e-05) & 1.72e-05 & (8.71e-06) & 1.19e-05 & (7.37e-06) \\ 
CO (34--33) & 3890.4 & 77.058 & 1.52e-05 & (9.72e-06) & 1.34e-05 & (6.48e-06) & 8.87e-06 & (5.39e-06) \\ 
CO (35--34) & 4003.1 & 74.889 & 1.18e-05 & (7.21e-06) & 1.02e-05 & (4.73e-06) & 6.46e-06 & (3.87e-06) \\ 
CO (36--35) & 4115.6 & 72.842 & 8.96e-06 & (5.22e-06) & 7.60e-06 & (3.37e-06) & 4.60e-06 & (2.71e-06) \\ 
CO (37--36) & 4228.0 & 70.907 & 6.52e-06 & (3.64e-06) & 5.43e-06 & (2.32e-06) & 3.15e-06 & (1.84e-06) \\ 
CO (38--37) & 4340.1 & 69.074 & 4.46e-06 & (2.39e-06) & 3.65e-06 & (1.50e-06) & 2.03e-06 & (1.17e-06) \\ 
CO (39--38) & 4452.2 & 67.336 & 2.72e-06 & (1.41e-06) & 2.19e-06 & (8.72e-07) & 1.17e-06 & (6.69e-07) \\ 
CO (40--39) & 4564.0 & 65.686 & 1.25e-06 & (6.22e-07) & 9.89e-07 & (3.81e-07) & 5.10e-07 & (2.88e-07)\\
\hline
           \noalign{\smallskip}
\end{tabular}
$$
\end{center}
\label{tbCO20}
\end{table*}

 \begin{table*}
\caption[]{ Intensities of CO lines (erg\,cm${}^{-2}$\,s${}^{-1}$\,sr${}^{-1}$) for shocks with velocity $\Vs$~=~30~\kms and the magnetic field strengths listed in Table~1. Results are given for models M1, which include grain-grain processing, and M2 (in parentheses), which neglect grain-grain processing. The preshock density is $n{}_{\rm H}$ = 10${}^5$ cm${}^{-3}$.}

\begin{center}
$$
\begin{tabular}{lllllllll}
\hline
           \noalign{\smallskip}

Transition & Freq. (GHz) & $\lambda$ ($\mu$m) & v30b1.5, M1 & (v30b1.5, M2) & v30b2, M1 & (v30b2, M2) & v30b2.5, M1 & (v30b2.5, M2)\\
           \noalign{\smallskip}
\hline
           \noalign{\smallskip}
           
CO (1--0) & 115.27 & 2600.7 & 4.01e-08 & (2.64e-07) & 9.93e-08 & (3.42e-07) & 1.65e-07 & (4.14e-07) \\ 
CO (2--1) & 230.54 & 1300.4 & 7.45e-07 & (3.35e-06) & 1.41e-06 & (4.14e-06) & 2.05e-06 & (4.88e-06) \\ 
CO (3--2) & 345.80 & 866.96 & 3.19e-06 & (1.37e-05) & 5.06e-06 & (1.65e-05) & 7.19e-06 & (1.92e-05) \\ 
CO (4--3) & 461.04 & 650.25 & 8.29e-06 & (3.68e-05) & 1.22e-05 & (4.39e-05) & 1.72e-05 & (5.12e-05) \\ 
CO (5--4) & 576.27 & 520.23 & 1.68e-05 & (7.71e-05) & 2.49e-05 & (9.47e-05) & 3.57e-05 & (1.12e-04) \\ 
CO (6--5) & 691.47 & 433.55 & 2.89e-05 & (1.33e-04) & 4.42e-05 & (1.71e-04) & 6.54e-05 & (2.07e-04) \\ 
CO (7--6) & 806.65 & 371.65 & 4.49e-05 & (2.01e-04) & 7.03e-05 & (2.69e-04) & 1.07e-04 & (3.30e-04) \\ 
CO (8--7) & 921.80 & 325.22 & 6.56e-05 & (2.86e-04) & 1.04e-04 & (3.88e-04) & 1.61e-04 & (4.79e-04) \\ 
CO (9--8) & 1036.9 & 289.12 & 9.13e-05 & (3.90e-04) & 1.44e-04 & (5.26e-04) & 2.26e-04 & (6.46e-04) \\ 
CO (10--9) & 1152.0 & 260.24 & 1.22e-04 & (5.06e-04) & 1.92e-04 & (6.74e-04) & 3.00e-04 & (8.15e-04) \\ 
CO (11--10) & 1267.0 & 236.61 & 1.55e-04 & (6.22e-04) & 2.43e-04 & (8.13e-04) & 3.77e-04 & (9.65e-04) \\ 
CO (12--11) & 1382.0 & 216.93 & 1.88e-04 & (7.22e-04) & 2.93e-04 & (9.23e-04) & 4.48e-04 & (1.07e-03) \\ 
CO (13--12) & 1496.9 & 200.27 & 2.18e-04 & (7.91e-04) & 3.35e-04 & (9.86e-04) & 5.03e-04 & (1.12e-03) \\ 
CO (14--13) & 1611.8 & 186.00 & 2.42e-04 & (8.26e-04) & 3.66e-04 & (1.00e-03) & 5.40e-04 & (1.12e-03) \\ 
CO (15--14) & 1726.6 & 173.63 & 2.55e-04 & (8.14e-04) & 3.80e-04 & (9.65e-04) & 5.50e-04 & (1.05e-03) \\ 
CO (16--15) & 1841.3 & 162.81 & 2.63e-04 & (7.89e-04) & 3.86e-04 & (9.15e-04) & 5.49e-04 & (9.77e-04) \\ 
CO (17--16) & 1956.0 & 153.27 & 2.62e-04 & (7.45e-04) & 3.81e-04 & (8.46e-04) & 5.34e-04 & (8.86e-04) \\ 
CO (18--17) & 2070.6 & 144.78 & 2.55e-04 & (6.87e-04) & 3.68e-04 & (7.65e-04) & 5.07e-04 & (7.87e-04) \\ 
CO (19--18) & 2185.1 & 137.20 & 2.43e-04 & (6.23e-04) & 3.48e-04 & (6.81e-04) & 4.73e-04 & (6.89e-04) \\ 
CO (20--19) & 2299.6 & 130.37 & 2.27e-04 & (5.57e-04) & 3.24e-04 & (5.98e-04) & 4.35e-04 & (5.96e-04) \\ 
CO (21--20) & 2413.9 & 124.19 & 2.07e-04 & (4.88e-04) & 2.94e-04 & (5.16e-04) & 3.91e-04 & (5.06e-04) \\ 
CO (22--21) & 2528.2 & 118.58 & 1.86e-04 & (4.23e-04) & 2.64e-04 & (4.40e-04) & 3.47e-04 & (4.25e-04) \\ 
CO (23--22) & 2642.3 & 113.46 & 1.65e-04 & (3.62e-04) & 2.34e-04 & (3.72e-04) & 3.05e-04 & (3.54e-04) \\ 
CO (24--23) & 2756.4 & 108.76 & 1.45e-04 & (3.08e-04) & 2.05e-04 & (3.11e-04) & 2.65e-04 & (2.93e-04) \\ 
CO (25--24) & 2870.3 & 104.44 & 1.25e-04 & (2.60e-04) & 1.78e-04 & (2.59e-04) & 2.28e-04 & (2.40e-04) \\ 
CO (26--25) & 2984.2 & 100.46 & 1.08e-04 & (2.18e-04) & 1.54e-04 & (2.14e-04) & 1.95e-04 & (1.96e-04) \\ 
CO (27--26) & 3097.9 & 96.772 & 9.19e-05 & (1.81e-04) & 1.32e-04 & (1.76e-04) & 1.66e-04 & (1.58e-04) \\ 
CO (28--27) & 3211.5 & 93.348 & 7.78e-05 & (1.50e-04) & 1.12e-04 & (1.43e-04) & 1.39e-04 & (1.27e-04) \\ 
CO (29--28) & 3325.0 & 90.162 & 6.53e-05 & (1.23e-04) & 9.40e-05 & (1.16e-04) & 1.17e-04 & (1.02e-04) \\ 
CO (30--29) & 3438.4 & 87.190 & 5.43e-05 & (1.00e-04) & 7.85e-05 & (9.33e-05) & 9.66e-05 & (8.07e-05) \\ 
CO (31--30) & 3551.6 & 84.410 & 4.49e-05 & (8.08e-05) & 6.51e-05 & (7.44e-05) & 7.94e-05 & (6.35e-05) \\ 
CO (32--31) & 3664.7 & 81.805 & 3.67e-05 & (6.47e-05) & 5.34e-05 & (5.89e-05) & 6.46e-05 & (4.95e-05) \\ 
CO (33--32) & 3777.6 & 79.359 & 2.97e-05 & (5.12e-05) & 4.33e-05 & (4.60e-05) & 5.19e-05 & (3.81e-05) \\ 
CO (34--33) & 3890.4 & 77.058 & 2.37e-05 & (4.00e-05) & 3.45e-05 & (3.55e-05) & 4.11e-05 & (2.90e-05) \\ 
CO (35--34) & 4003.1 & 74.889 & 1.85e-05 & (3.06e-05) & 2.70e-05 & (2.68e-05) & 3.19e-05 & (2.16e-05) \\ 
CO (36--35) & 4115.6 & 72.842 & 1.41e-05 & (2.29e-05) & 2.06e-05 & (1.98e-05) & 2.42e-05 & (1.57e-05) \\ 
CO (37--36) & 4228.0 & 70.907 & 1.03e-05 & (1.64e-05) & 1.51e-05 & (1.40e-05) & 1.76e-05 & (1.10e-05) \\ 
CO (38--37) & 4340.1 & 69.074 & 7.11e-06 & (1.11e-05) & 1.04e-05 & (9.36e-06) & 1.20e-05 & (7.26e-06) \\ 
CO (39--38) & 4452.2 & 67.336 & 4.37e-06 & (6.68e-06) & 6.39e-06 & (5.58e-06) & 7.31e-06 & (4.27e-06) \\ 
CO (40--39) & 4564.0 & 65.686 & 2.02e-06 & (3.03e-06) & 2.95e-06 & (2.50e-06) & 3.35e-06 & (1.89e-06) \\ 

\hline
           \noalign{\smallskip}
\end{tabular}
$$
\end{center}
\label{tbCO30}
\end{table*}

 \begin{table*}
\caption[]{ Intensities of CO lines (erg\,cm${}^{-2}$\,s${}^{-1}$\,sr${}^{-1}$) for shocks with velocity $\Vs$~=~40~\kms and the magnetic field strengths listed in Table~1. Results are given for models M1, which include grain-grain processing, and M2 (in parentheses), which neglect grain-grain processing. The preshock density is $n{}_{\rm H}$ = 10${}^5$ cm${}^{-3}$.}

\begin{center}
$$
\begin{tabular}{lllllllll}
\hline
           \noalign{\smallskip}

Transition & Freq. (GHz) & $\lambda$ ($\mu$m) & v40b2, M1 & (v40b2, M2) & v40b2.5, M1 & (v40b2.5, M2) & v40b3, M1 & (v40b3, M2)\\
           \noalign{\smallskip}
\hline
           \noalign{\smallskip}
CO (1--0) & 115.27 & 2600.7 & 3.42e-08 & (3.41e-07) & 1.00e-07 & (4.18e-07) & 1.73e-07 & (4.86e-07) \\ 
CO (2--1) & 230.54 & 1300.4 & 6.14e-07 & (4.33e-06) & 1.45e-06 & (5.02e-06) & 2.22e-06 & (5.69e-06) \\ 
CO (3--2) & 345.80 & 866.96 & 2.73e-06 & (1.76e-05) & 5.30e-06 & (1.98e-05) & 7.52e-06 & (2.21e-05) \\ 
CO (4--3) & 461.04 & 650.25 & 7.90e-06 & (4.67e-05) & 1.30e-05 & (5.22e-05) & 1.76e-05 & (5.82e-05) \\ 
CO (5--4) & 576.27 & 520.23 & 1.81e-05 & (9.75e-05) & 2.68e-05 & (1.11e-04) & 3.63e-05 & (1.25e-04) \\ 
CO (6--5) & 691.47 & 433.55 & 3.51e-05 & (1.68e-04) & 4.79e-05 & (1.99e-04) & 6.58e-05 & (2.30e-04) \\ 
CO (7--6) & 806.65 & 371.65 & 6.04e-05 & (2.53e-04) & 7.70e-05 & (3.11e-04) & 1.07e-04 & (3.66e-04) \\ 
CO (8--7) & 921.80 & 325.22 & 9.53e-05 & (3.58e-04) & 1.14e-04 & (4.49e-04) & 1.61e-04 & (5.35e-04) \\ 
CO (9--8) & 1036.9 & 289.12 & 1.40e-04 & (4.91e-04) & 1.59e-04 & (6.17e-04) & 2.27e-04 & (7.36e-04) \\ 
CO (10--9) & 1152.0 & 260.24 & 1.93e-04 & (6.46e-04) & 2.11e-04 & (8.08e-04) & 3.03e-04 & (9.57e-04) \\ 
CO (11--10) & 1267.0 & 236.61 & 2.51e-04 & (8.09e-04) & 2.67e-04 & (1.00e-03) & 3.84e-04 & (1.18e-03) \\ 
CO (12--11) & 1382.0 & 216.93 & 3.11e-04 & (9.61e-04) & 3.23e-04 & (1.18e-03) & 4.64e-04 & (1.37e-03) \\ 
CO (13--12) & 1496.9 & 200.27 & 3.66e-04 & (1.08e-03) & 3.71e-04 & (1.31e-03) & 5.32e-04 & (1.50e-03) \\ 
CO (14--13) & 1611.8 & 186.00 & 4.12e-04 & (1.15e-03) & 4.09e-04 & (1.38e-03) & 5.84e-04 & (1.56e-03) \\ 
CO (15--14) & 1726.6 & 173.63 & 4.39e-04 & (1.17e-03) & 4.29e-04 & (1.38e-03) & 6.09e-04 & (1.54e-03) \\ 
CO (16--15) & 1841.3 & 162.81 & 4.57e-04 & (1.16e-03) & 4.40e-04 & (1.36e-03) & 6.22e-04 & (1.50e-03) \\ 
CO (17--16) & 1956.0 & 153.27 & 4.61e-04 & (1.11e-03) & 4.39e-04 & (1.30e-03) & 6.17e-04 & (1.42e-03) \\ 
CO (18--17) & 2070.6 & 144.78 & 4.52e-04 & (1.04e-03) & 4.26e-04 & (1.21e-03) & 5.98e-04 & (1.31e-03) \\ 
CO (19--18) & 2185.1 & 137.20 & 4.33e-04 & (9.61e-04) & 4.05e-04 & (1.11e-03) & 5.68e-04 & (1.19e-03) \\ 
CO (20--19) & 2299.6 & 130.37 & 4.07e-04 & (8.71e-04) & 3.79e-04 & (1.00e-03) & 5.31e-04 & (1.07e-03) \\ 
CO (21--20) & 2413.9 & 124.19 & 3.74e-04 & (7.73e-04) & 3.46e-04 & (8.86e-04) & 4.85e-04 & (9.36e-04) \\ 
CO (22--21) & 2528.2 & 118.58 & 3.37e-04 & (6.76e-04) & 3.11e-04 & (7.74e-04) & 4.37e-04 & (8.13e-04) \\ 
CO (23--22) & 2642.3 & 113.46 & 3.01e-04 & (5.86e-04) & 2.76e-04 & (6.69e-04) & 3.89e-04 & (6.99e-04) \\ 
CO (24--23) & 2756.4 & 108.76 & 2.65e-04 & (5.02e-04) & 2.42e-04 & (5.74e-04) & 3.43e-04 & (5.96e-04) \\ 
CO (25--24) & 2870.3 & 104.44 & 2.30e-04 & (4.27e-04) & 2.11e-04 & (4.88e-04) & 2.99e-04 & (5.04e-04) \\ 
CO (26--25) & 2984.2 & 100.46 & 1.99e-04 & (3.61e-04) & 1.82e-04 & (4.12e-04) & 2.59e-04 & (4.23e-04) \\ 
CO (27--26) & 3097.9 & 96.772 & 1.70e-04 & (3.03e-04) & 1.56e-04 & (3.45e-04) & 2.23e-04 & (3.53e-04) \\ 
CO (28--27) & 3211.5 & 93.348 & 1.45e-04 & (2.52e-04) & 1.32e-04 & (2.88e-04) & 1.90e-04 & (2.93e-04) \\ 
CO (29--28) & 3325.0 & 90.162 & 1.22e-04 & (2.08e-04) & 1.11e-04 & (2.38e-04) & 1.61e-04 & (2.41e-04) \\ 
CO (30--29) & 3438.4 & 87.190 & 1.03e-04 & (1.71e-04) & 9.31e-05 & (1.95e-04) & 1.35e-04 & (1.97e-04) \\ 
CO (31--30) & 3551.6 & 84.410 & 8.53e-05 & (1.39e-04) & 7.72e-05 & (1.59e-04) & 1.12e-04 & (1.60e-04) \\ 
CO (32--31) & 3664.7 & 81.805 & 7.03e-05 & (1.12e-04) & 6.34e-05 & (1.28e-04) & 9.22e-05 & (1.28e-04) \\ 
CO (33--32) & 3777.6 & 79.359 & 5.73e-05 & (8.96e-05) & 5.15e-05 & (1.02e-04) & 7.50e-05 & (1.02e-04) \\ 
CO (34--33) & 3890.4 & 77.058 & 4.61e-05 & (7.05e-05) & 4.12e-05 & (8.06e-05) & 6.01e-05 & (7.99e-05) \\ 
CO (35--34) & 4003.1 & 74.889 & 3.63e-05 & (5.44e-05) & 3.23e-05 & (6.22e-05) & 4.72e-05 & (6.13e-05) \\ 
CO (36--35) & 4115.6 & 72.842 & 2.79e-05 & (4.09e-05) & 2.47e-05 & (4.68e-05) & 3.61e-05 & (4.59e-05) \\ 
CO (37--36) & 4228.0 & 70.907 & 2.06e-05 & (2.96e-05) & 1.82e-05 & (3.38e-05) & 2.66e-05 & (3.31e-05) \\ 
CO (38--37) & 4340.1 & 69.074 & 1.43e-05 & (2.02e-05) & 1.26e-05 & (2.30e-05) & 1.84e-05 & (2.24e-05) \\ 
CO (39--38) & 4452.2 & 67.336 & 8.88e-06 & (1.22e-05) & 7.73e-06 & (1.40e-05) & 1.13e-05 & (1.35e-05) \\ 
CO (40--39) & 4564.0 & 65.686 & 4.13e-06 & (5.58e-06) & 3.58e-06 & (6.36e-06) & 5.22e-06 & (6.14e-06) \\

\hline
           \noalign{\smallskip}
\end{tabular}
$$
\end{center}
\label{tbCO40}
\end{table*}

\begin{table*}
\caption[]{Intensities of OH lines (erg\,cm${}^{-2}$\,s${}^{-1}$\,sr${}^{-1}$) observable with the PACS instrument on the Herschel Space Observatory, for shocks with velocities $\Vs$~=~20~\kms (top), $\Vs$~=~30~\kms (middle), and $\Vs$~=~40~\kms (bottom) and the magnetic field strengths listed in Table 1. Results are given for models M1, which include grain-grain processing, and M2 (in parentheses), which neglect grain-grain processing. The preshock density is $n{}_{\rm H}$ = 10${}^5$ cm${}^{-3}$.}

\begin{center}
$$
\begin{tabular}{lllllllll}
\hline
           \noalign{\smallskip}

Transition & $\lambda$ ($\mu$m) &  $ E_{\rm up}$ (K)& v20b1, M1 & (v20b1, M2) & v20b1.5, M1 & (v20b1.5, M2) & v20b2, M1 & (v20b2, M2)\\
           \noalign{\smallskip}
\hline
           \noalign{\smallskip}	
1834.8 GHz & 163.39 &  269.8 &1.65e-08 & (1.76e-08) & 2.58e-08& (2.51e-08) & 3.92e-08& (3.03e-08) \\
1839.0 GHz& 163.01 &  270.2 & 2.42e-08& (2.54e-08) & 3.75e-08& (3.58e-08) & 5.65e-08& (4.35e-08) \\
2510.0 GHz & 119.44 &  120.5 & 1.79e-07& (2.71e-07) & 2.88e-07& (4.89e-07) & 4.38e-07& (4.20e-07) \\
2514.0 GHz& 119.23 &  120.7 & 9.78e-08& (1.65e-07) & 1.60e-07& (3.14e-07) & 2.44e-07& (2.50e-07) \\
3035.0 GHz& 98.76 &  415.9 & 3.11e-08& (3.08e-08)& 4.61e-08& (4.01e-08) & 6.78e-08& (5.11e-08) \\
3036.0 GHz& 98.74 &  415.5 & 1.00e-08& (1.00e-08) & 1.50e-08& (1.31e-08) & 2.22e-08& (1.68e-08) \\
3111.0 GHz& 96.36 &  269.8 & 4.01e-09& (4.28e-09) & 6.28e-09& (6.11e-09) & 9.54e-09& (7.36e-09) \\
3114.0 GHz& 96.27 &  270.2 & 5.86e-09& (6.16e-09) & 9.07e-09& (8.67e-09) & 1.37e-08& (1.05e-08) \\
3544.0 GHz& 84.60 &  290.5 & 8.10e-08& (8.30e-08) & 1.24e-07& (1.13e-07) & 1.86e-07& (1.41e-07) \\
3551.0 GHz& 84.42 &  291.2 & 1.94e-08& (2.01e-08) & 3.00e-08& (2.76e-08) & 4.52e-08& (3.44e-08) \\
3786.0 GHz& 79.18 &  181.7 & 1.01e-07& (1.19e-07) & 1.61e-07& (1.88e-07) & 2.47e-07& (2.03e-07) \\
3789.0 GHz& 79.12 &  181.9 & 1.11e-07& (1.27e-07) & 1.76e-07& (1.96e-07) & 2.68e-07& (2.17e-07) \\
4210.0 GHz& 71.22 &  617.9 & 1.51e-08& (1.45e-08) & 2.13e-08& (1.81e-08) & 3.04e-08& (2.31e-08) \\
4212.0 GHz& 71.17 &  617.6 & 2.92e-09& (2.81e-09) & 4.17e-09& (3.53e-09) & 5.96e-09& (4.52e-09) \\
4592.0 GHz& 65.28 &  510.9 & 2.78e-08& (2.71e-08) & 4.05e-08& (3.46e-08) & 5.87e-08& (4.43e-08) \\
4603.0 GHz& 65.13 &  512.1 & 4.03e-09& (3.94e-09) & 5.90e-09& (5.05e-09) & 8.58e-09& (6.47e-09) \\
\hline
           \noalign{\smallskip}
 \end{tabular}
$$
\end{center}
\label{tb2}
\end{table*}

 \begin{table*}

\begin{center}
$$
\begin{tabular}{lllllllll}
\hline
           \noalign{\smallskip}

Transition & $\lambda$ ($\mu$m)&  $E_{\rm up}$ (K) & v30b1.5, M1& (v30b1.5, M2) & v30b2, M1& (v30b2, M2) & v30b2.5, M1 & (v30b2.5, M2)\\
           \noalign{\smallskip}
\hline
           \noalign{\smallskip}
1834.8 GHz & 163.39 &  269.8 & 1.33e-07& (5.28e-08) & 6.54e-08& (5.38e-08) & 5.46e-08& (5.24e-08) \\
1839.0 GHz& 163.01 &  270.2 & 1.98e-07& (7.76e-08) & 9.67e-08& (7.86e-08) &8.00e-08 & (7.62e-08) \\
2510.0 GHz& 119.44 & 120.5 & 4.38e-07& (6.17e-07) & 6.95e-07& (6.58e-07) & 5.94e-07& (6.68e-07) \\
2514.0 GHz& 119.23 &  120.7 & 7.47e-07& (3.46e-07) & 3.77e-07& (3.76e-07) & 3.26e-07& (3.88e-07) \\
3035.0 GHz& 98.76 &  415.9  & 2.66e-07& (1.00e-07)& 1.27e-07& (9.94e-08) & 1.02e-07& (9.38e-08) \\
3036.0 GHz& 98.74 &  415.5 & 8.54e-08& (3.23e-08) & 4.08e-08& (3.21e-08) & 3.28e-08& (3.04e-08) \\
3111.0 GHz& 96.36 &  269.8 & 3.24e-08& (1.28e-08) & 1.59e-08& (1.31e-08) & 1.34e-08& (1.27e-08) \\
3114.0 GHz& 96.27 &  270.2 & 4.80e-08& (1.88e-08) & 2.34e-08& (1.90e-08) & 1.94e-08& (1.84e-08) \\
3544.0 GHz& 84.60 &  290.5 & 6.69e-07& (2.59e-07) & 3.24e-07& (2.61e-07) & 2.66e-07& (2.50e-07) \\
3551.0 GHz& 84.42 &  291.2 & 1.59e-07& (6.20e-08) & 7.73e-08& (6.26e-08) & 6.38e-08& (6.04e-08) \\
3786.0 GHz& 79.18 &  181.7 & 7.87e-07& (3.27e-07) & 3.93e-07& (3.40e-07) & 3.34e-07& (3.39e-07) \\
3789.0 GHz& 79.12 &  181.9 & 8.78e-07& (3.59e-07) & 4.36e-07& (3.71e-07) & 3.67e-07& (3.67e-07) \\
4210.0 GHz& 71.22 &  617.9 & 1.37e-07& (4.97e-08) & 6.34e-08& (4.83e-08) & 4.89e-08& (4.45e-08) \\
4212.0 GHz& 71.17 &  617.6  & 2.64e-08& (9.61e-09) & 1.22e-08& (9.35e-09) & 9.50e-09& (8.65e-09) \\
4592.0 GHz& 65.28&  510.9  & 2.45e-07& (9.07e-08) & 1.15e-07& (8.90e-08) & 9.07e-08& (8.31e-08) \\
4603.0 GHz& 65.13 &  512.1 & 3.53e-08& (1.31e-08) & 1.66e-08& (1.29e-08) & 1.31e-08& (1.21e-08) \\
\hline
           \noalign{\smallskip}
\end{tabular}
$$
\end{center}
\end{table*}

\begin{table*}

\begin{center}
$$
\begin{tabular}{lllllllll}
\hline
           \noalign{\smallskip}

Transition & $\lambda$ ($\mu$m)&  $ E_{\rm up}$  (K)& v40b2, M1 & (v40b2, M2) & v40b2.5, M1 & (v40b2.5, M2) & v40b3, M1 & (v40b3, M2)\\
           \noalign{\smallskip}
\hline
           \noalign{\smallskip}
	
1834.8 GHz & 163.39 &  269.8 & 6.25e-06& (2.49e-07) & 4.66e-07& (1.67e-07) & 2.05e-07& (1.36e-07) \\
1839.0 GHz& 163.01 &  270.2 & 9.23e-06& (3.69e-07) & 6.95e-07& (2.47e-07) & 3.05e-07& (2.01e-07) \\
2510.0 GHz& 119.44 &  120.5 & 6.49e-05& (2.67e-06) & 4.84e-06& (1.86e-06) & 2.15e-06& (1.58e-06) \\
2514.0 GHz& 119.23 &  120.7 & 3.68e-05& (1.45e-06) & 2.60e-06& (1.02e-06) & 1.16e-06& (8.84e-07) \\
3035.0 GHz& 98.76 & 415.9  & 1.23e-05& (4.90e-07)& 9.37e-07& (3.25e-07) & 4.06e-07& (2.60e-07) \\
3036.0 GHz& 98.74 &  415.5 & 4.21e-06& (1.57e-07) & 3.01e-07& (1.04e-07) & 1.30e-07& (8.37e-08) \\
3111.0 GHz& 96.36 &  269.8 & 1.52e-06& (6.06e-08) & 1.13e-07& (4.07e-08) & 4.99e-08& (3.32e-08) \\
3114.0 GHz& 96.27 &  270.2 & 2.24e-06& (8.94e-08) & 1.68e-07& (5.99e-08) & 7.38e-08& (4.87e-08) \\
3544.0 GHz& 84.60 &  290.5 & 3.11e-05& (1.24e-06) & 2.35e-06& (8.30e-07) &1.03e-06& (6.70e-07) \\
3551.0 GHz& 84.42 &  291.2 & 7.93e-06& (2.96e-07) & 5.59e-07& (1.98e-07) & 2.44e-07& (1.60e-07) \\
3786.0 GHz& 79.18 &  181.7 & 3.54e-05& (1.49e-06) & 2.74e-06& (1.02e-06) & 1.22e-06& (8.42e-07) \\
3789.0 GHz& 79.12 &  181.9 & 3.90e-05& (1.66e-06) & 3.06e-06& (1.12e-06) & 1.36e-06& (9.26e-07) \\
4210.0 GHz& 71.22 &  617.9 & 6.52e-06& (2.48e-07) & 4.85-07& (1.63e-07) & 2.07e-07& (1.29e-07) \\
4212.0 GHz& 71.17 &  617.6  & 1.29e-06& (4.79e-08) &9.34e-08 & (3.15e-08) &4.00e-08 & (2.50e-08) \\
4592.0 GHz& 65.28&  510.9  & 1.18e-05& (4.79e-08) & 8.66e-07& (2.96e-07) & 3.73e-07& (2.36e-07) \\
4603.0 GHz& 65.13 &  512.1 & 1.79e-06& (6.46e-08) & 1.25e-07& (4.27e-08) & 5.37e-08& (3.40e-08) \\
           \noalign{\smallskip}
\hline
  \end{tabular}
$$
\end{center}
\end{table*}

  \begin{table*}
\caption[]{Intensities of ortho-H${}_2$O lines (erg\,cm${}^{-2}$\,s${}^{-1}$\,sr${}^{-1}$) observable with the PACS (top) and HIFI (bottom) instruments on the Herschel Space Observatory, for shocks with velocity $\Vs$~=~20~\kms and the magnetic field strengths listed in Table 1. Results are given for models M1, which include grain-grain processing, and M2 (in parentheses), which neglect grain-grain processing. The preshock density is $n{}_{\rm H}$ = 10${}^5$ cm${}^{-3}$.}

\begin{center}
$$
\begin{tabular}{lllllllll}
\hline
           \noalign{\smallskip}

Transition & $\lambda$ ($\mu$m) &  $ E_{\rm up}$ (K)& v20b1, M1 & (v20b1, M2) & v20b1.5, M1 & (v20b1.5, M2) & v20b2, M1 & (v20b2, M2)\\
           \noalign{\smallskip}
\hline
           \noalign{\smallskip}
5500.1 GHz& 54.506 &  732.1 &1.31e-06 & (2.25e-06) & 1.56e-06 & (2.12e-06) & 1.38e-06 & (1.44e-06) \\
5437.8 GHz& 55.131 &  1274. & 5.49e-06 & (8.49e-06) & 6.44e-06 & (7.07e-06) & 5.49e-06 & (4.75e-06) \\
5276.5 GHz& 56.816 &  1324 & 1.97e-05 & (3.03e-05) &2.30e-05  & (2.43e-05) & 1.95e-05 & (1.64e-05) \\
5107.3 GHz& 58.699 &  550.4 & 1.10e-04 & (1.98e-04) & 1.29e-04 & (1.94e-04) & 1.16e-04 & (1.33e-04) \\
4764.0 GHz& 62.928 &  1553 & 2.74e-07 & (4.08e-07) & 3.20e-07 & (3.36e-07) & 2.67e-07 & (2.24e-07) \\
4734.3 GHz& 63.323 &  1071 & 3.62e-05 & (5.86e-05) & 4.26e-05 & (4.87e-05) & 3.72e-05 & (3.32e-05) \\
4600.4 GHz& 65.166 &  795.5 & 2.57e-05 & (4.19e-05) & 3.03e-05 & (3.89e-05) & 2.61e-05 & (2.61e-05) \\
4535.9 GHz& 66.092 &  1013 & 9.11e-06 & (1.44e-05) & 1.07e-05 & (1.29e-05) & 9.17e-06 & (8.64e-06) \\
4512.4 GHz& 66.437 &  410.7 & 1.54e-04 & (2.99e-04) & 1.76e-04 & (2.47e-04) & 1.74e-04 & (1.85e-04) \\
4456.6 GHz& 67.268 &  410.7 & 7.77e-06 & (1.24e-05) & 1.08e-05 & (2.21e-05) & 6.45e-06 & (1.24e-05) \\
4240.2 GHz& 70.702 &  1274 & 7.40e-07 & (1.14e-06) & 8.68e-07 & (9.53e-07) & 7.39e-07 & (6.40e-07) \\
4166.9 GHz& 71.946 &  843.5 & 6.01e-05 & (9.94e-05) & 7.09e-05 & (8.79e-05) & 6.23e-05 & (5.96e-05) \\
4000.2 GHz& 74.944 & 1126 & 2.34e-06 & (3.70e-06) & 2.75e-06 & (3.13e-06) & 2.36e-06 & (2.11e-06) \\
3977.0 GHz& 75.380 &  305.3 & 2.90e-04 & (4.45e-04) & 3.37e-04 & (4.79e-04) & 2.63e-04 & (3.07e-04) \\
3971.0 GHz& 75.495 & 1806& 3.32e-07 & (4.91e-07) & 3.83e-07 & (3.59e-07) & 3.20e-07 & (2.45e-07) \\
3807.3 GHz& 78.742 & 432.2& 1.30e-04 & (2.33e-04) & 1.53e-04 & (2.21e-04) & 1.36e-04 & (1.53e-04) \\
3654.6 GHz& 82.031 & 643.5& 1.07e-04 & (1.81e-04) & 1.26e-04 & (1.70e-04) & 1.11e-04 & (1.15e-04) \\
3536.7 GHz& 84.766 & 1013& 1.61e-06 & (2.56e-06) & 1.90e-06 & (2.29e-06) & 1.62e-06 & (1.53e-06) \\
3167.6 GHz& 94.643 & 795.5& 3.28e-06 & (5.34e-06) &3.87e-06  & (4.97e-06) & 3.34e-06 & (3.33e-06) \\
3165.5 GHz& 94.704 & 702.3& 1.26e-06 & (2.24e-06) & 1.51e-06 & (2.23e-06) & 1.34e-06 & (1.50e-06) \\
3013.2 GHz& 99.492 & 468.1& 1.97e-04 & (3.45e-04) & 2.33e-04 & (3.41e-04) & 2.04e-04 & (2.31e-04) \\
2970.8 GHz& 100.91 & 574.7& 2.80e-05 & (4.71e-05) & 3.32e-05 & (4.70e-05) & 2.84e-05 & (3.13e-05) \\
2774.0 GHz& 108.07 & 194.1& 6.17e-04 & (1.30e-03) & 7.25e-04 & (1.52e-03) & 6.04e-04 & (1.00e-03) \\
2664.6 GHz& 112.51 & 1340& 3.64e-07 & (5.72e-07) & 4.24e-07 & (4.42e-07) & 3.66e-07 & (3.04e-07) \\
2640.5 GHz& 113.54 & 323.5& 4.32e-04 & (7.79e-04) & 5.14e-04 & (8.34e-04) & 4.39e-04 & (5.55e-04) \\
2462.9 GHz& 121.72 & 550.4& 5.88e-06 & (9.46e-06) & 7.77e-06 & (1.16e-05) & 5.79e-06 & (7.11e-06) \\
2344.3 GHz& 127.88 & 1126& 7.24e-07 & (1.14e-06) & 8.51e-07 & (9.68e-07) & 7.30e-07 & (6.52e-07) \\
2264.1 GHz& 132.41 & 432.2& 2.81e-05 & (4.18e-05) & 3.53e-05 & (6.21e-05) & 2.45e-05 & (3.65e-05) \\
2221.8 GHz& 134.93 & 574.7& 1.01e-05 & (1.70e-05) & 1.20e-05 & (1.70e-05) & 1.03e-05 & (1.13e-05) \\
2196.3 GHz& 136.49 & 410.7& 2.79e-05 & (4.79e-05) & 3.62e-05 & (8.13e-05) & 2.36e-05 & (4.71e-05) \\
1918.5 GHz& 156.26 & 642.4& 6.92e-07 & (1.17e-06) & 8.33e-07 & (1.13e-06) & 7.12e-07 & (7.49e-07) \\
1867.7 GHz& 160.51 & 732.1& 9.50e-07 & (1.63e-06) & 1.13e-06 & (1.54e-06) & 9.99e-07 & (1.04e-06) \\
1716.8 GHz& 174.62 & 196.8& 6.14e-04 & (1.34e-03) & 7.21e-04 & (1.31e-03) & 6.77e-04 & (9.16e-04) \\
1669.9 GHz& 179.53 & 114.4& 9.57e-04 & (2.22e-03) & 1.09e-03 & (2.06e-03) & 1.05e-03 & (1.49e-03) \\
1661.0 GHz& 180.49 & 194.1& 1.93e-04 & (4.63e-04) & 2.30e-04 & (5.06e-04) & 2.03e-04 & (3.41e-04) \\
\hline
1162.9 GHz& 257.79 & 305.3& 5.65e-05 & (1.39e-04) & 7.12e-05 & (1.64e-04) & 6.68e-05 & (1.11e-04) \\
1153.1 GHz& 259.98 & 249.4& 1.22e-04 & (2.54e-04) & 1.49e-04 & (3.43e-04) & 1.12e-04 & (2.17e-04) \\
1097.4 GHz& 273.19 & 249.4& 8.72e-05 & (1.96e-04) & 1.04e-04 & (1.71e-04) & 1.07e-04 & (1.25e-04) \\
556.94 GHz& 538.28 & 61.0& 5.61e-05 & (1.32e-04) & 6.01e-05 & (1.21e-04) & 5.73e-05 & (8.89e-05) \\

\hline
           \noalign{\smallskip}
\end{tabular}
$$
\end{center}
\label{tb3}
\end{table*}


  \begin{table*}
\caption[]{Intensities of ortho-H${}_2$O lines (erg\,cm${}^{-2}$\,s${}^{-1}$\,sr${}^{-1}$) observable with the PACS (top) and HIFI (bottom) instruments on the Herschel Space Observatory, for shocks with velocity $\Vs$~=~30~\kms and the magnetic field strengths listed in Table 1. Results are given for models M1, which include grain-grain processing, and M2 (in parentheses), which neglect grain-grain processing. The preshock density is $n{}_{\rm H}$ = 10${}^5$ cm${}^{-3}$.
}

\begin{center}
$$
\begin{tabular}{lllllllll}
\hline
           \noalign{\smallskip}

Transition & $\lambda$ ($\mu$m) &   $ E_{\rm up}$  (K)& v30b1.5, M1& (v30b1.5, M2) & v30b2, M1& (v30b2, M2) & v30b2.5, M1 & (v30b2.5, M2)\\
           \noalign{\smallskip}
\hline
           \noalign{\smallskip}
5500.1 GHz& 54.506 &  732.1 & 2.16e-06 & (5.36e-06) & 2.80e-06 & (6.34e-06) & 3.60e-06 & (6.74e-06) \\
5437.8 GHz& 55.131&  1274 & 8.87e-06 & (2.26e-05) & 1.21e-05 & (2.60e-05) & 1.60e-05 & (2.68e-05) \\
5276.5 GHz& 56.816 &  1324 & 3.13e-05 & (8.00e-05) & 4.30e-05 & (8.99e-05) & 5.69e-05 & (9.14e-05) \\
5107.3 GHz& 58.699 &  550.4 & 1.75e-04 & (4.39e-04) & 2.13e-04 & (4.97e-04) & 2.72e-04 & (5.15e-04) \\
4764.0 GHz& 62.928 &  1553 & 4.44e-07 & (1.13e-06) & 6.15e-07 & (1.32e-06) & 8.16e-07 & (1.36e-06) \\
4734.3 GHz& 63.323 &  1071 & 5.78e-05 & (1.47e-04) & 7.78e-05 & (1.66e-04) & 1.02e-04 & (1.70e-04) \\
4600.4 GHz& 65.166 &  795.5 & 4.28e-05 & (1.05e-04) & 5.60e-05 & (1.25e-04) & 7.23e-05 & (1.33e-04) \\
4535.9 GHz& 66.092 &  1013 & 1.50e-05 & (3.76e-05) & 2.00e-05 & (4.43e-05) & 2.62e-05 & (4.65e-05) \\
4512.4 GHz& 66.437 &  410.7 & 2.35e-04 & (5.69e-04) & 2.82e-04 & (5.97e-04) & 3.60e-04 & (5.97e-04) \\
4456.6 GHz& 67.268 &  410.7  & 1.89e-05 & (3.95e-05) & 2.66e-05 & (6.85e-05) & 3.09e-05 & (8.70e-05) \\
4240.2 GHz& 70.702 &  1274 & 1.20e-06 & (3.04e-06) & 1.64e-06 & (3.51e-06) & 2.16e-06 & (3.62e-06) \\
4166.9 GHz& 71.946 &  843.5 & 9.74e-05 & (2.45e-04) & 1.29e-04 & (2.83e-04) & 1.68e-04 & (2.95e-04) \\
4000.2 GHz& 74.944 & 1126& 3.76e-06 & (9.59e-06) & 5.09e-06 & (1.10e-05) & 6.71e-06 & (1.13e-05) \\
3977.0 GHz& 75.380 &  305.3 & 5.48e-04 & (1.11e-03) & 6.56e-04 & (1.35e-03) & 7.45e-04 & (1.49e-03) \\
3971.0 GHz& 75.495 & 1806 & 5.24e-07 & (1.32e-06) & 7.29e-07 & (1.42e-06) & 9.63e-07 & (1.41e-06) \\
3807.3 GHz& 78.742 & 432.2 & 2.22e-04 & (5.14e-04) & 2.80e-04 & (6.19e-04) & 3.49e-04 & (6.68e-04) \\
3654.6 GHz& 82.031 & 643.5 & 1.76e-04 & (4.35e-04) & 2.27e-04 & (5.11e-04) & 2.92e-04 & (5.41e-04) \\
3536.7 GHz& 84.766 & 1013 & 2.65e-06 & (6.66e-06) & 3.55e-06 & (7.86e-06) & 4.64e-06 & (8.23e-06) \\
3167.6 GHz& 94.643 & 795.5 & 5.47e-06 & (1.35e-05) & 7.15e-06 & (1.60e-05) & 9.23e-06 & (1.69e-05) \\
3165.5 GHz& 94.704 & 702.3 & 2.11e-06 & (5.21e-06) & 2.69e-06 & (6.29e-06) & 3.43e-06 & (6.79e-06) \\
3013.2 GHz& 99.492 & 468.1 & 3.31e-04 & (7.96e-04) & 4.16e-04 & (9.46e-04) & 5.23e-04 & (1.01e-03) \\
2970.8 GHz& 100.91 & 574.7 & 4.87e-05 & (1.14e-04) & 6.22e-05 & (1.40e-04) & 7.79e-05 & (1.52e-04) \\
2774.0 GHz& 108.07 & 194.1 & 1.01e-03 & (2.62e-03) & 1.15e-03 & (3.14e-03) & 1.39e-03 & (3.42e-03) \\
2664.6 GHz& 112.51 & 1340 & 5.73e-07 & (1.46e-06) & 7.81e-07 & (1.58e-06) & 1.03e-06 & (1.58e-06) \\
2640.5 GHz& 113.54 & 323.5 & 7.37e-04 & (1.76e-03) & 8.96e-04 & (2.12e-03) & 1.11e-03 & (2.29e-03) \\
2462.9 GHz& 121.72 & 550.4 & 1.30e-05 & (2.68e-05) & 1.89e-05 & (4.43e-05) & 2.29e-05 & (5.58e-05) \\
2344.3 GHz& 127.88 & 1126 & 1.16e-06 & (2.96e-06) & 1.57e-06 & (3.39e-06) & 2.07e-06 & (3.49e-06) \\
2264.1 GHz& 132.41 & 432.2 & 5.71e-05 & (1.27e-04) & 7.08e-05 & (1.80e-04) & 8.42e-05 & (2.07e-04) \\
2221.8 GHz& 134.93 & 574.7 & 1.76e-05 & (4.13e-05) & 2.24e-05 & (5.05e-05) & 2.81e-05 & (5.47e-05) \\
2196.3 GHz& 136.49 & 410.7 & 5.57e-05 & (1.36e-04) & 6.77e-05 & (1.98e-04) & 7.99e-05 & (2.33e-04) \\
1918.5 GHz& 156.26 & 642.4 & 1.23e-06 & (2.87e-06) & 1.66e-06 & (3.67e-06) & 2.08e-06 & (4.13e-06) \\
1867.7 GHz& 160.51 & 732.1 & 1.57e-06 & (3.89e-06) & 2.03e-06 & (4.60e-06) & 2.62e-06 & (4.90e-06) \\
1716.8 GHz& 174.62 & 196.8 & 9.24e-04 & (2.48e-03) & 1.09e-03 & (2.70e-03) & 1.37e-03 & (2.79e-03) \\
1669.9 GHz& 179.53 & 114.4 & 1.38e-03 & (3.73e-03) & 1.58e-03 & (3.81e-03) & 1.96e-03 & (3.89e-03) \\
1661.0 GHz& 180.49 & 194.1 & 2.93e-04 & (8.45e-04) & 3.41e-04 & (9.58e-04) & 4.19e-04 & (1.03e-03) \\
\hline
1162.9 GHz& 257.79 & 305.3 & 8.11e-05 & (2.61e-04) & 9.72e-05 & (3.20e-04) & 1.35e-04 & (3.48e-04) \\
1153.1 GHz& 259.98 & 249.4 & 2.09e-04 & (5.63e-04) & 2.44e-04 & (7.36e-04) & 2.97e-04 & (8.35e-04) \\
1097.4 GHz& 273.19 & 249.4 & 1.27e-04 & (3.43e-04) & 1.52e-04 & (3.55e-04) & 2.00e-04 & (3.50e-04) \\
556.94 GHz& 538.28 & 61.0 & 7.98e-05 & (2.06e-04) & 8.45e-05 & (2.00e-04) & 9.89e-05 & (2.04e-04) \\
\hline
           \noalign{\smallskip}

\end{tabular}
$$
\end{center}
\label{tb4}
\end{table*}

 \begin{table*}
\caption[]{Intensities of ortho-H${}_2$O lines (erg\,cm${}^{-2}$\,s${}^{-1}$\,sr${}^{-1}$) observable with the PACS (top) and HIFI (bottom) instruments on the Herschel Space Observatory, for shocks with velocity $\Vs$~=~40~\kms and the magnetic field strengths listed in Table 1. Results are given for models M1, which include grain-grain processing, and M2 (in parentheses), which neglect grain-grain processing. The preshock density is $n{}_{\rm H}$ = 10${}^5$ cm${}^{-3}$.}

\begin{center}
$$
\begin{tabular}{lllllllll}
\hline
           \noalign{\smallskip}

Transition & $\lambda$ ($\mu$m) &  $E_{\rm up}$   (K)& v40b2, M1 & (v40b2, M2) & v40b2.5, M1 & (v40b2.5, M2) & v40b3, M1 & (v40b3, M2)\\
           \noalign{\smallskip}
\hline
           \noalign{\smallskip}

5500.1 GHz& 54.506 &  732.1  & 5.58e-06 & (9.15e-06) & 3.78e-06 & (1.08e-05) & 4.89e-06 & (1.19e-05) \\
5437.8 GHz& 55.131&  1274 & 1.47e-05 & (3.87e-05) & 1.53e-05 & (4.56e-05) & 2.14e-05 & (4.94e-05) \\
5276.5 GHz& 56.816 &  1324  & 4.82e-05 & (1.35e-04) & 5.37e-05 & (1.56e-04) & 7.51e-05 & (1.67e-04) \\
5107.3 GHz& 58.699 &  550.4  & 3.39e-04 & (7.11e-04) & 2.75e-04 & (7.70e-04) & 3.44e-04 & (8.04e-04) \\
4764.0 GHz& 62.928 &  1553 & 7.07e-07 & (1.96e-06) & 7.73e-07 & (2.35e-06) & 1.09e-06 & (2.55e-06) \\
4734.3 GHz& 63.323 &  1071 & 9.55e-05 & (2.46e-04) & 9.84e-05 & (2.83e-04) & 1.35e-04 & (3.03e-04) \\
4600.4 GHz& 65.166 &  795.5 & 9.22e-05 & (1.83e-04) & 7.40e-05 & (2.17e-04) & 9.76e-05 & (2.37e-04) \\
4535.9 GHz& 66.092 &  1013 & 2.84e-05 & (6.51e-05) & 2.59e-05 & (7.73e-05) & 3.51e-05 & (8.41e-05) \\
4512.4 GHz& 66.437 &  410.7 & 5.35e-04 & (8.84e-04) & 3.80e-04 & (9.42e-04) & 4.63e-04 & (9.85e-04) \\
4456.6 GHz& 67.268 &  410.7  & 1.27e-04 & (8.96e-05) & 4.90e-05 & (1.41e-04) & 5.51e-05 & (1.74e-04) \\
4240.2 GHz& 70.702 &  1274 & 1.98e-06 & (5.21e-06) & 2.07e-06 & (6.15e-06) & 2.88e-06 & (6.65e-06) \\
4166.9 GHz& 71.946 &  843.5 & 1.83e-04 & (4.15e-04) & 1.66e-04 & (4.81e-04) & 2.22e-04 & (5.19e-04) \\
4000.2 GHz& 74.944 & 1126& 6.26e-06 & (1.62e-05) & 6.44e-06 & (1.89e-05) & 8.87e-06 & (2.03e-05) \\
3977.0 GHz& 75.380 &  305.3 & 1.64e-03 & (2.05e-03) & 9.85e-04 & (2.40e-03) & 1.10e-03 & (2.62e-03) \\
3971.0 GHz& 75.495 & 1806 & 7.12e-07 & (2.20e-06) & 8.98e-07 & (2.48e-06) & 1.27e-06 & (2.62e-06) \\
3807.3 GHz& 78.742 & 432.2 & 6.95e-04 & (8.97e-04) & 4.06e-04 & (1.09e-03) & 4.95e-04 & (1.21e-03) \\
3654.6 GHz& 82.031 & 643.5 & 3.82e-04 & (7.41e-04) & 3.01e-04 & (8.65e-04) & 3.90e-04 & (9.37e-04) \\
3536.7 GHz& 84.766 & 1013 & 5.03e-06 & (1.15e-05) & 4.58e-06 & (1.37e-05) & 6.22e-06 & (1.49e-05) \\
3167.6 GHz& 94.643 & 795.5 & 1.18e-05 & (2.34e-05) & 9.46e-06 & (2.78e-05) & 1.25e-05 & (3.02e-05) \\
3165.5 GHz& 94.704 & 702.3 & 5.90e-06 & (8.94e-06) & 3.70e-06 & (1.07e-05) & 4.67e-06 & (1.18e-05) \\
3013.2 GHz& 99.492 & 468.1 & 8.28e-04 & (1.36e-03) & 5.70e-04 & (1.59e-03) & 7.08e-04 & (1.73e-03) \\
2970.8 GHz& 100.91 & 574.7 & 1.47e-04 & (2.03e-04) & 8.82e-05 & (2.48e-04) & 1.10e-04 & (2.76e-04) \\
2774.0 GHz& 108.07 & 194.1 & 2.17e-03 & (4.21e-03) & 1.56e-03 & (4.71e-03) & 1.79e-03 & (5.06e-03) \\
2664.6 GHz& 112.51 & 1340 & 8.39e-07 & (2.40e-06) & 9.72e-07 & (2.69e-06) & 1.35e-06 & (2.85e-06) \\
2640.5 GHz& 113.54 & 323.5 & 1.78e-03 & (3.00e-03) & 1.24e-03 & (3.45e-03) & 1.48e-03 & (3.73e-03) \\
2462.9 GHz& 121.72 & 550.4 & 7.19e-05 & (6.04e-05) & 3.27e-05 & (9.70e-05) & 3.95e-05 & (1.21e-04) \\
2344.3 GHz& 127.88 & 1126 & 1.94e-06 & (5.02e-06) & 1.99e-06 & (5.84e-06) & 2.74e-06 & (6.28e-06) \\
2264.1 GHz& 132.41 & 432.2 & 1.94e-04 & (2.46e-04) & 1.06e-04 & (3.13e-04) & 1.23e-04 & (3.53e-04) \\
2221.8 GHz& 134.93 & 574.7 & 5.05e-05 & (7.31e-05) & 3.16e-05 & (8.86e-05) & 3.94e-05 & (9.80e-05) \\
2196.3 GHz& 136.49 & 410.7 & 1.60e-04 & (2.56e-04) & 9.58e-05 & (3.20e-04) & 1.10e-04 & (3.57e-04) \\
1918.5 GHz& 156.26 & 642.4 & 5.16e-06 & (5.30e-06) & 2.55e-06 & (7.21e-06) & 3.22e-06 & (8.64e-06) \\
1867.7 GHz& 160.51 & 732.1 & 4.07e-06 & (6.64e-06) & 2.75e-06 & (7.87e-06) & 3.55e-06 & (8.66e-06) \\
1716.8 GHz& 174.62 & 196.8 & 1.62e-03 & (3.74e-03) & 1.39e-03 & (3.94e-03) & 1.69e-03 & (4.10e-03) \\
1669.9 GHz& 179.53 & 114.4 & 2.15e-03 & (5.35e-03) & 1.98e-03 & (5.35e-03) & 2.38e-03 & (5.47e-03) \\
1661.0 GHz& 180.49 & 194.1 & 5.39e-04 & (1.28e-03) & 4.43e-04 & (1.37e-03) & 5.20e-04 & (1.45e-03) \\
\hline
1162.9 GHz& 257.79 & 305.3 & 1.37e-04 & (3.89e-04) & 1.20e-04 & (4.39e-04) & 1.52e-04 & (4.74e-04) \\
1153.1 GHz& 259.98 & 249.4 & 5.07e-04 & (9.47e-04) & 3.38e-04 & (1.12e-03) & 3.89e-04 & (1.24e-03) \\
1097.4 GHz& 273.19 & 249.4 & 2.20e-04 & (5.04e-04) & 1.97e-04 & (5.22e-04) & 2.42e-04 & (5.29e-04) \\
556.94 GHz& 538.28 & 61.0 & 1.14e-04 & (2.84e-04) & 1.06e-04 & (2.71e-04) & 1.20e-04 & (2.74e-04) \\
\hline
           \noalign{\smallskip}

\end{tabular}
$$
\end{center}
\label{tb5}
\end{table*}


 \begin{table*}
\caption[]{Intensities of para-H${}_2$O lines (erg\,cm${}^{-2}$\,s${}^{-1}$\,sr${}^{-1}$) observable with the PACS (top) and HIFI (bottom) instruments on the Herschel Space Observatory, for shocks with velocity $\Vs$~=~20~\kms and the magnetic field strengths listed in Table 1. Results are given for models M1, which include grain-grain processing, and M2 (in parentheses), which neglect grain-grain processing. The preshock density is $n{}_{\rm H}$ = 10${}^5$ cm${}^{-3}$.}

\begin{center}
$$
\begin{tabular}{lllllllll}
\hline
           \noalign{\smallskip}

Transition & $\lambda$ ($\mu$m) &  $ E_{\rm up}$   (K)& v20b1, M1 & (v20b1, M2) & v20b1.5, M1 & (v20b1.5, M2) & v20b2, M1 & (v20b2, M2)\\
           \noalign{\smallskip}
\hline
           \noalign{\smallskip}
5322.5 GHz & 56.325 &  552.3 & 2.46e-05 & (4.10e-05) & 2.91e-05 & (4.81e-05) & 2.37e-05 & (3.10e-05) \\ 
5280.7 GHz & 56.771 &  1324 & 4.69e-06 & (7.15e-06) & 5.49e-06 & (5.91e-06) & 4.63e-06 & (3.96e-06) \\ 
5201.4 GHz & 57.636 &  454.3 & 6.05e-05 & (1.10e-04) & 7.08e-05 & (1.04e-04) & 6.41e-05 & (7.16e-05) \\ 
5194.9 GHz & 57.709 &  1270 & 9.33e-07 & (1.41e-06) & 1.09e-06 & (1.25e-06) & 9.06e-07 & (8.23e-07) \\ 
4997.6 GHz & 59.987 &  1021 & 1.76e-06 & (2.71e-06) & 2.06e-06 & (2.59e-06) & 1.71e-06 & (1.69e-06) \\ 
4850.3 GHz & 61.808 &  552.3 & 3.85e-07 & (6.35e-07) & 4.60e-07 & (7.56e-07) & 3.68e-07 & (4.84e-07) \\ 
4724.0 GHz & 63.457 &  1070 & 8.78e-06 & (1.40e-05) & 1.03e-05 & (1.18e-05) & 8.87e-06 & (8.00e-06) \\ 
4468.6 GHz & 67.089 &  410.4 & 6.05e-05 & (1.16e-04) & 7.04e-05 & (1.13e-04) & 6.54e-05 & (7.90e-05) \\ 
4218.4 GHz & 71.067 &  598.8 & 1.54e-05 & (2.56e-05) & 1.82e-05 & (2.67e-05) & 1.54e-05 & (1.76e-05) \\ 
4190.6 GHz & 71.539 &  843.8 & 1.76e-05 & (2.95e-05) & 2.08e-05 & (2.53e-05) & 1.85e-05 & (1.73e-05) \\ 
3798.3 GHz & 78.928 &  781.1 & 3.35e-06 & (5.31e-06) & 3.95e-06 & (5.66e-06) & 3.27e-06 & (3.66e-06) \\ 
3691.3 GHz & 81.215 &  1021 & 2.36e-07 & (3.63e-07) & 2.77e-07 & (3.48e-07) & 2.29e-07 & (2.27e-07) \\ 
3599.6 GHz & 83.283 &  642.7 & 3.16e-05 & (5.37e-05) & 3.74e-05 & (5.04e-05) & 3.29e-05 & (3.40e-05) \\ 
3331.5 GHz & 89.988 &  296.8 & 5.73e-05 & (9.59e-05) & 6.85e-05 & (1.03e-04) & 5.74e-05 & (6.87e-05) \\ 
3182.2 GHz & 94.209 &  877.8 & 2.05e-07 & (3.32e-07) & 2.44e-07 & (3.57e-07) & 2.05e-07 & (2.32e-07) \\ 
3135.0 GHz & 95.626 &  469.9 & 5.68e-05 & (9.94e-05) & 6.73e-05 & (1.01e-04) & 5.86e-05 & (6.75e-05) \\ 
2968.7 GHz & 100.98 &  195.9 & 1.86e-04 & (3.60e-04) & 2.14e-04 & (4.13e-04) & 1.68e-04 & (2.66e-04) \\ 
2884.3 GHz & 103.94 &  781.1 & 7.63e-07 & (1.21e-06) & 9.01e-07 & (1.29e-06) & 7.45e-07 & (8.34e-07) \\ 
2685.6 GHz & 111.63 &  598.8 & 1.77e-06 & (2.94e-06) & 2.09e-06 & (3.08e-06) & 1.77e-06 & (2.02e-06) \\ 
2631.0 GHz & 113.95 &  725.1 & 1.35e-06 & (2.40e-06) & 1.60e-06 & (1.98e-06) & 1.47e-06 & (1.40e-06) \\ 
2391.6 GHz & 125.35 &  319.5 & 1.12e-04 & (2.04e-04) & 1.33e-04 & (2.18e-04) & 1.14e-04 & (1.46e-04) \\ 
2365.9 GHz & 126.71 &  410.4 & 3.25e-06 & (5.59e-06) & 4.66e-06 & (8.41e-06) & 3.33e-06 & (4.95e-06) \\ 
2164.1 GHz & 138.53 &  204.7 & 2.83e-04 & (5.69e-04) & 3.36e-04 & (6.40e-04) & 2.82e-04 & (4.21e-04) \\ 
2074.4 GHz & 144.52 &  396.4 & 1.10e-05 & (1.84e-05) & 1.31e-05 & (2.17e-05) & 1.04e-05 & (1.39e-05) \\ 
2040.5 GHz & 146.92 &  552.3 & 5.41e-07 & (8.92e-07) & 6.47e-07 & (1.06e-06) & 5.18e-07 & (6.80e-07) \\ 
1919.4 GHz & 156.19 &  296.8 & 2.97e-05 & (5.47e-05) & 3.59e-05 & (7.92e-05) & 2.50e-05 & (4.86e-05) \\ 
1794.8 GHz & 167.03 &  867.3 & 1.81e-07 & (3.03e-07) & 2.16e-07 & (3.08e-07) & 1.87e-07 & (2.04e-07) \\ 
1602.2 GHz & 187.11 &  396.4 & 9.43e-06 & (1.59e-05) & 1.12e-05 & (1.86e-05) & 8.99e-06 & (1.20e-05) \\ 
\hline
1228.8 GHz & 243.97 &  195.9 & 3.48e-05 & (9.50e-05) & 4.45e-05 & (9.36e-05) & 4.62e-05 & (6.63e-05) \\ 
1113.3 GHz & 269.27 &  53.4 & 2.64e-04 & (6.40e-04) & 2.97e-04 & (5.82e-04) & 2.89e-04 & (4.25e-04) \\ 
987.93 GHz & 303.45 &  100.8 & 2.10e-04 & (5.15e-04) & 2.42e-04 & (4.97e-04) & 2.33e-04 & (3.55e-04) \\ 
752.03 GHz & 398.64 &  136.9 & 6.59e-05 & (1.75e-04) & 7.72e-05 & (1.71e-04) & 7.54e-05 & (1.21e-04) \\ 
\hline
           \noalign{\smallskip}
\end{tabular}
$$
\end{center}
\label{tb6}
\end{table*}



  \begin{table*}
\caption[]{Intensities of para-H${}_2$O lines (erg\,cm${}^{-2}$\,s${}^{-1}$\,sr${}^{-1}$) observable with the PACS (top) and HIFI (bottom) instruments on the Herschel Space Observatory, for shocks with velocity $\Vs$~=~30~\kms and the magnetic field strengths listed in Table 1. Results are given for models M1, which include grain-grain processing, and M2 (in parentheses), which neglect grain-grain processing. The preshock density is $n{}_{\rm H}$ = 10${}^5$ cm${}^{-3}$.}

\begin{center}
$$
\begin{tabular}{lllllllll}
\hline
           \noalign{\smallskip}

Transition & $\lambda$ ($\mu$m)&  $ E_{\rm up}$  (K) & v30b1.5, M1& (v30b1.5, M2) & v30b2, M1& (v30b2, M2) & v30b2.5, M1 & (v30b2.5, M2)\\
           \noalign{\smallskip}
\hline
           \noalign{\smallskip}
5322.5 GHz & 56.325 & 552.30 & 4.33e-05 & (1.03e-04) & 5.31e-05 & (1.34e-04) & 6.64e-05 & (1.48e-04) \\ 
5280.7 GHz & 56.771 & 1324.0 & 7.51e-06 & (1.93e-05) & 1.03e-05 & (2.21e-05) & 1.38e-05 & (2.26e-05) \\ 
5201.4 GHz & 57.636 & 454.30 & 9.73e-05 & (2.39e-04) & 1.19e-04 & (2.65e-04) & 1.50e-04 & (2.74e-04) \\ 
5194.9 GHz & 57.709 & 1270.3 & 1.54e-06 & (3.88e-06) & 2.08e-06 & (4.62e-06) & 2.74e-06 & (4.84e-06) \\ 
4997.6 GHz & 59.987 & 1021.0 & 2.97e-06 & (7.39e-06) & 3.94e-06 & (9.09e-06) & 5.15e-06 & (9.70e-06) \\ 
4850.3 GHz & 61.808 & 552.30 & 7.05e-07 & (1.62e-06) & 8.95e-07 & (2.20e-06) & 1.11e-06 & (2.51e-06) \\ 
4724.0 GHz & 63.457 & 1070.0 & 1.41e-05 & (3.60e-05) & 1.90e-05 & (4.10e-05) & 2.51e-05 & (4.21e-05) \\ 
4468.6 GHz & 67.089 & 410.40 & 9.46e-05 & (2.37e-04) & 1.12e-04 & (2.61e-04) & 1.42e-04 & (2.70e-04) \\ 
4218.4 GHz & 71.067 & 598.80 & 2.62e-05 & (6.37e-05) & 3.32e-05 & (7.92e-05) & 4.23e-05 & (8.59e-05) \\ 
4190.6 GHz & 71.539 & 843.80 & 2.82e-05 & (7.16e-05) & 3.73e-05 & (8.08e-05) & 4.87e-05 & (8.35e-05) \\ 
3798.3 GHz & 78.928 & 781.10 & 5.73e-06 & (1.41e-05) & 7.39e-06 & (1.81e-05) & 9.56e-06 & (1.97e-05) \\ 
3691.3 GHz & 81.215 & 1021.0 & 3.98e-07 & (9.91e-07) & 5.29e-07 & (1.22e-06) & 6.91e-07 & (1.30e-06) \\ 
3599.6 GHz & 83.283 & 642.70 & 5.17e-05 & (1.29e-04) & 6.68e-05 & (1.51e-04) & 8.63e-05 & (1.60e-04) \\ 
3331.5 GHz & 89.988 & 296.80 & 1.09e-04 & (2.28e-04) & 1.33e-04 & (2.99e-04) & 1.60e-04 & (3.30e-04) \\ 
3182.2 GHz & 94.209 & 877.80 & 3.47e-07 & (8.74e-07) & 4.48e-07 & (1.12e-06) & 5.89e-07 & (1.22e-06) \\ 
3135.0 GHz & 95.626 & 469.90 & 9.47e-05 & (2.32e-04) & 1.19e-04 & (2.78e-04) & 1.51e-04 & (2.99e-04) \\ 
2968.7 GHz & 100.98 & 195.90 & 3.21e-04 & (7.66e-04) & 3.69e-04 & (8.96e-04) & 4.19e-04 & (9.80e-04) \\ 
2884.3 GHz & 103.94 & 781.10 & 1.31e-06 & (3.23e-06) & 1.69e-06 & (4.12e-06) & 2.18e-06 & (4.49e-06) \\ 
2685.6 GHz & 111.63 & 598.80 & 3.02e-06 & (7.33e-06) & 3.83e-06 & (9.14e-06) & 4.87e-06 & (9.91e-06) \\ 
2631.0 GHz & 113.95 & 725.10 & 2.11e-06 & (5.34e-06) & 2.74e-06 & (5.74e-06) & 3.55e-06 & (5.85e-06) \\ 
2391.6 GHz & 125.35 & 319.50 & 1.92e-04 & (4.57e-04) & 2.34e-04 & (5.55e-04) & 2.89e-04 & (6.03e-04) \\ 
2365.9 GHz & 126.71 & 410.40 & 7.21e-06 & (1.60e-05) & 1.03e-05 & (2.66e-05) & 1.25e-05 & (3.32e-05) \\ 
2164.1 GHz & 138.53 & 204.70 & 4.71e-04 & (1.19e-03) & 5.54e-04 & (1.42e-03) & 6.74e-04 & (1.54e-03) \\ 
2074.4 GHz & 144.52 & 396.40 & 2.05e-05 & (4.55e-05) & 2.57e-05 & (6.10e-05) & 3.10e-05 & (6.94e-05) \\ 
2040.5 GHz & 146.92 & 552.30 & 9.90e-07 & (2.28e-06) & 1.26e-06 & (3.10e-06) & 1.56e-06 & (3.54e-06) \\ 
1919.4 GHz & 156.19 & 296.80 & 5.46e-05 & (1.33e-04) & 6.41e-05 & (1.80e-04) & 7.51e-05 & (2.08e-04) \\ 
1794.8 GHz & 167.03 & 867.30 & 3.01e-07 & (7.63e-07) & 3.89e-07 & (9.46e-07) & 5.08e-07 & (1.02e-06) \\ 
1602.2 GHz & 187.11 & 396.40 & 1.72e-05 & (3.89e-05) & 2.11e-05 & (5.08e-05) & 2.56e-05 & (5.67e-05) \\ 
\hline
1228.8 GHz & 243.97 & 195.90 & 4.90e-05 & (1.61e-04) & 5.99e-05 & (1.80e-04) & 8.46e-05 & (1.84e-04) \\ 
1113.3 GHz & 269.27 & 53.400 & 3.72e-04 & (1.03e-03) & 4.20e-04 & (1.01e-03) & 5.13e-04 & (1.02e-03) \\ 
987.93 GHz & 303.45 & 100.80 & 2.97e-04 & (8.56e-04) & 3.40e-04 & (8.95e-04) & 4.31e-04 & (9.18e-04) \\ 
752.03 GHz & 398.64 & 136.90 & 9.10e-05 & (2.81e-04) & 1.03e-04 & (2.98e-04) & 1.35e-04 & (3.04e-04) \\ 

\hline
           \noalign{\smallskip}
\end{tabular}
$$
\end{center}
\label{tb7}
\end{table*}




\begin{table*}
\caption[]{ Intensities of para-H${}_2$O lines (erg\,cm${}^{-2}$\,s${}^{-1}$\,sr${}^{-1}$) observable with the PACS (top) and HIFI (bottom) instruments on the Herschel Space Observatory, for shocks with velocity $\Vs$~=~40~\kms and the magnetic field strengths listed in Table 1. Results are given for models M1, which include grain-grain processing, and M2 (in parentheses), which neglect grain-grain processing. The preshock density is $n{}_{\rm H}$ = 10${}^5$ cm${}^{-3}$.}

\begin{center}
$$
\begin{tabular}{lllllllll}
\hline
           \noalign{\smallskip}

Transition & $\lambda$ ($\mu$m)&  $ E_{\rm up}$  (K) & v40b2, M1 & (v40b2, M2) & v40b2.5, M1 & (v40b2.5, M2) & v40b3, M1 & (v40b3, M2)\\
           \noalign{\smallskip}
\hline
           \noalign{\smallskip}

5322.5 GHz & 56.325 & 552.30 & 1.19e-04 & (1.86e-04) & 7.42e-05 & (2.27e-04) & 9.05e-05 & (2.50e-04) \\ 
5280.7 GHz & 56.771 & 1324.0 & 1.25e-05 & (3.28e-05) & 1.30e-05 & (3.84e-05) & 1.82e-05 & (4.14e-05) \\ 
5201.4 GHz & 57.636 & 454.30 & 1.94e-04 & (3.86e-04) & 1.55e-04 & (4.16e-04) & 1.93e-04 & (4.33e-04) \\ 
5194.9 GHz & 57.709 & 1270.3 & 2.91e-06 & (6.81e-06) & 2.67e-06 & (8.18e-06) & 3.68e-06 & (8.93e-06) \\ 
4997.6 GHz & 59.987 & 1021.0 & 6.42e-06 & (1.32e-05) & 5.17e-06 & (1.61e-05) & 6.97e-06 & (1.77e-05) \\ 
4850.3 GHz & 61.808 & 552.30 & 3.27e-06 & (3.06e-06) & 1.40e-06 & (4.07e-06) & 1.68e-06 & (4.80e-06) \\ 
4724.0 GHz & 63.457 & 1070.0 & 2.47e-05 & (6.05e-05) & 2.40e-05 & (7.00e-05) & 3.29e-05 & (7.51e-05) \\ 
4468.6 GHz & 67.089 & 410.40 & 1.96e-04 & (3.74e-04) & 1.46e-04 & (3.99e-04) & 1.79e-04 & (4.15e-04) \\ 
4218.4 GHz & 71.067 & 598.80 & 6.55e-05 & (1.12e-04) & 4.51e-05 & (1.35e-04) & 5.73e-05 & (1.49e-04) \\ 
4190.6 GHz & 71.539 & 843.80 & 5.10e-05 & (1.19e-04) & 4.78e-05 & (1.36e-04) & 6.40e-05 & (1.46e-04) \\ 
3798.3 GHz & 78.928 & 781.10 & 1.37e-05 & (2.56e-05) & 9.94e-06 & (3.15e-05) & 1.30e-05 & (3.48e-05) \\ 
3691.3 GHz & 81.215 & 1021.0 & 8.61e-07 & (1.77e-06) & 6.94e-07 & (2.16e-06) & 9.36e-07 & (2.38e-06) \\ 
3599.6 GHz & 83.283 & 642.70 & 1.07e-04 & (2.19e-04) & 8.77e-05 & (2.54e-04) & 1.14e-04 & (2.75e-04) \\ 
3331.5 GHz & 89.988 & 296.80 & 4.09e-04 & (4.30e-04) & 2.10e-04 & (5.44e-04) & 2.38e-04 & (6.11e-04) \\ 
3182.2 GHz & 94.209 & 877.80 & 8.47e-07 & (1.57e-06) & 5.99e-07 & (1.93e-06) & 7.86e-07 & (2.14e-06) \\ 
3135.0 GHz & 95.626 & 469.90 & 2.28e-04 & (3.96e-04) & 1.61e-04 & (4.63e-04) & 2.02e-04 & (5.04e-04) \\ 
2968.7 GHz & 100.98 & 195.90 & 7.60e-04 & (1.28e-03) & 5.30e-04 & (1.41e-03) & 5.83e-04 & (1.52e-03) \\ 
2884.3 GHz & 103.94 & 781.10 & 3.14e-06 & (5.83e-06) & 2.27e-06 & (7.18e-06) & 2.96e-06 & (7.95e-06) \\ 
2685.6 GHz & 111.63 & 598.80 & 7.73e-06 & (1.30e-05) & 5.23e-06 & (1.57e-05) & 6.63e-06 & (1.72e-05) \\ 
2631.0 GHz & 113.95 & 725.10 & 3.67e-06 & (8.53e-06) & 3.51e-06 & (9.32e-06) & 4.60e-06 & (9.83e-06) \\ 
2391.6 GHz & 125.35 & 319.50 & 5.05e-04 & (7.82e-04) & 3.29e-04 & (9.15e-04) & 3.92e-04 & (9.97e-04) \\ 
2365.9 GHz & 126.71 & 410.40 & 3.46e-05 & (3.41e-05) & 1.72e-05 & (5.11e-05) & 2.00e-05 & (6.20e-05) \\ 
2164.1 GHz & 138.53 & 204.70 & 1.05e-03 & (1.96e-03) & 7.60e-04 & (2.20e-03) & 8.84e-04 & (2.36e-03) \\ 
2074.4 GHz & 144.52 & 396.40 & 8.42e-05 & (8.54e-05) & 4.01e-05 & (1.11e-04) & 4.69e-05 & (1.29e-04) \\ 
2040.5 GHz & 146.92 & 552.30 & 4.64e-06 & (4.30e-06) & 1.96e-06 & (5.72e-06) & 2.37e-06 & (6.76e-06) \\ 
1919.4 GHz & 156.19 & 296.80 & 1.61e-04 & (2.38e-04) & 9.42e-05 & (2.91e-04) & 1.05e-04 & (3.26e-04) \\ 
1794.8 GHz & 167.03 & 867.30 & 6.52e-07 & (1.33e-06) & 5.12e-07 & (1.60e-06) & 6.72e-07 & (1.75e-06) \\ 
1602.2 GHz & 187.11 & 396.40 & 5.10e-05 & (7.11e-05) & 3.06e-05 & (8.73e-05) & 3.60e-05 & (9.67e-05) \\ 
\hline
1228.8 GHz & 243.97 & 195.90 & 8.03e-05 & (2.34e-04) & 7.51e-05 & (2.53e-04) & 9.65e-05 & (2.63e-04) \\ 
1113.3 GHz & 269.27 & 53.400 & 5.30e-04 & (1.43e-03) & 5.21e-04 & (1.39e-03) & 6.21e-04 & (1.40e-03) \\ 
987.93 GHz & 303.45 & 100.80 & 4.52e-04 & (1.22e-03) & 4.20e-04 & (1.24e-03) & 5.10e-04 & (1.27e-03) \\ 
752.03 GHz & 398.64 & 136.90 & 1.37e-04 & (3.95e-04) & 1.26e-04 & (4.05e-04) & 1.54e-04 & (4.15e-04) \\ 
\hline
           \noalign{\smallskip}
  \end{tabular}
$$
\end{center}
\label{tb8}
\end{table*}


\end{appendix} 

\end{document}